\documentclass[aps,twocolumn,floats,superscriptaddress,prd,nofootinbib]{revtex4-1}
\usepackage{graphicx, epsfig, bm, amsmath}

\usepackage{color}
\usepackage{hyperref}
\usepackage{bbold}
\begin{document}


\newcommand{\lcdm}{$\Lambda$CDM}

\newcommand{\gpr}{G^{\prime}}

\newcommand{\fnl}{f_{\rm NL}}
\newcommand{\curv}{{\cal R}}
\newcommand{\amp}{{\cal C}}
\newcommand{\ampold}{{\cal C}}

\definecolor{darkgreen}{cmyk}{0.85,0.2,1.00,0.2}
\newcommand{\peter}[1]{\textcolor{red}{[{\bf PA}: #1]}}
\newcommand{\cora}[1]{\textcolor{darkgreen}{[{\bf CD}: #1]}}
\newcommand{\wh}[1]{\textcolor{blue}{[{\bf WH}: #1]}}
\newcommand{\eugene}[1]{\textcolor{green}{[{\bf EL}: #1]}}
\newcommand{\gB}{g_B}
\newcommand{\WP}{W}
\newcommand{\XP}{X}
\newcommand{\B}{B^{\rm Bulk}}
\newcommand{\damp}{{\cal D}} 
\newcommand{\kstep}{k_{f}}
\newcommand{\xstep}{x_{f}}
\newcommand{\etastep}{\eta_{f}}
\newcommand{\ellfeat}{\ell_{f}}
\newcommand{\epsilonstep}{\epsilon_{0}}
\newcommand{\phifeat}{\phi_{f}}
\newcommand{\feat}{f}
\newcommand{\sh}{*}
\newcommand{\GSR}{{\cal R}}
\newcommand{\feature}{}
\newcommand{\reduced}{\frac{6}{5}f_{\rm NL}}

\newcommand{\aap}{Astron. Astrophys.}


\pagestyle{plain}

\title{Non-Gaussianity from Step Features in the Inflationary Potential}

\author{Peter Adshead}
\affiliation{Kavli Institute for Cosmological Physics,  Enrico Fermi Institute, University of Chicago, Illinois 60637, USA}
        
\author{Cora Dvorkin}
\affiliation{Kavli Institute for Cosmological Physics,  Enrico Fermi Institute, University of Chicago, Illinois 60637, USA}
\affiliation{Department of Physics, University of Chicago, Illinois 60637, USA}
\affiliation{School of Natural Sciences, Institute for Advanced Study, Princeton, New Jersey 08540, USA}

\author{  Wayne Hu}
\affiliation{Kavli Institute for Cosmological Physics,  Enrico Fermi Institute, University of Chicago, Illinois 60637, USA}
\affiliation{Department of Astronomy \& Astrophysics, University of Chicago, Illinois 60637, USA}
        
\author{Eugene A. Lim}
\affiliation{Department of Applied Maths and Theoretical Physics, University of Cambridge, Cambridge CB3 0HA, U.K.}

\begin{abstract}

We provide analytic solutions for the power spectrum and bispectrum of curvature fluctuations produced by a step feature in the inflaton potential, valid in the limit that the step is short and sharp. In this limit, the bispectrum is strongly scale dependent and its effective non-linearity attains a large oscillatory amplitude. The perturbations to the curvature power spectrum, on the other hand, remain a small component on top of the usual spectrum of fluctuations generated by slow roll. We utilize our analytic solutions to assess the observability of the predicted non-Gaussian signatures and show that, if present, only very sharp steps on scales larger than $\sim 2$ Gpc are likely to be able to be detected by Planck.  Such  features  are not only consistent with WMAP7 data, but can also improve its likelihood by $2\Delta \ln L \approx 12$ for two extra parameters, the step location and height.
If this improvement were due to a slow roll violating step as considered here, a bispectrum or corresponding polarization power spectrum detection would provide definitive checks as to its primordial origin.

\end{abstract}

\maketitle

\section{Introduction}
\label{sec:intro}

In this paper we employ the generalized slow roll (GSR) approach \cite{Choe:2004zg,Dvorkin:2009ne,Dvorkin:2010dn,Dvorkin:2011ui} to consider the bispectrum of primordial fluctuations generated by a transient violation of slow roll due to a step feature in the inflaton potential. While earlier works in the field have considered the non-Gaussianity from these models, here we explore a hitherto unexplored region of the parameter space where the step in the potential becomes sharp. This region of parameter space is  near the so-called ``decoupling" limit where the perturbations in the power spectrum due to the step remain small while higher order $N$-point functions are parametrically enhanced \cite{Flauger:2010ja, Leblond:2010yq}. 

Features in the inflationary potential have a long history \cite{Starobinsky:1992ts}. They came into vogue as a possible explanation for the apparent low multipole glitch at $\ell \sim 20 -40$ in the angular spectrum of the cosmic microwave background  (CMB) radiation \cite{Peiris:2003ff,Covi:2006ci,Hamann:2007pa, Mortonson:2009qv}. If such a feature arises as a result of a glitch in the primordial potential then corresponding features should arise in the electric-type polarization pattern of the CMB \cite{Mortonson:2009qv}.  More generally, failure to detect consistent patterns in the polarization may even be considered evidence against a single adiabatic degree of freedom as the source of the primordial perturbations  \cite{Dvorkin:2011ui}. 

Beyond two-point statistics such as the angular temperature and polarization power spectra, it is well known that such a glitch in the potential leads to higher $N$-point functions which are enhanced relative to those generated by the otherwise smooth background  \cite{Chen:2006xjb,Chen:2008wn, Hotchkiss:2009pj, Arroja:2011yu, Martin:2011sn}. However, as was shown in our earlier work \cite{Adshead:2011bw}, the bispectrum, or three-point function produced by a step feature that best fits the glitch in the WMAP data \cite{Mortonson:2009qv, Dvorkin:2011ui} falls short of detectability by a very wide margin.

In this work we investigate the space of parameters that correspond to the step becoming short and sharp. As we will see, in this limit, the perturbations in the power spectrum remain small while the reduced bispectrum  or effective non-linearity, $f_{\rm NL}(k)$ becomes large. Despite this seeming largeness, the bispectrum remains  difficult to detect due to 
its oscillatory form.   Nonetheless, for a sufficiently sharp step at sufficiently large scales, an observably large non-Gaussianity is compatible with current power spectrum constraints.

This paper is organized as follows:\ In \S \ref{sec:backandpow} we derive approximate solutions to the evolution of the inflaton on an inflationary potential that undergoes a sharp downward step. These solutions yield  analytic expressions  for the perturbation to the curvature power spectrum in \S \ref{sec:powerspec}. In \S \ref{sec:bispectrumformal} we  review the formalism used to calculate the bispectrum from which we obtain
approximate analytic solutions.
The constraints that the WMAP 7 year power spectrum data  place on the sharp step region of parameter space in  \S \ref{sec:likelihoodan} inform the prospects
 for the detection of the bispectrum in a cosmic variance limited CMB experiment
in \S \ref{sec:bispectrumSNR}. We summarize our findings in \S\ref{sec:discussion}. 
 
 Some  related but separate studies are presented in  the Appendixes. In Appendix \ref{sec:realspace}, we investigate the real space analogs of the power spectrum and bispectrum of curvature fluctuations generated by a sharp-step feature. In Appendix \ref{app:slowrollcorrections}, we calculate the leading order slow roll corrections to our analytic solutions. Finally in Appendix \ref{app:separability}, we demonstrate that our analytic solutions are approximately separable.

Throughout the paper we work in units where the reduced Planck mass $M_{\rm Pl} =(8\pi G)^{-1/2}= 1$ as well as $\hbar=c=1$. 

\section{Background Solution}\label{sec:backandpow}

We consider a homogeneous scalar field minimally coupled to gravity with a potential of the form 
\begin{align}\label{eqn:potential}
V(\phi) = V_{0}(\phi)\left[1+cF\left(\frac{\phifeat-\phi}{d}\right)\right],
\end{align}
where the potential $V_0$ supports slow roll inflation in the limit where $c \rightarrow 0$, and $F(x)$ is a function which transitions from {$-1$ to $+1$} as its argument passes through $x = 0$ {from above} with a characteristic width $\Delta x=1$.   Thus the potential characterizes a step of height $2c$ and half-width $d$.
When an explicit choice is required, we will take
\begin{equation}
F(x) = -\tanh (x),
\label{eqn:tanh}
\end{equation}
as an example in the following sections.

The scalar field obeys the usual equation of motion on the background
\begin{align}\label{eqn:KGeqn}
\ddot{\phi}+3H\dot{\phi} + \frac{d V}{d\phi} = 0,
\end{align}
where $H = \dot a/a$ is the Hubble parameter, and an overdot represents a derivative with respect to cosmic time.

In the limit in which the step is short, $c \ll 1$, its presence in the potential will have only a small effect on the evolution of the inflaton. Specifically, as long as the change in the potential as the field rolls across the step is small compared with the kinetic energy of the background field, then we can think of the effect of the step as a perturbation on the evolution of $\phi$ on the potential $V_{0}$. As the field crosses the step a potential energy $\Delta V \approx c V$ is converted into kinetic energy. Comparing this with the kinetic energy of the inflaton rolling on the background we find 
\begin{align}\label{eqn:pertcond}
\frac{2\Delta V}{{\dot\phi^2}} \approx \frac{ 4 c V}{\dot{\phi}^2} \approx 6\frac{c}{\epsilon_{H}},
\end{align}
where we have introduced the slow roll parameter
\begin{equation}\label{eqn:SReps}
\epsilon_H \equiv - {\dot H \over H^2} = \frac{\dot{\phi}^{2}}{2H^2}.
\end{equation}
Whenever 
\begin{equation}\label{eqn:pertcond2}
c \ll \epsilon_{H}/6, 
\end{equation}
we can treat the effect of the step as a small perturbation on the background described by $V_0$.

With this limit in mind, we look to solve Eq.~(\ref{eqn:KGeqn}) iteratively as follows: taking $6c \ll \epsilon_{H}$, we split the field into a piece zeroth order in $c$ and pieces which we take to be higher order in $c$
\begin{align}\label{eqn:pertdef}
\phi = \phi_{0} + \phi_{1}+\ldots, \quad \epsilon_H= \epsilon_0 + \epsilon_1+\ldots
\end{align}
In this expression, $\phi_{0}$ characterizes the behavior of the field in the absence of the step, i.e.~on the potential $V_{0}$,  while $\phi_{1}$ is the perturbation which is taken to be linear in $c$, and `$\ldots$' denotes terms higher order in an expansion in 
\begin{equation}
\ampold \equiv \frac{6c}{\epsilonstep}.
\end{equation}
While the condition in Eq.~(\ref{eqn:pertcond2}) requires that $\dot{\phi}_{1}$ remains small, it places no restriction on higher derivatives. In particular, we will see that by making the step sharp we can make the acceleration and jerk of the field $\phi_{1}$ arbitrarily large.

It proves useful to write the equation of motion for the perturbation, $\phi_1$, using the value of the background field $\phi_{0}$ as the independent variable. The condition $6c \ll \epsilonstep$ requires that $\dot\phi_{0}\neq0$ and $\phi_{0}(t)$ evolve monotonically in the vicinity of the step which ensures that $\phi_{0}$ is a suitable time variable. Given that on the background potential $V_{0}$, the slow roll approximation yields $\epsilonstep \approx (V'/V)^2/2$ and assuming that $\epsilonstep \approx $ const.  (and so not $\ll |V''/V|$), we obtain the equation of motion for $\phi_{1}$ as
\begin{align} \label{eqn:perteqn}
\frac{d}{d\phi_{0}}\Big(e^{-\frac{3 \phi_{0}}{\sqrt{2\epsilonstep}}}&\frac{d\phi_{1}}{d\phi_{0}}\Big)  = -\frac{\ampold}{4} \left(\frac{\sqrt{2\epsilonstep}}{2}F+\frac{d F}{d\phi_{0}}\right)e^{-\frac{3\phi_{0}}{\sqrt{2\epsilonstep}}},
\end{align}
where  the subscript $_0$ denotes quantities to be evaluated on the unperturbed ($c = 0$) background and we have dropped some terms that are suppressed by slow roll parameters evaluated on the unperturbed background.

For small sharp steps,  the $dF/d\phi_0$ term in Eq.~(\ref{eqn:perteqn}) dominates and
\begin{equation}
e^{- \frac{3\phi_0}{\sqrt{2\epsilonstep}}} \frac{d\phi_1}{d\phi_0}\approx  -\frac{\ampold}{4}
\int d \phi_0 \frac{ d F}{d\phi_0} e^{- \frac{3\phi_0}{\sqrt{2\epsilonstep}}} +{\rm const.}
\end{equation}
The exponential in the integrand is slowly varying across the width of the step and may
be evaluated at $\phifeat$ and taken out of the integral.   Furthermore the integration constant is set by the boundary
condition that $d\phi_1/d\phi_0=0$ before crossing the step and so
\begin{align} \label{eqn:smallphi1cond}
\frac{d\phi_{1}}{d\phi_{0}}
\approx &\frac{\ampold}{4} e^{-\frac{3}{\sqrt{2\epsilonstep}}(\phifeat-\phi_{0})}\left[ 1-F\left(\frac{\phifeat-\phi_{0}}{d}\right)\right].
\end{align}
In principle, we could further integrate to find $\phi_{1}$; however, this is not necessary for this work.

The final step is to reference this solution to conformal time rather than the background
field.   Since the perturbation $\phi_1$ is only important for a short period of time, the variation of the unperturbed ($c=0$) slow roll parameters is negligible. In this approximation, the evolution of the background in conformal time is very simple
\begin{align}
\frac{d\phi_{0}}{d\ln\eta} = \pm\sqrt{2\epsilonstep} \approx \mbox{const.},\nonumber\\
\phi_0-\phifeat = \pm \sqrt{2 \epsilonstep} \ln (\eta/\etastep) ,
\end{align}
where the sign should be chosen depending on the direction the field is rolling -- i.e.~whether one is in a large, or small field inflationary model and one is rolling toward or away from the origin. In this work, we have large field inflationary models in mind, e.g.\ $V_0 = \frac{1}{2}m^2\phi^2$ and will take the positive sign. 

With these relations, we can explicitly solve for the first order changes to the slow roll
parameters as a function of conformal time.   
The slow roll parameters are defined by Eq.~(\ref{eqn:SReps}) and
\begin{align}
\eta_H =  -\frac{\ddot{\phi}}{\dot{\phi}H}, \quad 
\delta_2 = \frac{\dddot{\phi}}{\dot{\phi} H^2} .
\end{align}
Using the results above, we divide the slow roll parameters into a background part and a part that is linear in the perturbation,   $\eta_H = \eta_0+\eta_1$ and $\delta_2 = \delta_{2,0}+\delta_{2,1}$,
\begin{align}
\epsilon_{1} = & \frac{\ampold}{2} \epsilonstep \left( \frac{\eta}{\etastep}\right)^{3}\left\{1-F\left[ - \frac{\sqrt{2\epsilonstep}}{d}\ln\left(\frac{\eta}{\etastep}\right)\right] \right\},\nonumber\\
\eta_{1} = & \frac{\epsilon_1'}{2\epsilonstep},
\qquad 
\delta_{2, 1} =   \eta_1'.
\label{eqn:slowroll}
\end{align}
Here and below  primes denote derivatives with respect to $\ln\eta$.
Notice that $\epsilon_{1}$ remains a tiny correction to the already small $\epsilonstep$, 
of order $\mathcal{O}(\amp)$, as expected.  On the other hand, in principle $\eta_{1}$ and $\delta_{2,1}$ can be arbitrarily large due to the presence of the $1/d$ factors from the derivatives. In what follows we will see that these give rise to a large and strongly scale dependent bispectrum.

In deriving these results, we have assumed that the perturbation series 
converges.  At first, it may seem that this requirement would impose a restriction on the width of the step, $d$, due to successively expanding the $dV/d\phi$ term in Eq.~(\ref{eqn:KGeqn}) at each order in perturbation theory. However, note that by iteration, these terms are always multiplied by the previous order in the perturbation series 
\begin{align} \label{eqn:perteqn2}
\frac{d}{d\phi_{0}}\Big(e^{-\frac{3 \phi_{0}}{\sqrt{2\epsilonstep}}}&\frac{d\phi_{n}}{d\phi_{0}}\Big)  = -\frac{\ampold}{4} \frac{d^n F}{d\phi_{0}^n}\phi_{n-1} e^{-\frac{3\phi_{0}}{\sqrt{2\epsilonstep}}} .
\end{align}
Around the step, the derivatives of $F$ carry factors of $d^{-n}$ for a
$\Delta\phi_0$ interval of order $d$.   Thus in this neighborhood, the integrals over $\phi_0$
give contributions of order $d$, $\phi_1$ is of order
$(\amp d)$, and  $\phi_n$ is of order $(\amp d)^n$.   With these relations $d\phi_n/d\phi_0$ is  order $\amp^n$ there and is independent of $d$ as $d \rightarrow 0$.   The corrections to the slow roll parameter are then also strongly convergent as long
as $\amp \ll 1$.

Another way of seeing why the series is convergent as we take $d \rightarrow 0$ is to realize that these perturbative corrections amount to small shifts in when the inflaton crosses
the sharp feature and hence unobservable changes in the temporal location of features in the
slow roll parameters. 

In fact for the purpose of computing the shapes of these features, we can always define the position of the step in some arbitrary choice of the
zero point in $\ln \eta$ or efolds to be the same as that of the background
to all orders.    The background  has an approximate time translation symmetry (broken only weakly by $\dot{H}\neq 0$) and so there is no dynamical consequence to this choice.  
 As $d\rightarrow 0$ the step becomes sharp and changes to the potential are no longer
well approximated by a Taylor series for finite perturbations in field value. 
 But making use of the approximate time translation invariance of the background, we have
 defined these away.   It then follows that the linear expansion in $\amp$ remains valid in the limit $d\rightarrow 0$. We shall see in the next sections that nonlinear scalings of this type can be used to extend the analytic solutions
to the power spectrum and bispectrum out to $\amp \sim 1$.

\section{Curvature Power Spectrum}\label{sec:powerspec}
In \S \ref{sec:analyticpower}, we use the background solution of the last section to 
derive an analytic form for the curvature power spectrum of the step potential.
We test this analytic form against numerical calculations in \S\ \ref{sec:numericalpower}
and show that it can be extended to order unity features with a nonlinear rescaling of 
parameters.

\subsection{Analytic power spectrum solution}\label{sec:analyticpower}

In the generalized slow roll approximation, the power spectrum of curvature fluctuations
$\Delta_\curv^2 = k^3 P_\curv/2\pi^2$  is given to leading order in the source function by \cite{Dvorkin:2009ne}
\begin{align}\label{eqn:GSRLfullpower}
\ln\Delta^{2}_{\GSR}(k) = G(\ln\eta_{\sh})+\int_{\eta_{\sh}}^{\infty}\frac{d\eta}{\eta}\WP(k\eta)G'(\ln\eta),
\end{align}
where the source function 
\begin{align}
G(\ln\eta) = -2\ln f + \frac{2}{3}(\ln f)' ,
\end{align}
with
\begin{align}
f = \frac{\sqrt{8\pi^2\epsilon_H}}{H}(aH\eta).
\end{align}
Recall that primes denote derivatives with respect to $\ln\eta$. To leading order in the slow roll approximation, note that $f^{-2}$ is simply the power spectrum. The window function, $W(x)$, is given by
\begin{align}
W(x) = \frac{3\sin(2x)}{2x^3}-\frac{3\cos(2x)}{x^2}-\frac{3\sin(2x)}{2x}.
\end{align}

To linear order in the step height $c$ or kinetic energy perturbation $\amp$, the power spectrum source is given by $G' = G'_0 + G'_1$ with
\begin{align}
G'_1(\ln\eta)= \frac{2}{3}\left(6\epsilon_1 -3\eta_1+\delta_{2,1}\right),
\end{align}
where $G'_0(\ln\eta)$ is the source function in the absence of the step. Compared to $\eta_1$ and $\delta_{2,1}$, $\epsilon_{1}$ is negligible and
\begin{align}\label{eqn:powersource}
 G'_1 \approx & -\frac{\ampold}{6}  \left[    \left( \frac{\eta}{\etastep}\right)^{3} F' \right]' .
\end{align}

We can now evaluate the power spectrum. Inserting Eq.~(\ref{eqn:powersource}) into Eq.~(\ref{eqn:GSRLfullpower}), integrating by parts and dropping the negligible boundary terms, we obtain
\begin{align}
\label{eqn:powerintegral}
 \ln\Delta^{2}_{\GSR}(k) = \ln\Delta^{2}_{\GSR,0}(k)+\frac{\ampold}{6}\int_{\eta_{\sh}}^{\infty}\frac{d\eta}{\eta}\WP'(k\eta)\left(\frac{\eta}{\etastep}\right)^{3} F'  .
\end{align}

In the $d\rightarrow 0$ limit, $F'$ becomes 2 times a delta function at $\ln \etastep$,
given our convention that the step height is $2c$ and so in this limit
\begin{align}
\lim_{d\rightarrow 0} \ln\Delta^{2}_{\GSR}(k) &= \ln\Delta^{2}_{\GSR,0}(k) + \frac{\ampold}{3}  W'(k\etastep) .
\end{align}
Notice that since $W'(x) = d W /d\ln x$,
\begin{align}
W'(x) = \left( -3 +\frac{9}{x^2}  \right) \cos 2x
+\left(15- \frac{9}{x^2} \right)\frac{\sin 2x}{2x},
\end{align}
and in this limit there are  oscillations of amplitude $\amp$ in the power spectrum out
to $k \rightarrow \infty$.
This behavior is reminiscent of the Fourier transform of a function with a sharp
feature in real space.   In Appendix \ref{sec:realspace}, we shall see that this
intuition is borne out by an analysis of the real space correlation function and that
the high $k$ oscillations correspond to a feature at $r=2\etastep$.

As we move away from the $d\rightarrow 0$ limit, the function $F'$ has finite width of order $\Delta\eta/\etastep \sim d/ \sqrt{\epsilonstep}$ whereas the oscillatory features in $W'$ are of order $\Delta \eta \sim 1/k$.   Therefore the delta function approximation holds for $k\etastep\ll \sqrt{\epsilonstep}/d$.    For larger values, we expect the integral to 
be smaller due to integration over the oscillations in $W'$.   In other words, the $d\rightarrow 0$ results should be multiplied by  some damping factor  that depends on $k\etastep d/\sqrt{\epsilonstep}$.

\begin{figure}[t]
\centerline{\psfig{file=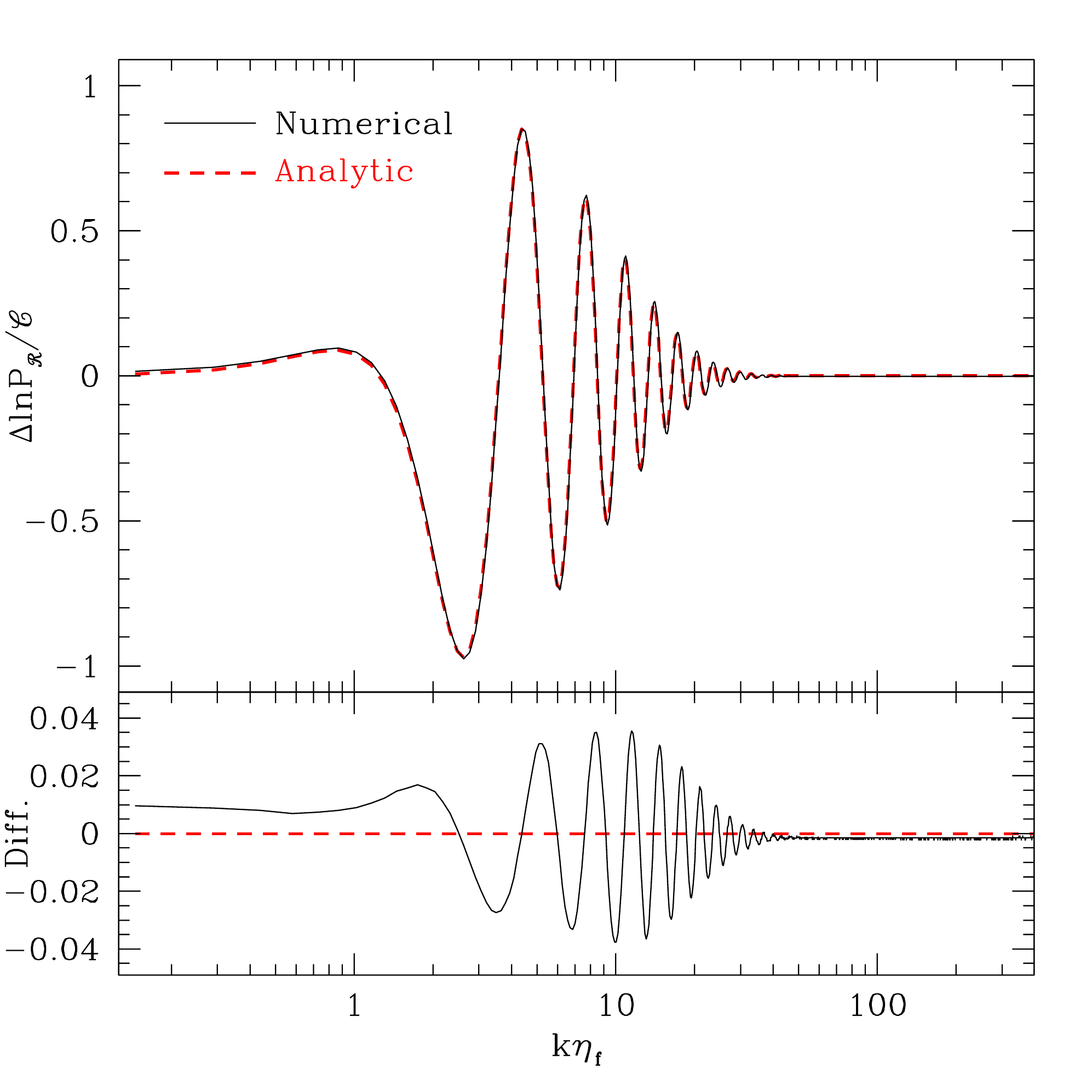, width=3.45in}}
\caption{\footnotesize Initial curvature power spectrum evaluated numerically (in black solid curves) compared with using the approximation in Eq.~(\ref{eqn:poweranalytic}) (in dashed red curves) for a small amplitude step $c=10^{-5}\ll\epsilonstep/6$ (or $\amp \ll 1$), where the approximation is expected to work,  and a step width, $d = 10^{-2}$, or $x_d \approx 4.3$. The lower panel shows the difference between the two curves, numerical - analytic. The remaining parameters used in both models are shown in Table \ref{table:modelparameters}.  
}
\label{plot:DeltaPofk_test_pertcond1a}
\end{figure}

\begin{figure}[t]
\centerline{\psfig{file=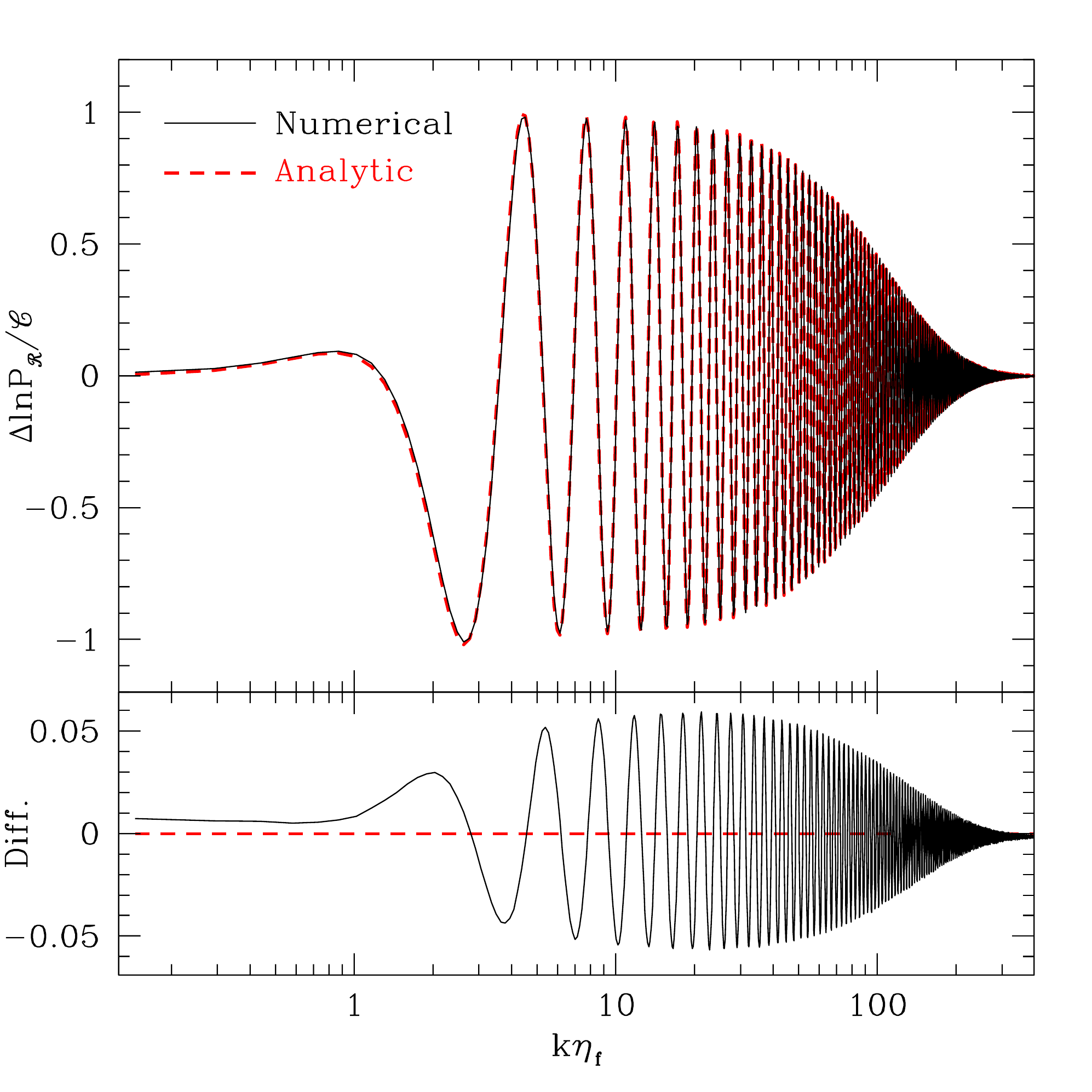, width=3.45in}}
\caption{\footnotesize 
Same as Fig. \ref{plot:DeltaPofk_test_pertcond1a} but using a smaller value of the step width. Here we take  $d = 10^{-3}$,  or  $x_d \approx 43$.
}
\label{plot:DeltaPofk_test_pertcond1b}
\end{figure}

\begin{figure}[t]
\centerline{\psfig{file=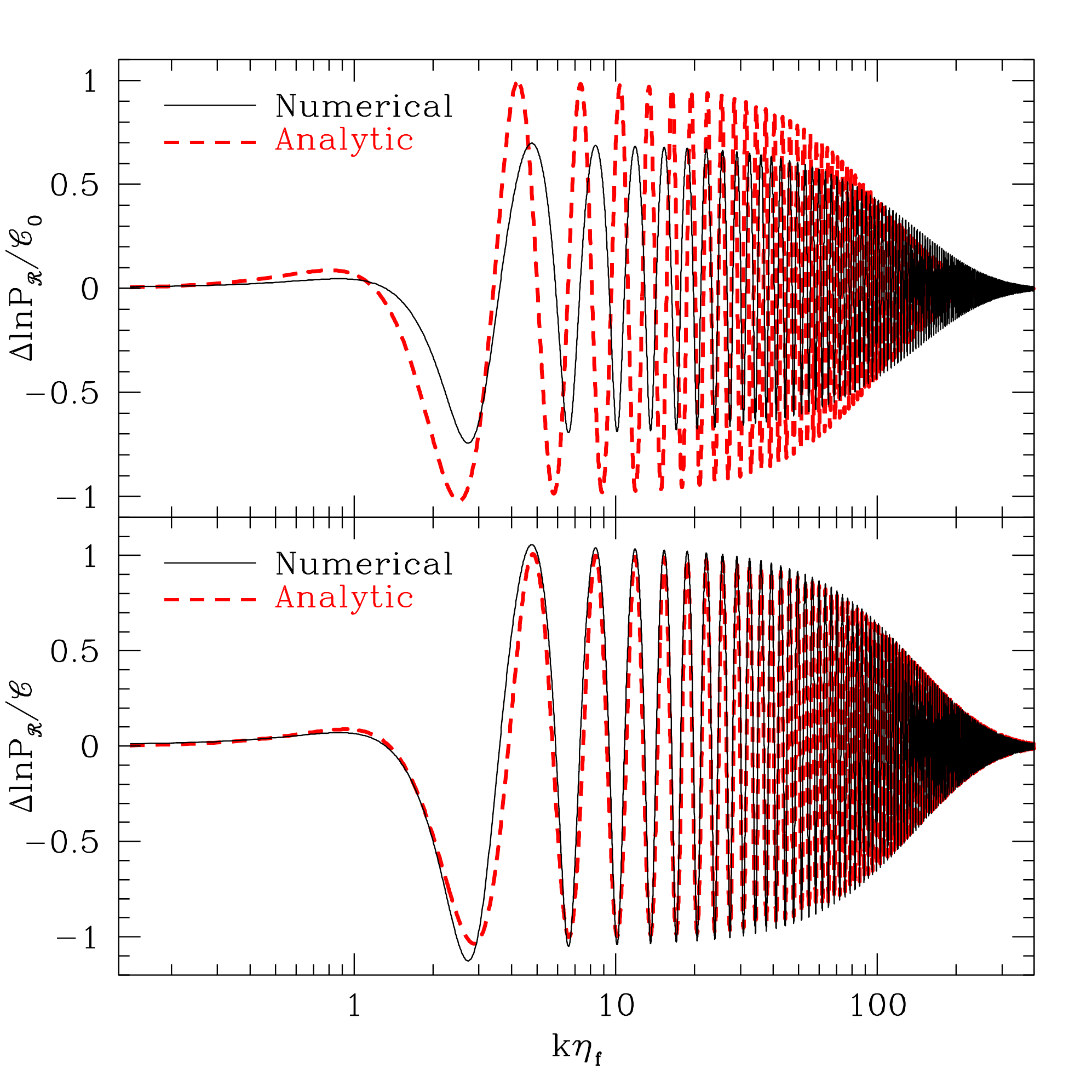, width=3.45in}}
\caption{\footnotesize Initial curvature power spectrum evaluated numerically (solid black curves) compared with the approximation in Eq.~(\ref{eqn:poweranalytic}) (dashed red curves) for a large amplitude $c =\epsilonstep/6$ (or $\amp_0=1$, $\amp=2/3$), where the linear expansion is violated by order unity terms.  Using the leading order expressions for the step location, oscillation amplitude and damping 
$\etastep$, $\amp = \amp_0$ and $x_d$ (upper panel) leads to phase and amplitude errors that are largely corrected by  non-linearly rescaling these parameters using
Eqs.~(\ref{eqn:nonlinearetaf}) and (\ref{eqn:nonlinearamp}) (lower panel).
}
\label{plot:DeltaPofk_test_pertcond2}
\end{figure}

In order to explicitly evaluate this damping function, we now assume the   
form for $F$ given in Eq.~(\ref{eqn:tanh}).  The required integrals then take the
form
\begin{align}
I = \int_{\eta_{\sh}}^{\infty}d\ln \eta\,f(k\eta)\,{\rm sech^2}\left[ \frac{\sqrt{2\epsilonstep}}{d}\ln\left(\frac{\etastep}{\eta}\right)\right],
\end{align}
where $f(k\eta)$ is a polynomial times a sine or cosine and we take $\eta_{\sh}$ to be a time well after the inflaton has crossed the feature. In the limit that $d\rightarrow 0$, the polynomial varies slowly while $\rm sech^2$ is non-zero, and consequently we can replace it by its value when $\rm sech^2$ is peaked, namely, $(k\eta)^n \rightarrow (k\etastep)^n$. We are then left to evaluate integrals such as
\begin{align}
I = \int_{\eta_{\sh}}^{\infty}d\ln \eta\,\sin(2k\eta){\rm sech^2}\left[ \frac{\sqrt{2\epsilonstep}}{d}\ln\left(\frac{\etastep}{\eta}\right)\right],
\end{align}
as well as the same thing with sine replaced by cosine. Working in the limit $\eta_{\sh}\rightarrow 0$ and changing variable to $y = - \frac{\sqrt{2\epsilonstep}}{d} \ln(\frac{\eta}{\etastep})$, the integral can be written
\begin{align}
I = \frac{d}{\sqrt{2\epsilonstep}}\int_{-\infty}^{\infty}dy\,\sin\left[ 2k \etastep \exp\left(-\frac{d}{\sqrt{2\epsilonstep}}y\right)\right]{\rm sech^2}y .
\end{align}
The integrand above only has support for $y\in (-1, 1)$ and, in the limit  $d\rightarrow 0$, the phase of the sine varies slowly, and thus we can expand,
\begin{align}\label{eqn:int}
I \approx\frac{d}{\sqrt{2\epsilonstep}}\int_{-\infty}^{\infty}dy\,\sin\left[2k \etastep\left(1-\frac{d}{\sqrt{2\epsilonstep}}y\right)\right]{\rm sech^2}y .
\end{align}
The phase error of the expression in brackets is  $\mathcal{O}(k \etastep(yd/ \sqrt{2\epsilonstep})^2) $ and,  given the support, the approximation 
is valid for $k \etastep \ll 2\pi\epsilonstep/d^2$. But, since $d/\sqrt{2\epsilonstep} \ll 1$, this only breaks down for values of $k \etastep$ that have many oscillations in the region $(-1,1)$ and are hence far into the region where the integral is already negligible. The integral in Eq.~(\ref{eqn:int}) can then be performed by integrating around the square contour in the complex plane, with vertices at $(R, 0)$, $(R, i\pi)$, $(-R, i\pi)$ and $(-R, 0)$ and taking the limit $R\rightarrow \infty$, obtaining,
\begin{align}\label{eqn:intapprox}
I \approx2\frac{d}{\sqrt{2\epsilonstep}}
\damp\left( \frac{k\etastep}{x_d}\right) 
\sin(2k\eta_f) ,
\end{align}
and
similarly for the terms with $\cos(2k\etastep)$.   Here
\begin{equation}
x_d = 
\frac{\sqrt{2\epsilonstep}}{\pi d}  ,
\label{eqn:xdzero}
\end{equation}
determines the value of $k\etastep$ at which damping starts to become 
important and
\begin{align}\label{eqn:damp}
\damp(y) = \frac{y}{\sinh y},
\end{align}
is the damping envelope for the tanh step profile.

 Generically then, the power spectrum takes the form
\begin{align}\label{eqn:poweranalytic}
& \ln\Delta^{2}_{\GSR}(k) = \ln\Delta^{2}_{\GSR,0}(k)+\frac{\ampold}{3}
\damp\left( \frac{k\etastep}{x_d}\right)  W' (k\etastep).  
\end{align}
Here, the oscillating window function $W'$ is modulated by the decaying envelope $\damp$ which is set by the details of the step. As $x \rightarrow 0$, $W'(x) \rightarrow 0$, and
no spurious superhorizon contributions during inflation are generated.  Interestingly, the amplitude of the perturbation is independent of the width of the step, while the range (in $k$ space) in which the perturbation persists  is independent of the height of the step. This analytic solution was derived in a different way by Stewart \cite{Stewart:2001cd} who also noted that different functions $F$ simply change the damping envelope. Similar conclusions were also reached in connection with steps arising from features due to duality cascades in brane inflation \cite{Bean:2008na}.
In Appendix \ref{sec:realspace}, we address the paradox that even if the feature scale $\etastep$ is greater
than the current horizon, the power spectrum can retain oscillations to arbitrarily high $k$.  

\subsection{Numerical comparison and nonlinear scaling}\label{sec:numericalpower}

In Figs.\ \ref{plot:DeltaPofk_test_pertcond1a} - \ref{plot:DeltaPofk_test_pertcond2}, we compare this approximation to the exact solution.   We choose a model with $V_0=m^2\phi^2/2$, and show curves for $d=0.01$ in Fig.\ \ref{plot:DeltaPofk_test_pertcond1a} and $d = 0.001$ in Fig.\ \ref{plot:DeltaPofk_test_pertcond1b}, to illustrate the damping behavior. We summarize the other parameters used in Table \ref{table:modelparameters}.  

\begin{table}
\begin{center}
\begin{tabular}{cc}
\hline
Parameter & Value\\
\hline
\hline
$m$ & $7.126\times10^{-6}$ \\
$\phi_f$  & $14.668$\\
$\epsilonstep$ & $0.00925$  \\
\hline
$100\Omega_bh^2$ & $2.231$ \\
$\Omega_ch^2$ & $0.1114$ \\
$\theta$ & $1.0401$ \\
$\tau$ & $0.084$ \\
\hline\\
\end{tabular}
\end{center}
\caption{Fiducial model parameters for a step at $\phi_f$ on a potential $V_0=m^2\phi^2/2$ with physical baryon and cold dark matter densities  $\Omega_b h^2$, $\Omega_c h^2$, 100 times the angular size of the sound horizon $\theta$, and the optical depth to reionization $\tau$.}
\label{table:modelparameters}
\end{table}

For  $c=10^{-5}\ll\epsilonstep/6$ (see Figs.\ \ref{plot:DeltaPofk_test_pertcond1a} and \ref{plot:DeltaPofk_test_pertcond1b}), the agreement is excellent both in the 
amplitude and phase of the results.  With $c=10^{-5}$ the inflaton crosses $\phi_f$ at $\eta=1454.6$ Mpc, which agrees to $0.1\%$ with $\etastep=1456.1$ Mpc, defined as the epoch when the inflaton crosses the feature for no step $c=0$.  {In the lower panels of Figs.\ \ref{plot:DeltaPofk_test_pertcond1a} and \ref{plot:DeltaPofk_test_pertcond1b} we plot the difference between our analytic approximation in Eq.\ (\ref{eqn:poweranalytic}) and a numerical evaluation of the spectrum. The error on our approximation here is at the level of 5\%, and is contained in a component that is out of phase by  $\pi/2$ with our approximation. As we point out in Appendix \ref{app:slowrollcorrections}, this error can be attributed to slow roll corrections to the modefunctions which are not captured by our leading order approximation at Eq.\ (\ref{eqn:GSRLfullpower}). Furthermore, the spectra also disagree on the percent level on scales far from the $k \sim 1/\etastep$. This disagreement is due to our analytic approximation of Eq. (\ref{eqn:GSRLfullpower}) by Eq.\ (\ref{eqn:poweranalytic}) and can be corrected in a straightforward manner by including subleading terms in the power spectrum source, $G'$, at Eq.\ (\ref{eqn:powersource}).}

For $c=\epsilonstep/6$ (see Fig.~\ref{plot:DeltaPofk_test_pertcond2}, upper panel), we see both 
an error in the amplitude, especially at the first few oscillations, as well as a difference in 
phase.   As discussed in the previous section, the phase error simply represents a 
difference between when the inflaton crosses the feature at $\phi_f$ relative to the $c=0$
model.  Here the inflaton crosses the feature at  $\eta=1271.7$ Mpc.   If we instead define $\etastep$ in the analytic expression Eq.~(\ref{eqn:poweranalytic}) to be 
\begin{equation}
\etastep \equiv \eta(\phi_f), 
\label{eqn:nonlinearetaf}
\end{equation}
we eliminate the phase error. We can also substantially improve the fit by rescaling of the value of $\epsilonstep$ to include the first order perturbation as the field crosses the step,
\begin{align}
\epsilon_H(\eta_f) \approx \epsilonstep + 3c .
\end{align}
Note that the first order expression suffices since its accuracy is guaranteed by energy
conservation so long as $c \ll 1$.   In particular we can define the non-linear amplitude
and damping as
\begin{align}
\amp &\equiv \frac{6 c}{\epsilonstep + 3 c} ,\nonumber\\
x_d &\equiv \frac{\sqrt{2\epsilonstep+6c}} {\pi d}.
\label{eqn:nonlinearamp}
\end{align}

We demonstrate these non-linear corrections with the general form 
of Eq.~(\ref{eqn:poweranalytic}) in the lower panel of Fig. \ref{plot:DeltaPofk_test_pertcond2}. The parameter $\amp$ in the lower panel is related to the original linearized version $\amp_0= 6c/\epsilonstep$ in the upper panel by
\begin{align}
\amp=\frac{\amp_0}{1+{\amp_0/2}}.
\label{eqn:amp0}
\end{align}
We employ these definitions of $\etastep$, $\amp$ and $x_d$ from this point forward.
In Table \ref{tab:stepparams}, we compile a guide to the step parameter notation used in 
the rest of this paper.

\begin{table}
\begin{center}
\begin{tabular}{clc}
\hline
Parameter & Definition & Eq.\\
\hline
\hline
$ F $ & Step shape function& (\ref{eqn:potential})\\
$\phi_f$ & Potential step position & (\ref{eqn:potential})\\
$\etastep$ & Step crossing time & (\ref{eqn:nonlinearetaf}) \\
$c$ & Potential step height & (\ref{eqn:potential}) \\
$\amp$ & Kinetic energy perturbation & (\ref{eqn:nonlinearamp})\\
$\amp_0$ & Linearized $\amp$ & (\ref{eqn:amp0})\\
$A_C$  & Angular power spectrum amplitude & (\ref{eqn:AC})\\
$A_B$  & Angular bispectrum amplitude & (\ref{eqn:bispecamp}) \\
$d$ & Potential step width & (\ref{eqn:potential}) \\
$x_d$ & Dimensionless damping scale & (\ref{eqn:nonlinearamp})\\
$\ell_d$ & Angular damping multipole & (\ref{eqn:ld}) \\ 
$\damp$ & Damping function & (\ref{eqn:damp}) \\
\hline\\
\end{tabular}
\end{center}
\caption{Step parameters and their defining equations.}
\label{tab:stepparams}
\end{table}

\section{Curvature Bispectrum}\label{sec:bispectrumformal}

In \S \ref{sec:methods} we briefly describe the method we use to calculate the bispectrum. For more detail, we refer the reader to  \cite{Maldacena:2002vr, Weinberg:2005vy, Adshead:2011bw}.  We use this method to derive an analytic approximation in \S \ref{sec:GSRzero}.  In \S \ref{sec:ksqueezedlimit}, we discuss limiting cases and consistency checks for the approximation and in \S \ref{sec:numericalbi} we compare it
with direct numerical computation of the exact solution.

\subsection{The action and the in-in formalism}
\label{sec:methods}

In this paper to move beyond the linear equations of motion, we work in comoving gauge in which the time slicing is chosen so that the inflaton is unperturbed. In this gauge the scalar degree of freedom is the comoving curvature perturbation, while the remaining physical metric degrees of freedom are the two polarizations of the transverse traceless tensor perturbation.  Tensor perturbations are not enhanced due to the presence of the step \cite{Adams:2001vc} and thus we neglect them here. We make use of the interaction picture, where we choose our basis so that the fields diagonalize the Hamiltonian arising from the quadratic action,
\begin{align}\label{eqn:quadaction}
S_{2} = \frac{1}{2}\int dt d^{3}x\,a^{3}2\epsilon_{H}\left[\mathcal{\dot{R}}^{2} -\frac{ (\partial\mathcal{R})^2}{a^{2}}\right].
\end{align}
In this picture the Hamiltonian arising from cubic and higher order terms in the Lagrangian, which we shall refer to as the interaction Hamiltonian, $H_{I}$, evolve the states. At leading order, the tree-level bispectrum at time $t_\sh$ is then given by \cite{Maldacena:2002vr}
\begin{align}\label{eqn:bispec}\nonumber
& \langle\hat{\mathcal{R}}_{\bf k_{1}}(t_{\sh})\hat{\mathcal{R}}_{\bf k_{2}}(t_{\sh})\hat{\mathcal{R}}_{\bf k_{3}}(t_{\sh})\rangle = \\& 2\Re\left[ -i\int^{t_{\sh}}_{-\infty} dt \langle \hat{\mathcal{R}}_{\bf k_{1}}(t_{\sh})\hat{\mathcal{R}}_{\bf k_{2}}(t_{\sh})\hat{\mathcal{R}}_{\bf k_{3}}(t_{\sh})H_{I}(t)\rangle\right].
\end{align}
At leading order in fluctuations, the interaction Hamiltonian in Eq.~(\ref{eqn:bispec}) arises from the cubic action. In this work we are interested in potentials in which the inflaton undergoes a sharp transient acceleration but inflation is not interrupted. Consequently, $\epsilon_{H}\ll1$ everywhere and, to a very good approximation, the cubic action is given by
\begin{align}\nonumber\label{eqn:truncatedaction}
S_{3} \approx & \int dt d^{3}x \, \bigg[a^{3}\epsilon_H (\dot\epsilon_H - \dot\eta_H)\mathcal{R}^{2}\dot{\mathcal{R}} \nonumber\\
&\qquad 
- \frac{d}{dt}\left(a^{3}\epsilon_H(\epsilon_H-\eta_H)\mathcal{R}^{2}\dot{\mathcal{R}}\right)\bigg].
\end{align}
Note that, after switching to conformal time, $\eta = \int_{t}^{t_{end}}dt'/a(t)$ (defined to be positive during inflation), and performing some integration by parts, the action in Eq.~(\ref{eqn:truncatedaction}) can be written as
\begin{align}\label{eq:S3}
S_{3} = \int d^{3}xd\eta \,2a^{2}\epsilon_{H}(\eta_{H} - \epsilon_H)\mathcal{R} \left[\mathcal{R}'^2 -(\partial\mathcal{R})^2 \right]
\end{align}
and thus (for this interaction), 
\begin{align}
\mathcal{L}_3 = 2(\eta_{H} - \epsilon_H)\mathcal{R}\mathcal{L}_2.
\end{align}
Naively, in order that we trust our perturbative evaluation of correlation functions using the interaction picture, we might demand that $\mathcal{L}_{3} < \mathcal{L}_{2}$ so that the evolution of the operators is well described by the equations of motion arising from the quadratic action. Thus we require at least
\begin{align}\label{eq:pertcond}
{2}(\eta_{H} - \epsilon_{H})\mathcal{R} < 1.
\end{align}
We shall see that in the context of the step model, this constraint places a lower limit on the
step width $d$ that we can consider using perturbative techniques.

In conformal time,  the relevant interaction Hamiltonian that follows from the action in Eq.~(\ref{eqn:truncatedaction}) is given by
\begin{align}
H_{I}(\eta) = - \int&\frac{d^{3}q_a}{(2\pi)^3}\frac{d^{3}q_b}{(2\pi)^3}\frac{d^{3}q_c}{(2\pi)^3} (2\pi)^{3}\delta^{3}({\bf q}_a+{\bf q}_b+{\bf q}_c)\nonumber\\ 
& \times\Bigg[ \frac{a^{2}\epsilon_{H}}{3\eta^2}   (\epsilon_{H} - \eta_{H})' \left(  \hat{\mathcal{R}}_{{\bf q}_a}\hat{\mathcal{R}}_{{\bf q}_b}\hat{\mathcal{R}}_{{\bf q}_c}\right)' \\
\nonumber
& -\frac{d}{d\eta}\left(\frac{a^{2} \epsilon_{H}}{3\eta}(\epsilon_{H}-\eta_{H})(\hat{\mathcal{R}}_{{\bf q}_a}\hat{\mathcal{R}}_{{\bf q}_b}\hat{\mathcal{R}}_{{\bf q}_c})'\right)\Bigg],
\end{align}
where recall $' \equiv d/d\ln\eta$. In this expression, the fields $\mathcal{R}$ are interaction picture fields chosen to diagonalize the Hamiltonian derived from the quadratic action in Eq.~(\ref{eqn:quadaction}). 

We define the bispectrum through
\begin{equation}
\langle \hat{\curv}_{{\bf k}_1} \hat{\curv}_{{\bf k}_2} \hat{\curv}_{{\bf k}_3}\rangle
= (2\pi)^3 \delta({\bf k}_1+ {\bf k}_2+{\bf k}_3) B_{\curv}(k_1,k_2,k_3),
\end{equation}
where
\begin{align}\nonumber\label{eqn:bispectrum1st}
& B_{\curv}(k_1,k_2,k_3) =  4 \Re\Bigg\{ i\mathcal{R}_{k_{1}}(\eta_{\sh})\mathcal{R}_{k_{2}}(\eta_{\sh})\mathcal{R}_{k_{3}}(\eta_{\sh})\\ \nonumber
&\times \Bigg[ \int_{\eta_{\sh}}^{\infty} {d\eta \over \eta^2}\,  {a^{2} \epsilon_{H}}(\epsilon_{H} - \eta_{H})' (\mathcal{R}^{*}_{k_{1}}\mathcal{R}^{*}_{ k_{2}}\mathcal{R}^{*}_{ k_{3}})'\\
& +{a^{2}\epsilon_{H} \over \eta_\sh}(\epsilon_{H}-\eta_{H})(\mathcal{R}^{*}_{k_{1}}\mathcal{R}^{*}_{ k_{2}}\mathcal{R}^{*}_{k_{3}})' \Big|_{\eta = \eta_{\sh}} \Bigg]
\Bigg\}
\end{align}
{and $\Re$ denotes the real part.}
We choose $\eta_{\sh}$ to be a time well after the inflaton has crossed the feature, and $\epsilon_{1} = \eta_{1} = \delta_{2,1} = 0$. Consequently, the second term in Eq.~(\ref{eqn:bispectrum1st}) is tiny, and we neglect it in what follows.
The derivation of Eq.~(\ref{eqn:bispectrum1st}) is described in detail in \cite{Adshead:2011bw}, and we refer the reader to this work for details.

\subsection{Analytic bispectrum solutions}
\label{sec:GSRzero}

As described in \cite{Adshead:2011bw}, the GSR approach proceeds by iteratively correcting the evolution of the mode function for the effect of deviations from de Sitter space.   As we will see in this section, in the limit that $d\rightarrow 0$ the modefunctions, $\curv_{k}$,  are well treated as unperturbed, and the non-Gaussian features are well described by integrals involving only the unperturbed modefunctions. We have already seen that, even in the limit $d\rightarrow 0$, the power spectrum, which may be thought of as the square modulus of the modefunctions, gains a correction that is $\mathcal{O}(c/\epsilon_{H})$.  Thus at linear order, the corrections involving the perturbed mode functions will be small compared to those arising from the perturbed slow roll parameters.

There is an additional weak constraint on $d$ from Eq.~(\ref{eq:pertcond}) due to the perturbative expansion. 
The perturbations in the curvature themselves do not become large as the field crosses the feature, and thus we may approximate $\mathcal{R}\sim 10^{-5}$. Since $\epsilon_{H}$ remains small, we can neglect it and focus on $\eta_{H}$. Near the step, from above, we know that
\begin{align}\label{eq:cd_condition}
\eta_{H} = \frac{\ddot{\phi}}{H^2\dot{\phi}} \sim \frac{c}{\sqrt{\epsilonstep}}\frac{1}{d}.
\end{align}
Thus, with $\epsilonstep \sim 0.01$, this implies the limit on the ratio
\begin{align}
\frac{c}{d} < 10^{4}.
\end{align}
In the limit of infinitesimal width $d\rightarrow0$ with finite height, while the field fluctuations remain small, they are no longer well characterized by their tree level $N$-point functions.
On the other hand, this is not a particularly restrictive constraint and when considering the
 impact on the primary CMB anisotropy is equivalent to taking $d \rightarrow 0$.   
Nevertheless, it does cure the weak logarithmic divergence in the real space correlation function
discussed in Appendix \ref{sec:realspace}.  

In the GSR approach, 
as described in detail in \cite{Adshead:2011bw},  we can approximate 
Eq.~(\ref{eqn:bispectrum1st}) for the bispectrum as
\begin{eqnarray}
B_{\curv}(k_1,k_2,k_3)  &\approx & {(2 \pi)^4  \over k_1^3 k_2^3 k_3^3} {\Delta_{\cal R}(k_1) 
\Delta_{\cal R}(k_2)
\Delta_{\cal R}(k_3)\over 4} \nonumber\\&&\quad
\Big[ -I_0(K) k_1 k_2 k_3 - I_1(K) \sum_{i \ne j} k_i^2 k_j 
\nonumber\\&&\quad + I_2(K) K (k_1^2+k_2^2 + k_3^2) \Big],
\label{eqn:gsrlbi}
\end{eqnarray}
where $K = k_{1}+k_{2}+k_{3}$ is the perimeter of the triangle in momentum space.
The principle advantage of Eq.~(\ref{eqn:gsrlbi}) is that it involves integrals universal in $K$, 
\begin{equation}\label{eqn:Integrals}
I_n(K) =G_B(\ln\eta_\sh)  W_n(0) + \int_{\eta_{\sh}}^{\infty} {d\eta \over \eta}  G_B'(\ln\eta) W_n(K\eta) ,
\end{equation}
where
\begin{equation}
W_0(x) = x \sin x , \,\,\,\,\, W_1(x) = \cos x, \,\,\,\,\, W_2(x) = \frac{\sin x}{x},
\end{equation}
and the bispectrum source  is given by
\begin{equation}
G_B=   \left( {\epsilon_H - \eta_H \over f} \right).
\end{equation} 
The function $\Delta_{\cal R}(k) $ is given by the root of the power spectrum, $\Delta_{\cal R}(k)  = \sqrt{\Delta^2_{\cal R}(k) }$.

We have already obtained analytic solutions for the power spectrum $\Delta_{\cal R}(k)$ above, and it remains to evaluate the integrals $I_0$, $I_1$ and $I_2$.   Following the analysis for the power spectrum, we split the bispectrum source into a background piece and feature piece, but here the background piece generates a negligible bispectrum.  Likewise the boundary
term in Eq.~(\ref{eqn:Integrals}) is negligible and the $\epsilon_H$ term in the source
is suppressed by a factor of $\epsilon_0$ in its contribution.     
Integrating twice by parts, 
we then obtain from Eq.~(\ref{eqn:slowroll})
\begin{equation}
I_n(K) = \frac{\amp}{4 f_0}\int_{\eta_{\sh}}^{\infty} {d\eta \over \eta} (F-1) \left( {\eta \over \eta_f}\right)^3 W_n''.
\end{equation} 
We can bring this into the form of Eq.~(\ref{eqn:powerintegral}) with one more integration by
parts that defines a new window function
\begin{equation}
X_n(K\eta) = -\int_0^\eta  {d\tilde\eta \over \tilde\eta}  \left( {\tilde\eta \over \eta_f}\right)^3
W_n''(K\tilde\eta).
\label{eqn:analyticbi}
\end{equation}
Using the approximation for the tanh step in Eq.~(\ref{eqn:intapprox}) we obtain
\begin{align}\label{eqn:bispecintegrals}
I_n(K) = & \frac{\amp}{2 f_0}\damp
\left( \frac{K\etastep}{2 x_d} \right) 
X_n({K}{\etastep}) ,
\end{align}
where recall $x_d$ was defined in Eq.~(\ref{eqn:xdzero}) such that $x_d \rightarrow \infty$ as $d \rightarrow 0$.   Explicitly
\begin{align}\nonumber\label{eqn:biwindows}
X_0(x) = &-\frac{\left(x^4-9 x^2+54\right) \cos x}{x^2}\\ \nonumber&\quad  +\frac{\left(2 x^4 -27 x^2+54\right) \sin x}{x^3},\\\nonumber
X_1(x) = & \frac{3 \left(x^2-6\right) \cos x}{x^2} +\frac{\left(x^2-6\right) \left(x^2-3\right) \sin x}{x^3},\\
X_2(x) = & -\frac{\left(x^2-9\right) \cos x}{x^2}+\frac{\left(4 x^2-9\right) \sin x}{x^3}.
\end{align}

All of these window functions vanish as $x^2$ as $x \rightarrow 0$, which means that we do not generate any spurious superhorizon effects.  Notice also that $X_0$ diverges as $x^2$ for large $x$ while $X_1$ diverges linearly and  $X_3$ approaches a constant in this limit. This implies that in the limit $d \rightarrow 0$ the quantity in
brackets in Eq.~(\ref{eqn:gsrlbi}) increases without bound as $k\rightarrow \infty$. 
For small but finite $d$, 
 we can estimate the location and height of the peak non-Gaussianity. 
In this limit the bispectrum is dominated by the quadratic term $I_0$,
\begin{align}
I_0 \sim \left( K\etastep \right)^2
 {\frac{\amp}{2f_0}}\damp \left(  \frac{K\etastep}{2 x_d}  \right)\cos(K\etastep),
\label{eqn:I0limit}
\end{align}
whose envelope behaves as $x^{3}\exp(-x)$ near its peak $x=3$ or 
\begin{align}\label{eqn:peakamp}
K_{\rm peak} = 6 \frac{x_d}{\etastep}.
\end{align}
For a fixed step position, i.e.~a fixed value of $\etastep$, as $d\rightarrow 0$, the peak of the reduced bispectrum moves to larger and larger values of $K$.   Simultaneously reducing the width of the feature and moving it to larger scales, i.e.\ increasing $\etastep$, fixes the position of the peak of the reduced bispectrum, while increasing its amplitude proportional to the square of the ratio of step positions   and increasing the frequency of its oscillations.
Thus the sharper the feature, the larger the non-Gaussianity as anticipated in \cite{Bean:2008na, Chen:2011zf, Chen:2011tu}. 
 
Indeed naively, it might seem that the non-Gaussianity strongly diverges with decreasing $d$ due to the
high $k$-modes.    Here one must be careful in defining what is meant by a large non-Gaussianity.  
Note that  the bispectrum itself, i.e.~the amplitude of individual triangles in $k$-space, does not diverge.  It is only when written in terms of the reduced bispectrum or effective $f_{\rm NL}$, that the apparent quadratic divergence appears
\begin{align}
\reduced(k_1,k_2,k_3) \equiv &
\frac{B_\curv(k_1,k_2,k_3)}{P_\curv(k_1)P_\curv(k_2) + {\rm perm.}},
\end{align}
where ``perm." refers to cyclic permutation of the indices.  
To see where this arises, we can rewrite the reduced bispectrum in terms of the quantity
 defined by \cite{Chen:2006xjb,Chen:2008wn} 
\begin{align}
 \frac{\mathcal{G}(k_1,k_2,k_3)}{k_1 k_2 k_3}
= & \frac{k_1^2 k_2^2 k_3^2}{(2\pi)^4\tilde A_{s}^2}B_{\curv}(k_1,k_2,k_3) \\
 \approx &
\frac{1}{4}\left( \frac{k_1^3+ k_2^3 + k_3^3}{k_1 k_2 k_3}  \right)
\reduced(k_1,k_2,k_3), \nonumber
\end{align}
where $\tilde A_{s}$ is a constant taken to be $\tilde A_{s} = \Delta_\curv^2$ for nearly
scale invariant spectra.
The extra factors of $k$ account for the volume factors in $k$ space or equivalently the number of $k$-space triangles available.  Divergence in the reduced bispectrum is
 usually associated with an equivalent divergence in the non-Gaussianity by assuming that
 the individual triangles add coherently.   For oscillatory bispectra, counting triangles through $k$ factors leads to a misleading sense of both the UV divergence and the observability of the
 non-Gaussianity.
 
 A direct means of seeing this fallacy is to evaluate the non-Gaussianity in the real space
 three-point correlation function.   We show in Appendix 
\ref{sec:realspace} that, similar to the power spectrum oscillations,  the oscillating and diverging
reduced bispectrum is associated with sharp
features in the
three-point  correlation on scales $r \sim \etastep$ whose amplitude diverges no 
more than logarithmically.  The oscillations in the bispectrum prevent triangles from coherently adding up to a large non-Gaussianity in real space. Furthermore we have seen that there is a mild limitation on our derivation of $c/d < 10^4$ from the validity of the perturbative expansion of the action which prevents even  the logarithmic divergence from being manifest.

\subsection{Squeezed limit}\label{sec:ksqueezedlimit}

It is well known that the bispectrum of curvature fluctuations produced by a single scalar field during inflation satisfies a consistency relation which relates the bispectrum in the squeezed limit to the slope of the power spectrum \cite{Maldacena:2002vr, Creminelli:2004yq}. Since we have analytic solutions for both the bispectrum and power spectrum, this provides us with a non-trivial test of our results. 

In the limit where one of the momenta is much smaller than the other two, $k_{S}\ll k_{L}$, the consistency relation implies for the reduced bispectrum
\begin{align}\label{eqn:sqreduced}
\reduced (k_{\rm S}, k_{\rm L}, k_{\rm L}) &= 
-\frac{1}{2}\left.\frac{d \ln \Delta^2_{\curv}}{d\ln k}\right|_{k_{\rm L}} \nonumber\\
&\approx 
2  \frac{\mathcal{G}(k_{\rm S}, k_{\rm L}, k_{\rm L})}{k_{\rm L}^3}, 
\end{align} 
where the approximation holds for nearly scale invariant power spectra.
Equation (\ref{eqn:gsrlbi}) then implies that, in this limit,
\begin{align} \label{eqn:bisq}
\left.\frac{d\ln\Delta^2_{\curv}}{d\ln k}\right|_{k_{\rm L}} = f_{0}\left[2I_1-4I_2\right]_{K = 2k_L}.
\end{align}
Now, from Eq.~(\ref{eqn:poweranalytic}), and dropping terms of order $\mathcal{O}(d/\sqrt{2\epsilonstep})$ -- which amounts to ignoring the variation of the envelope function -- we find
\begin{align}\label{eqn:sqpower}
\left.\frac{d\ln\Delta^2_{\curv}}{d\ln k}\right|_{k_{\rm L}}&  =   n_s(k_{\rm L})-1
+\frac{\amp}{3} \damp\left( \frac{k \etastep}{x_d} \right) W''(k\etastep).
\end{align}
Comparing Eq.~(\ref{eqn:sqpower}) to Eq.~(\ref{eqn:bisq}) using  Eq.~(\ref{eqn:biwindows}), and ignoring the slow roll suppressed terms, we find that in this approximation, the consistency relation is satisfied.

A second non-trivial check of our results is that corrections to the squeezed limit of the  bispectrum in Eq.~(\ref{eqn:gsrlbi}) are proportional to $(k_{\rm S}/k_{\rm L})^2$, where $k_{\rm S}$ is the side that is being taken to zero. This is in accordance with the behavior of the so called ``not so squeezed limit" \cite{Creminelli:2011rh}.

\begin{figure}[t]
\centerline{\psfig{file=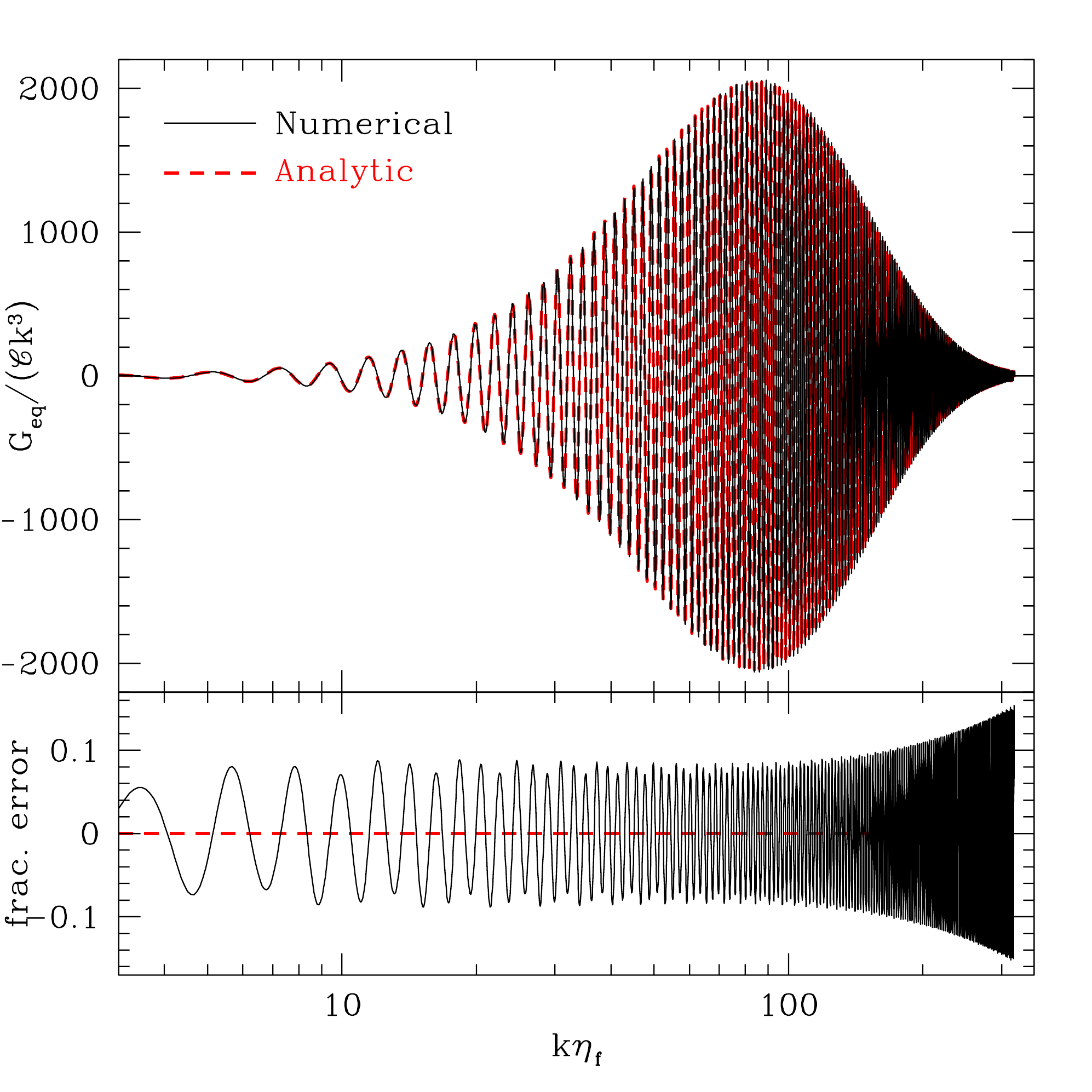, width=3.45in}}
\caption{\footnotesize Equilateral bispectrum computed using the analytic approximation in Eq.~(\ref{eqn:gsrlbi}) (red dashed) compared with a full numerical evaluation of the integral in Eq.~(\ref{eqn:bispectrum1st}) for a small amplitude $c=10^{-5}$ (or $\amp \ll 1$). 
The lower panel shows the difference between the curves, expressed as a fraction of the
envelope of the analytic approximation in Eq.~(\ref{eqn:bienvelope}). }
\label{fig:cem5}
\end{figure}

\begin{figure}[t]
\centerline{\psfig{file=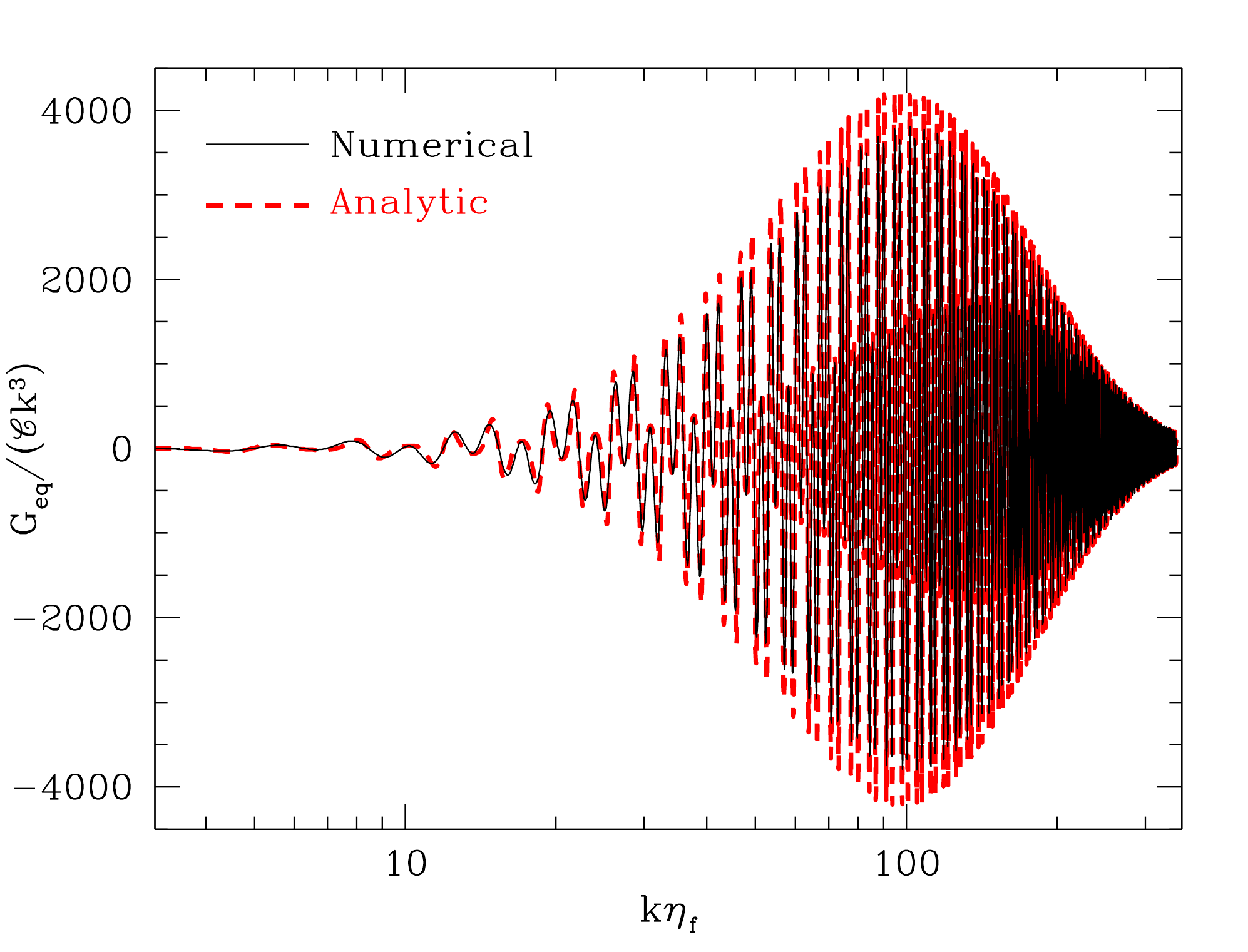, width=3.45in}}
\caption{\footnotesize Equilateral bispectrum computed using the analytic approximation in Eq.~(\ref{eqn:gsrlbi}) (red dashed) compared with a full numerical evaluation of the integral in Eq.~(\ref{eqn:bispectrum1st}) for  a large amplitude  $c \approx \epsilonstep/6$ (or $\amp_0=1, \amp=2/3$).  Here the location, amplitude and damping scale of the feature have been nonlinearly scaled according to Eq.~(\ref{eqn:nonlinearetaf}) and Eq.~(\ref{eqn:nonlinearamp}).}
\label{fig:c1rescale}
\end{figure}

\subsection{Numerical comparison} \label{sec:numericalbi}

We begin with a comparison for a small step amplitude.
In Fig.~\ref{fig:cem5}  we plot the bispectrum resulting from a step with height $c = 10^{-5}$ and width $d = 0.001$.  We show the result of evaluating Eq.~(\ref{eqn:bispectrum1st}) numerically as well as the result of using the analytic approximation in Eq.~(\ref{eqn:gsrlbi}). Overall, in both amplitude and frequency the approximation is excellent.   
We show a more detailed comparison in the lower panel, where we divide the
difference by the  envelope of the analytic expectation
\begin{equation}
\frac{9}{8}\frac{\Delta^{3}_{\curv, 0}(k)}{\tilde A_{S}^{3/2}}(k\etastep)^2 \damp\left( \frac{3 k\etastep}{2 x_d} \right),
\label{eqn:bienvelope}
\end{equation}
 {where the amplitude of the powerspectrum is taken throughout this section to be $\tilde A_{S} = 2.45\times 10^{-9}$.}
The good overall agreement hides an oscillatory 10\% residual error.

 Given that we are perturbing in $\amp \sim 0.6\%$ here,  this cannot be an ${\cal O}(\amp^2)$ correction.  In Appendix \ref{app:slowrollcorrections}, we demonstrate that these corrections are largely due to slow roll corrections to the modefunctions. On superhorizon scales, these corrections lead to a phase shift in the growing mode of the curvature perturbation,  resulting in a correction on the 10\% level to the bispectrum. In fact, as pointed out by \cite{Burrage:2011hd}, leading order slow roll corrections generically lead to  10\% rather than $\mathcal{O}(\epsilon_0)\sim 1\%$ corrections as one might naively guess. Additionally, note that the error begins to increase for modes that are far into the damping window. These momenta have many oscillations during the period where the bispectrum source in  Eq.~(\ref{eqn:Integrals}) is non-zero and are thus in the region where our approximation in Eq.~(\ref{eqn:intapprox}) is beginning to break down. Given that this is occurring in the region where the bispectrum is strongly damped, it will have little effect on our results.

As we increase the step height, the leading order approximation for $\etastep$,  $\amp$ and $x_d$ in
Eq.~(\ref{eqn:analyticbi}) breaks down.   Just as in the power spectrum, these
errors in phase and amplitude are largely corrected by the nonlinear rescalings
of Eq.~(\ref{eqn:nonlinearetaf}) and Eq.~(\ref{eqn:nonlinearamp}) as shown in 
Fig.\ \ref{fig:c1rescale}.
 While residual errors of $20\%$-$40\%$ remain, they are largely contained in a $\pi/2$ out of phase component rather than in an overall amplitude error, and are related to the nonlinear analogue of the corrections discussed in Appendix \ref{app:slowrollcorrections}.   Furthermore, the modulation of the oscillations comes from
the power spectrum prefactors of the analytic expression causing the inner and outer
envelope effect seen in Fig.~\ref{fig:c1rescale}.  In the signal-to-noise calculation that follows, such a modulation mainly cancels out and in fact the  use of the simple unmodulated $\amp \rightarrow 0$ form of $\mathcal{G}/\amp$ in Fig.~\ref{fig:cem5} suffices out to $\amp \sim 1$ for our purposes.

\begin{figure}[t]
\centerline{\psfig{file=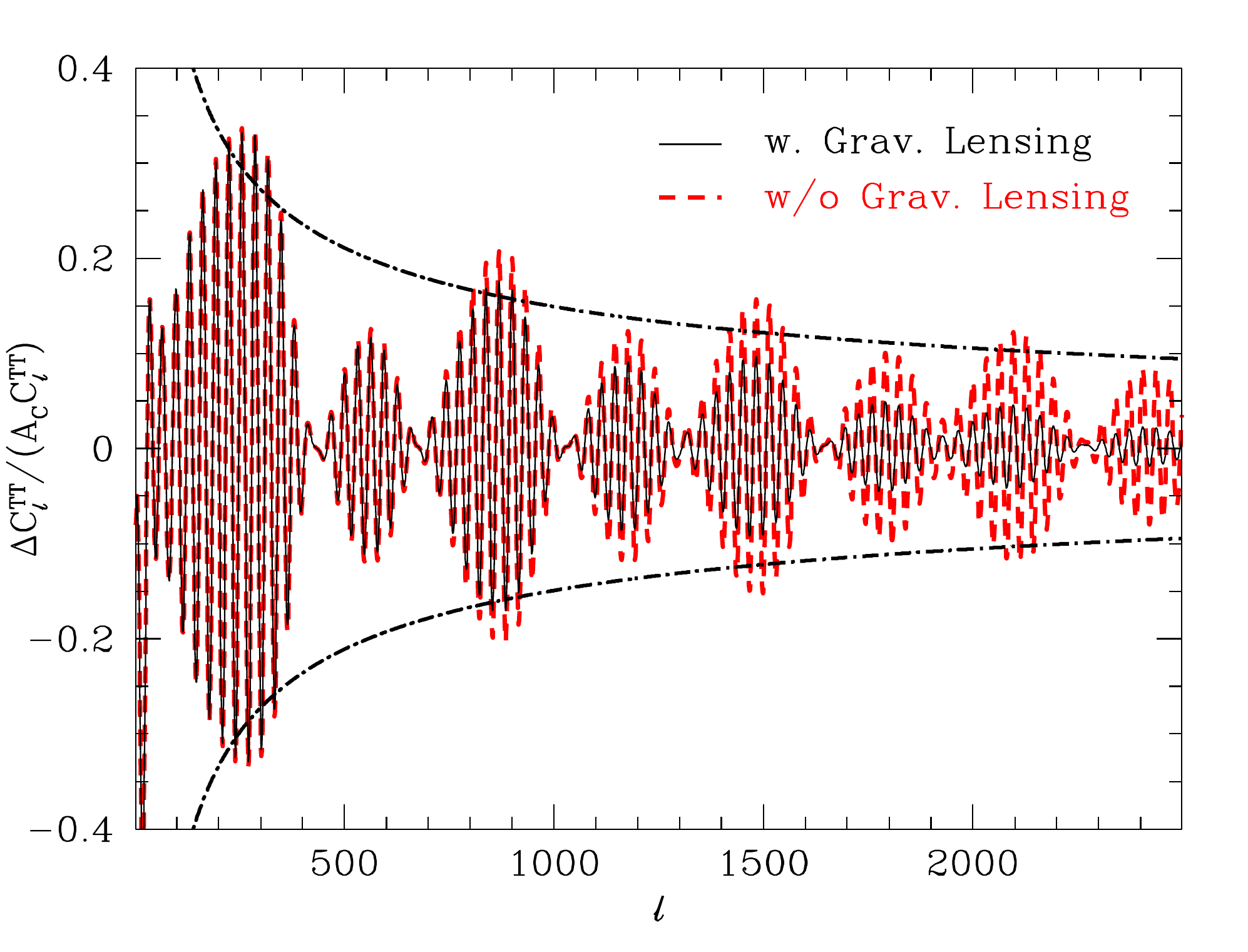, width=3.45in}}
\caption{\footnotesize  Fractional difference of the temperature power spectrum of a model with $\eta_f=1.44$ Gpc, $c=10^{-5}$ and $d=10^{-5}$, relative to a model with $c=0$ per unit power spectrum feature amplitude $A_C$ (see Eq.~\ref{eqn:AC}). The case without gravitational lensing is shown in red dashed lines, and when gravitational lensing is taken into account is shown in black solid lines. In black dot-dashed lines we show the expected $\ell^{-1/2}$ envelope from Eq.~(\ref{eqn:projectionenvelope}). }
\label{plot:DeltaClTT_c1dm5_dm5_kf6p9dm4_unlens_vs_lens}
\end{figure}

\begin{figure}[t]
\centerline{\psfig{file=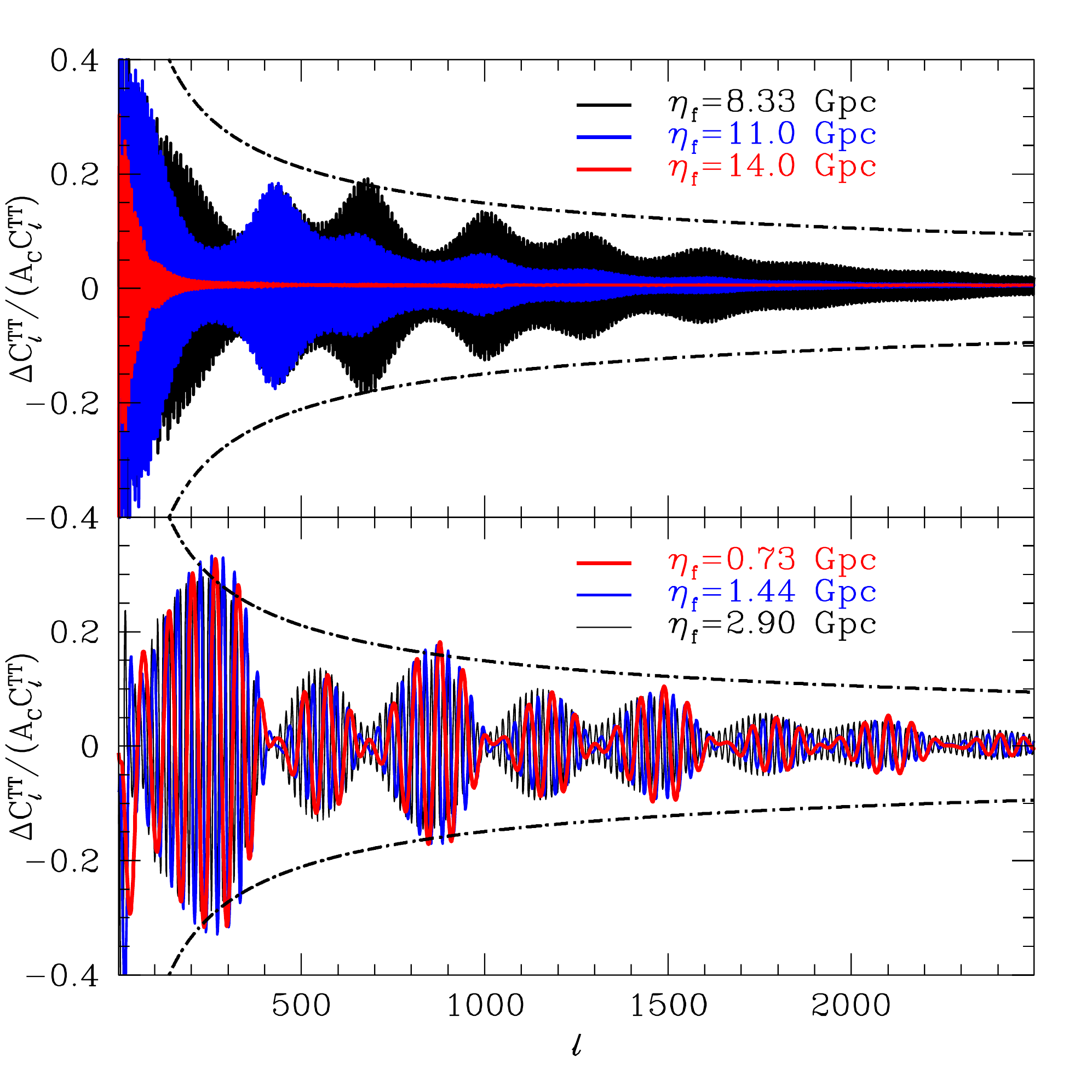, width=3.45in}}
\caption{\footnotesize Fractional difference of the temperature power spectrum (relative to a model with $c=0$) of a model with $c=10^{-5}$ and $d=10^{-5}$ and different step positions. Gravitational lensing is taken into account in these examples. Note that as $D/\eta_f\rightarrow 1$ (in the upper panel), the nodes shift, and the shape changes significantly. In black dot-dashed lines we show the expected $\ell^{-1/2}$ envelope from Eq.~(\ref{eqn:projectionenvelope}). }
\label{plot:DeltaClTT_c1dm5_dm5_kfseries_nolensing}
\end{figure}

\section{CMB Power Spectrum}\label{sec:likelihoodan}

In \S \ref{sec:clscaling} we discuss the phenomenology of a sharp step in the inflaton potential on the CMB temperature power spectrum and derive scaling relations based on projection effects.    We then consider WMAP constraints on the location and height of the step in \S \ref{sec:constraints}.

\subsection{Scaling relations}
\label{sec:clscaling}

The oscillatory features in the curvature power spectrum transfer onto the temperature anisotropy spectrum in a manner that reflects projection onto the recombination surface,
evolution through the acoustically oscillating plasma, and gravitational lensing after
recombination.   
These same effects impact the angular bispectrum as well and so it is useful to obtain
some physical intuition in the simpler case of the power spectrum.

Projection effects damp the amplitude of the $k$-space oscillations for modes where
$k\etastep \gg 1$.  
It is simple to derive scaling relations for this effect in the flat-sky approximation.
Ignoring for the moment the acoustic evolution by taking the Sachs-Wolfe limit, the temperature field $a(\hat{\bf n}) =
\Delta T/T$, in the angular direction $\hat{\bf n}$, is given by
\begin{equation}
a(\hat{\bf n}) = -{1\over 5} \curv({\bf x}= D{\hat {\bf n}})
= -{1 \over 5} \int {d^3 k \over (2\pi)^3} \curv_{\bf k} e^{i {\bf k} \cdot D{\hat{\bf  n}}},
\label{eqn:flatskySW}
\end{equation}
where $D$ is the distance to recombination.  The power spectrum then becomes
\begin{equation}\label{eqn:SWcell}
C_\ell = {1 \over 5^2 D^2} \int{ d k_{ \parallel} \over 2\pi} P_\curv({\bf k}=({\bf l}/D,k_{\parallel})),
\end{equation}
where $\parallel$ is the direction along the line of sight, orthogonal to the
plane of the sky.

Taking the $k\etastep \gg 1$ and $d\rightarrow 0$ limits of the change to the power spectrum in 
Eq.~(\ref{eqn:poweranalytic}) and assuming $\Delta_{{\cal R},0}^2 \approx $const. so that
\begin{equation}
C_{\ell,0} = \frac{2\pi}{\ell^2} \frac{\Delta_{{\cal R},0}^2}{5^2},
\label{eqn:clscaleinvariant}
\end{equation}
 the fractional change in the power spectrum is given by
 \begin{equation}
\frac{\Delta C_\ell}{C_\ell} \approx 
-\amp P\left(\frac{2 \ell \etastep}{D}\right)
\label{eqn:projection}
\end{equation}
with the projection factor 
\begin{align}
P(x) = &\int_1^\infty {d z \over z^2} \frac{1}{\sqrt{z^2-1}} \cos(x z) 
\nonumber\\
\approx & \sqrt{\frac{\pi}{2x}}  \cos(x+\pi/4),
\end{align}
where the approximation is in the $x \gg 1$ or $\ell \gg D/2\etastep$ limit and can be proven
by considering that $d^2 P/dx^2=(\pi/2) Y_0(x)$.  Rapid oscillations in $k$-space are therefore suppressed in $\ell$-space by a factor of $(\ell/\ell_f)^{-1/2}$ where
$\ell_f = D/\etastep$.  
It is useful then to scale the numerical results to this expectation by defining the scaling
factor
\begin{equation}
A_C = \amp \left( \frac{\etastep}{1 {\rm Gpc}} \right)^{-1/2}.
\label{eqn:AC}
\end{equation} 
Note that dividing out by the factor of $A_C$ appropriately rescales the oscillations to be order unity at low $\ell$  and damp as $\ell^{-1/2}$ according to the envelope
\begin{equation}
 \sqrt{\frac{\pi}{2}} \left(\frac{\ell}{D/(1 {\rm Gpc})} \right)^{-1/2}.
 \label{eqn:projectionenvelope}
 \end{equation}
In Fig.~\ref{plot:DeltaClTT_c1dm5_dm5_kf6p9dm4_unlens_vs_lens} we show a comparison between the full numerical result and these expectations for a model with
$V_0 = m^2 \phi^2/2$, and parameters defined in Table \ref{table:modelparameters}.  With this model, 
$\eta_f=1.44$ Gpc and $D=14.18$ Gpc.
We take for illustrative purposes $c=10^{-5}$ and $d=10^{-5}$.

There are two notable differences between the full result vs.\ the flat sky Sachs-Wolfe scaling.
The first is that the oscillations are modulated by the acoustic transfer.   This reflects
the fact that at nodes in the acoustic oscillations between the acoustic peaks there is no
transfer of power to local temperature fluctuations.    The second notable effect is
a stronger damping starting at $\ell \sim 10^3$.   This is due to gravitational lensing
as demonstrated in  Fig.~\ref{plot:DeltaClTT_c1dm5_dm5_kf6p9dm4_unlens_vs_lens}.
Without lensing the envelope follows the $\ell^{-1/2}$ scaling of Eq.~(\ref{eqn:projectionenvelope}) as expected.
For this reason, when considering the bispectrum where lensing effects are more
difficult to calculate, we will always take an $\ell_{\rm max}=2000$ where lensing effects
become order unity for the power spectrum oscillations.   

It is also interesting to explore the scaling with the feature scale $\etastep$.   
For a model with features at the WMAP power spectrum glitches $\etastep \approx  1.44$ Gpc. For changes by a factor of
a few around this value, the results scale as expected as shown in 
Fig.~\ref{plot:DeltaClTT_c1dm5_dm5_kfseries_nolensing}.  However, as the feature
scale approaches the distance to recombination, projection effects take on a very
different character.   As discussed in Appendix \ref{sec:realspace}, sky curvature prevents
a superhorizon feature scale from leaving any imprint on the CMB.

For this reason, when considering the bispectrum where we calculate in the flat sky approximation, we restrict ourselves to $\etastep \lesssim 10$ Gpc.

\subsection{Constraints} \label{sec:constraints}

The WMAP7 temperature and polarization power spectra place constraints on the height, width, and location of a step in the inflaton potential.    These constraints then limit
the observability of corresponding features in the bispectrum.   In order to limit models
to reasonable cosmologies we also add the following data sets:  
BICEP and QUAD, which include polarization constraints \cite{Chiang:2009xsa,Brown:2009uy},  UNION2\footnote{\url{http://www.supernova.lbl.gov/Union}} supernovae data, the SHOES measurement of $H_0=(74.2\pm3.6)$ km/s/Mpc \cite{SHOES} and a big bang nucleosynthesis constraint of $\Omega_bh^2=0.022\pm0.002$ \cite{Burles:2000zk}. We include the effect of gravitational lensing on the CMB.

With these data sets and their likelihood functions, we perform a MCMC likelihood analysis
on the joint step and cosmological parameters.
We parametrize the initial curvature power spectrum as
\begin{align}
\ln \Delta_\curv^2(k) = & \ln A_s + (n_s-1)\ln \left( \frac{k}{0.05 {\rm Mpc^{-1}}}\right)\\
&+ \frac{A_C}{3}  
\left( \frac{\etastep}{1 {\rm Gpc}} \right)^{1/2} \damp\left( \frac{k\etastep}{x_d} \right) W'(k\etastep). \nonumber
\end{align}
Since we are interested in the $x_d \rightarrow \infty$ limit, we take a sufficiently small
width $d$ in which the damping behavior falls outside of the WMAP range of observation.
The initial conditions are then described by 4 parameters $\{ A_s, n_s, A_C,\ln \etastep \}$,
to which we add 4 cosmological parameters $\{ \Omega_bh^2,\Omega_ch^2,\theta,\tau\}$ in a flat
$\Lambda$CDM context.   
We use $A_C$ as the normalization parameter since
we expect its errors to be roughly independent of $\etastep$.

{
We take flat  priors on each of these 8 parameters.   For all but $\ln\etastep$
the prior range is an uninformative $> 45\sigma$ of the posterior in each. }
We start with a coarse analysis in a wide range in $0.4 < \etastep/{\rm Gpc} < 12$.  
Errors on $A_C$ throughout the whole range are approximately $\sigma(A_C) \sim 0.03-0.06$, consistent with the scaling arguments above, but for certain discrete ranges of $\etastep$ there is a preference for non-zero
mean values.   These multiple maxima make a global MCMC analysis highly inefficient.

\begin{figure}[t]\centerline{\psfig{file=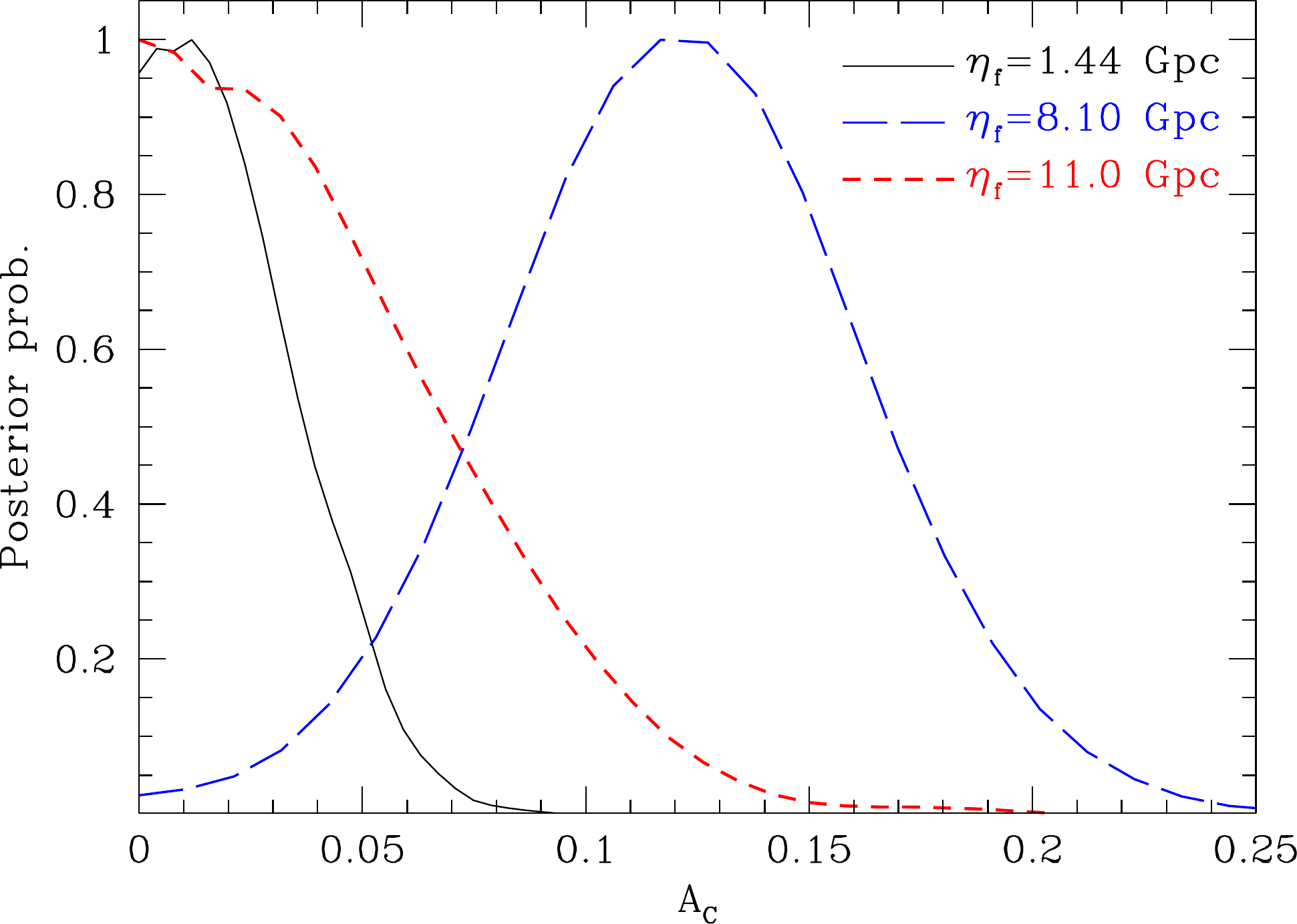, width=3.45in}}
\caption{\footnotesize Posterior probability distribution of the oscillation amplitude $A_C$
in the angular power spectrum for representative value of $\etastep=1.44$, {8.1} and $11$ Gpc.   
The constraint  weakens as $\etastep$ approaches the horizon for typical
cases.   For  $\etastep \sim 8.1$Gpc and other specific values, the peak of the posterior is shifted to $A_C\sim 0.1$
with comparable distribution width.}
\label{plot:posterior}
\end{figure}

In Fig.~\ref{plot:posterior}, we instead show the result of separate MCMC chains at fixed values of $\etastep$
away from these special regions at widely separated $\etastep=1.44$,  and 
11 Gpc.   The results here are consistent with $A_C=0$ and yield 1-sided 95\% CL bounds of $A_C < 0.05, 0.10$, respectively.

\begin{figure}[t]\centerline{\psfig{file=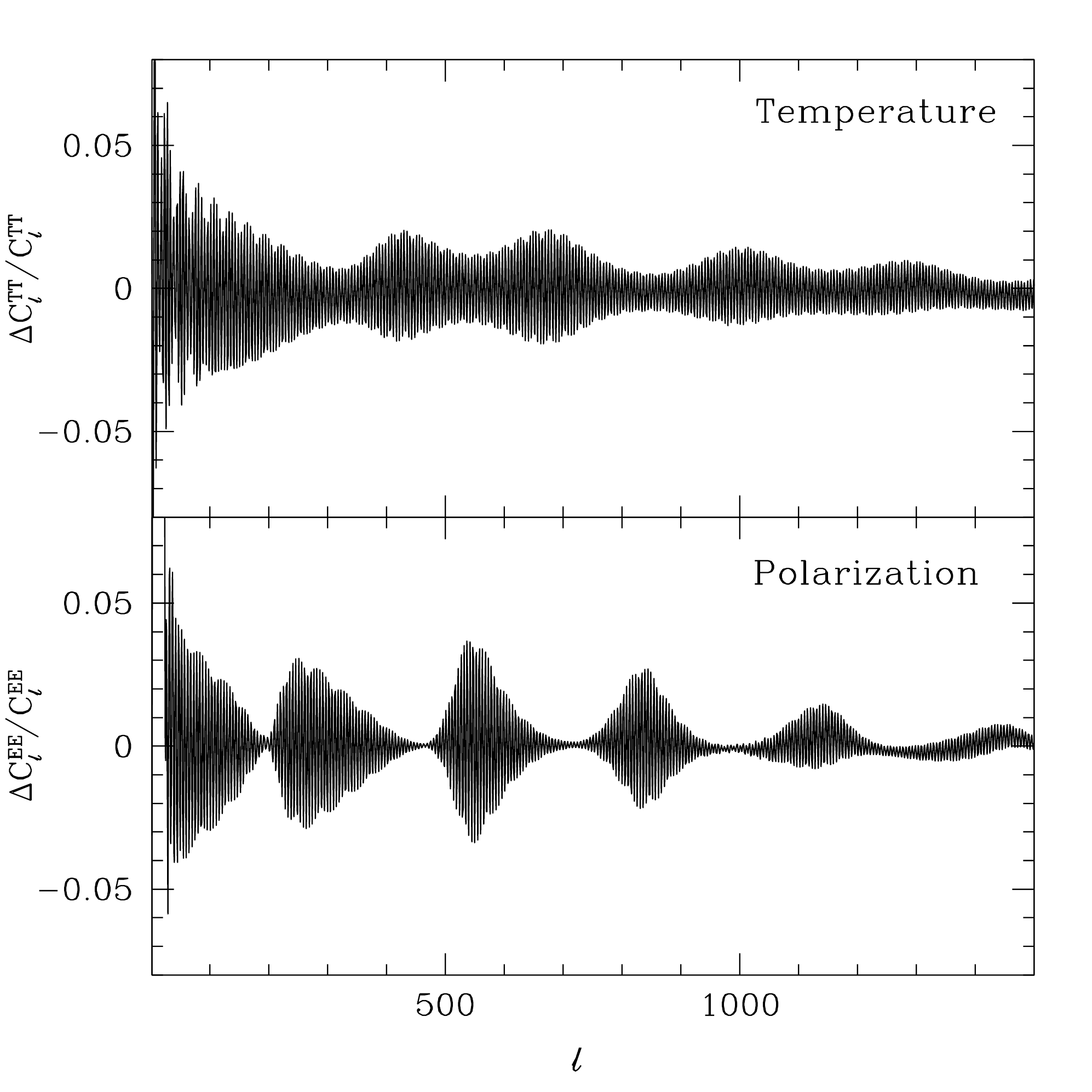, width=3.45in}}
\caption{\footnotesize Fractional angular power spectrum differences between the maximum likelihood sharp-step model ($A_C=0.11, \etastep=8.163$ Gpc)
and pure power law model $(A_C=0)$.   The WMAP data
prefer a few percent oscillation in the temperature power spectrum (top panel) around the first acoustic peak by 
$2\Delta \ln L = 11.6$.   This model predicts matching $E$-polarization power spectrum
oscillations (bottom panel) modulated by an orthogonal acoustic phase.}
\label{plot:ml}
\end{figure}

Around specific values of $\etastep$ a model with $A_C \approx 0.1$ is actually a better
fit to the data than $A_C=0$.  In Fig.~\ref{plot:posterior}, we show an example with
$\etastep=8.1$ Gpc.
 Preference for high  frequency oscillations in the power spectrum around the first acoustic peak in the WMAP data have previously been noted by ref.\ \cite{Martin:2003sg} and more recently by refs.\ \cite{Flauger:2009ab, Meerburg:2011gd} in different contexts.
In Fig.~\ref{plot:ml}, we plot an example from the 2D chain.  Here 
$\etastep=8.163$ Gpc and 
$A_C =0.11$.  
Note that
this model improves the total likelihood by $2\Delta\ln L = 11.5$ for the two extra parameters.  
Moreover, this improvement comes from the WMAP likelihood with
 $2\Delta\ln L_{\rm WMAP} = 11.6$.    
 Importantly, it does not come from the low multipole moments nor the glitches at $\ell =20-40$ but mainly oscillations {with a period of $\Delta \ell \approx 5.4$}  around the first 
 acoustic peak 
 $100 \le \ell \le 300$ with contributions continuing out to the beam scale. {Small
 variations in $\etastep$ around this value also produce similar improvements as long
 as the angular scale $\etastep/D$ remains fixed via allowed adjustments of other cosmological parameters, mainly $\Omega_m h^2$.}
 
 {Other specific values of $\etastep$ show comparable likelihood improvements}, for example at $\etastep=5.135$ Gpc and $A_C=0.09$,  $2\Delta\ln L = 9.9$.
 While this level of improvement for 2 parameters is notable, it is possible that it represents
 fitting of excess noise at the few percent level in $C_\ell$ that is not well modeled in the
 likelihood function.
 
 If these improvements are in fact signal and not noise, then there are sharp predictions
 that can be verified with future data.    We shall see in the next section that the bispectrum
 is observably large so long as the oscillations persist undamped to $\ell \gtrsim 500$.   Furthermore there must be matching features in the 
 polarization power spectra (see Fig.~\ref{plot:ml}).  The oscillations in the polarization power spectra are comparable but larger at peak as they are less affected by projection and modulated by an orthogonal acoustic transfer.  Detection of this matching signal in polarization would make a convincing case for a primordial origin of the improvement.

\section{CMB Bispectrum}\label{sec:bispectrumSNR}

We can now use our analytic results to estimate the signal-to-noise ratio (SNR) in the 
bispectrum.
We review the SNR calculation for a cosmic variance limited data set
in \S~\ref{sec:cv} and approximate methods for its estimation in \S~\ref{sec:biapprox}. 
From this approximate method, we derive scaling relations for its dependence on the amplitude, width and location of the step feature  in \S~\ref{sec:scaling} and calibrate them against numerical calculations in \S~\ref{sec:binumerical}.
\subsection{Cosmic variance} \label{sec:cv}

The temperature or angular bispectrum is defined as the three-point function 
of the spherical harmonic coefficients $a_{\ell m}$ of the temperature
anisotropy
\begin{equation}
B_{\ell_1 \ell_2 \ell_3} =\!\!\!\! \sum_{m_1 m_2 m_3} \left(
\begin{array}{ccc}
\ell_1 & \ell_2 & \ell_3 \\
m_1 & m_2 & m_3 
\end{array}
\right) \langle a_{\ell_1 m_1} a_{\ell_2 m_2} a_{\ell_3 m_3} 
\rangle .
\end{equation}
The cosmic variance of the Gaussian part of the field puts an irreducible
limit on the SNR of 
\begin{equation}
\left( { S \over N}\right)^2 = \sum_{\ell_3 \ge \ell_2 \ge \ell_1} {B^2_{\ell_1\ell_2\ell_3} \over C_{\ell_1} C_{\ell_2} C_{\ell_3}
d_{\ell_1\ell_2\ell_3}}\, ,
\label{eqn:sn}
\end{equation}
where
\begin{equation}
d_{\ell_1 \ell_2\ell_3} = 
[1+ \delta_{\ell_1\ell_2} + \delta_{\ell_2\ell_3} + \delta_{\ell_3 \ell_1} +
2\delta_{\ell_1 \ell_2} \delta_{\ell_2 \ell_3}],
\end{equation}
accounts for permuted contractions of repeated $\ell$'s 
and the angular power spectrum is defined by
\begin{equation}
\langle a^*_{\ell m} a_{\ell ' m'} \rangle = \delta_{\ell \ell'} \delta_{m m'} C_\ell \,.
\end{equation}
We thus require an efficient means of predicting the angular bispectrum given the
curvature bispectrum.

The bispectrum arising from the approximations in Eq.~(\ref{eqn:gsrlbi})  and Eq.~(\ref{eqn:intapprox}) looks especially formidable to project onto the angular sky due to the inseparable damping function $\damp(x)$. In Appendix \ref{app:separability}, we demonstrate that, to a good approximation, this bispectrum can be cast into an approximately separable form, which will make the full projection onto the angular sky a much more tractable problem. Still, this separable form requires computational intensive
operations making an exploration of the whole parameter space difficult.
We leave evaluation of this to future work.

\subsection{Approximations} \label{sec:biapprox}

Here we are interested only in an order of magnitude estimate for the SNR
through a crude computation of the angular bispectrum from the
curvature bispectrum.  
We therefore take the flat-sky approach
and the Sachs-Wolfe limit for the temperature anisotropy.

 In the flat sky approximation, the angular bispectrum is defined
 by the three-point function of the Fourier moments
 of the temperature field given by $a({\bf l})$
 \begin{equation}
\langle a({\bf l}_1) a({\bf l}_2) a({\bf l}_3) \rangle 
= (2\pi)^2 \delta({\bf l}_1+ {\bf l}_2+{\bf l}_3) B_{(\ell_1,\ell_2,\ell_3)} .
\end{equation}
 For $\ell_1$, $\ell_2$, $\ell_3 \gg 1$, it is related to the all-sky bispectra as
 \cite{Hu:2000ee}
 \begin{eqnarray}\label{eqn:flattofull}
B_{\ell_1 \ell_2 \ell_3} &=& \sqrt{ (2\ell_1+1)(2\ell_2+1)(2\ell_3+1)\over 4\pi} 
\left(
\begin{array}{ccc}
\ell_1 & \ell_2 & \ell_3 \\
0 & 0 & 0 
\end{array} \right) \nonumber\\
&& \times
B_{(\ell_1,\ell_2,\ell_3)}. 
\end{eqnarray}

Under the flat Sachs-Wolfe approximation of Eq.~(\ref{eqn:flatskySW}), the
bispectrum becomes
\begin{align}
 B_{(\ell_1,\ell_2,\ell_3)}
=&  -{2 \over 5^3 D^4}\int_{0}^\infty{ d k_{1 \parallel} \over 2\pi}
 \int_{-\infty}^\infty
{ d k_{2 \parallel} \over 2\pi} 
B_{\curv}(k_1,k_2,k_3),
\end{align}
where
\begin{align}
{\bf k}_1 = ({\bf l}_{1}/D,k_{1\parallel}), \quad
{\bf k}_2 = ({\bf l}_{2}/D,k_{2\parallel}), \quad
{\bf k}_3 = -{\bf k}_{1}-{\bf k}_{2}.
\end{align}

It is useful to note
 that with the correspondence between all-sky and flat-sky expressions \cite{Hu:2000ee}
 \begin{equation}
 \left( { S \over N}\right)^2 \approx 4\pi
 \int  \frac{d^2\ell_1}{(2\pi)^2} \int  \frac{d^2 \ell_2}{(2\pi)^2} { B_{(\ell_1,\ell_2,\ell_3)}^2 \over 6 C_{\ell_1}C_{\ell_2}C_{\ell_3}}.
\end{equation}
Scaling relations for the SNR can thus be derived from scaling relationships for the flat sky spectra combined with a counting of triangles in the available phase space.

\subsection{Scaling arguments}\label{sec:scaling}

Before we numerically compute Eq.~(\ref{eqn:sn}), we can estimate how strongly we expect the SNR to scale as we add triangles. 
\subsubsection{Local bispectrum}

As a proof of technique, let us first examine the case of a local non-Gaussianity where the 
reduced bispectrum is a constant or
\begin{align}
B_\curv(k_{1}, k_{2}, k_{3}) =\frac{6}{5} f_{\rm NL}\left[ P_\curv(k_{1})P_\curv(k_{2})+{\rm perm.}\right].
\end{align}
The signal is dominated by squeezed configurations $\ell_{\rm S} \ll \ell_{\rm L}$ and the above arguments imply that 
\begin{align}
B^{\rm local}_{(\ell_{\rm S}, \ell_{\rm L},\ell_{\rm L})} \propto \frac{\Delta^2_{\curv}\left({\ell_{\rm L}}/{D}\right)\Delta^2_{\curv}\left({\ell_{\rm S}}/{D}\right)}{\ell_{\rm L}^2\ell_{\rm S}^2}.
\end{align}
Consequently, the SNR for triangles with long and short sides between
$\ell_{\rm min}$ and $\ell_{\rm max}$ for a scale invariant power spectrum will go as
\begin{align}
\left(\frac{S}{N}\right)^2  \propto 
\int d^2 \ell_S \int  d^2\ell_L \frac{\ell_{\rm L}^4 \ell_{\rm S}^2 }{\ell_{\rm L}^4\ell_{\rm S}^4} 
\propto \ell_{\rm max}^2\ln\left(\frac{\ell_{\rm max}}{\ell_{\rm min}}\right),
\end{align}
where we have used  Eq.~(\ref{eqn:clscaleinvariant}) to take $C_\ell \propto \ell^{-2}$ in the
same flat-sky, Sachs-Wolfe limit.

The integrals can be thought of as counting triangles:
 there are $\ell_L^2$ ways of choosing one of the long sides of the triangles and likewise
 $\ell_S^2$ ways of choosing the short side.   The final long side is determined by requiring the triangle close and so the total number of triangles becomes $\ell_L^2 \ell_S^2$. 
 
The naively infrared divergent integral is regulated by the lowest multipole available, which in our cosmic variance limited calculation is $\ell_{\rm min}=2$. This leaves an overall scaling of $\ell_{\rm max}^2$, in agreement with the cosmic variance limit of a well known result in the literature that the error on $f_{\rm NL}$ drops as $N_{\rm pix}^{-1/2}$, where $N_{\rm pix} = f_{\rm sky}\ell_{\rm max}^2$ is the number of  observed pixels \cite{Creminelli:2006gc, Babich:2004yc}. 

\subsubsection{Feature bispectrum}

We can now look at our feature bispectrum. There are now three cases we need to consider, the SNR being dominated by contributions from:
\begin{enumerate}

\item equilateral type shapes: $k_{1}\sim k_{2} \sim k_{3}$;

\item flat type shapes: $k_{1} \sim k_{2} \sim 2 k_{3}$; and

\item squeezed type shapes, as in the local case $k_{3}\ll k_{1}\sim k_{2}$.

\end{enumerate}
We consider these separately below, estimating how their contribution to the SNR scales with the number of modes.

\subsubsection*{Equilateral type}

Consider the leading contribution to the bispectrum of Eq.~(\ref{eqn:gsrlbi}) assuming a scale invariant spectrum, and taking $f_{0}^{-1} = \Delta_{\curv}(k)$. In the UV or large-$k$ limit, the bispectrum is dominated by the term quadratic in the perimeter of the momentum triangle
\begin{align}\nonumber\label{eqn:uvkspacebi}
 B_\curv(k_1,k_2,k_3)  \approx & \frac{\amp}{2}  {(2 \pi)^4 } {\Delta_{\cal R}^4 
\over 4}\damp\left(
\frac{K\etastep}{2x_{d}}\right)\\ & \times
\Big[\frac{1}{(k_1 k_2 k_3)^2}\left(K\etastep\right)^2\cos\left(K\etastep\right) \Big].
\end{align}
The effect of the damping envelope ${\cal D}(x)$ is to impose a limit on the maximum scale that can contribute to the SNR, $\ell_{\rm max}$. Neglecting the prefactors and taking $\ell_{f} = D/\etastep$, we expect the projection of the bispectrum in Eq.~(\ref{eqn:uvkspacebi}) onto the flat-sky to scale as
\begin{align}\label{eqn:snrequil}
B^{\feature}_{(\ell, \ell, \ell)} \propto \frac{1}{\ell^4}\left(\frac{\ell}{\ellfeat}\right)  = \ellfeat^{-1}\ell^{-3},
\end{align}
where the $(K\etastep)^2$ supplies two factors of $(\ell/\ell_f)$ and the two integrals
over $k_\parallel$ suppress the result by $(\ell/\ell_f)^{-1}$ due to the oscillatory integrands
as in the power spectrum.
Finally, the integrals over $\ell_1$ and $\ell_2$ provide a factor of $\ell_{\rm max}^4$
 and yield
\begin{align}
\left(\frac{S}{N}\right)_{\rm eq}^2  
\propto \ell_{\rm max}^4 \frac{\ellfeat^{-2}\ell_{\max}^{-6}}{\ell_{\max}^{-6}} \propto \frac{\ell_{\rm max}^{4}}{\ellfeat^{2}}.
\end{align}
The $\ell_{\rm max}^4$ factor can be thought of as a counting of equilateral type triangles: in three dimensions, there are $\ell_{\rm max}^2$ ways to choose the first side, another $\ell_{\rm max}^2$ ways to choose second side while the third side is determined by requiring the triangle close, yielding $\sim\ell_{\rm max}^4$ triangles.
\subsubsection*{Flat type}
Flat triangles scale equivalently to equilateral type triangles,
\begin{align}
B^{\feature}_{(\ell, \ell, 2\ell)} \propto \ellfeat^{-1}\ell^{-3},
\end{align}
however, due to the restriction that all modes be co-linear, this reduces the number of triangles available. Flat triangles correspond to the restriction  $|\ell_{1} -\ell_{2} - \ell_{3} |< \Delta L$, where $L = \ell_1+\ell_2+\ell_3$ and $\Delta L$ is the tolerance. With this restriction, we expect the number of triangles to scale as $\ell_{\rm max}^4 \sqrt{\Delta L/\ell_{\rm max}}$. The curved-sky imposes a minimum $\Delta L = 1$, and thus we expect the number of triangles to scale as $\ell_{\rm max}^{3.5}$ and the contribution of flat triangles to the SNR scales as
\begin{align}
\left(\frac{S}{N}\right)_{\rm flat}^2  
\propto \ell_{\rm max}^{3.5} \frac{\ellfeat^{-2}\ell_{\rm max}^{-6}}{\ell_{\max}^{-6}} \sim \ellfeat^{-2}\ell_{\rm max}^{3.5}.
\end{align}
Thus flat triangles contribute less to the total SNR than equilateral triangles
for $\ell_{\max} \gg \ell_f$.

\subsubsection*{Squeezed type}
Considering now squeezed triangles, in the limit $k_{\rm S} \ll k_{\rm L}$, and taking
the leading order term
\begin{align}\nonumber
B_\curv(k_L,k_L,k_S)  \approx& -\frac{\amp}{2}{(2 \pi)^4} {\Delta_{\cal R}^4\over 4} \damp\left(
\frac{k_{\rm L}\etastep}{x_d}\right)\\& \times
 2 k_{\rm L} \etastep \sin\left(2k_{\rm L}\etastep \right)
 k_{\rm L}^{-3}
 k_{\rm S}^{-3},
 \label{eqn:squeezedscaling}
 \end{align}
we estimate that this bispectrum contributes to the signal in squeezed angular space configurations $\ell_{S} \ll \ell_{L}$ as,
\begin{align}\label{eqn:squeezedbi}
B^{\feature}_{(\ell_{\rm S}, \ell_{\rm L},\ell_{\rm L})} \propto
\ell_{\rm S}^{-5/2} \ell_{\rm L}^{-3/2}  .
\end{align}
Compared to the result in Eq.~(\ref{eqn:snrequil}), for a fixed $\ell_{\rm S}$, the signal in squeezed triangles falls slower in the UV relative to the signal in equilateral triangles. However, considering the SNR,
\begin{equation}
\left(\frac{S}{N}\right)_{\rm sq}^2 \propto \int d^2 \ell_{\rm S} \int d^2 \ell_{\rm L} 
\ell_{\rm L} \ell_{\rm S}^{-3},
\end{equation}
we obtain
\begin{align}
\left(\frac{S}{N}\right)^2  \propto  \frac{\ell_{\rm max}^3}{\ell_{\rm min}},
\end{align}
which implies that the contribution of squeezed triangles compared to the equilateral type goes to zero as $\ell_{\rm max} \rightarrow \infty$. Interestingly, while squeezed triangles dominate the signal at high $\ell$, they do not dominate the signal to noise. The reason for this is two-fold. Squeezed triangles in themselves suffer from higher cosmic variance which, for a given triangle, eliminates the scaling advantage. In addition, there are many less triangles which contribute to the squeezed limit.

\subsection{Scaling and numerical results} \label{sec:binumerical}

To summarize the scaling relations of the previous section,
for small values of $\amp$, 
\begin{equation}
\left(\frac{S}{N} \right)^2 \propto \amp^{2} \frac{\ell_{\rm max}^4}{\ell_f^2}
 = 
\amp^{2} \frac{\ell_{\rm max}^4\etastep^2}{D^2},
\end{equation}
as $d \rightarrow 0$ and for $\ell_{\rm max}\gg \ell_f$. 
With this behavior in mind, we define the angular bispectrum amplitude,
\begin{align}\label{eqn:bispecamp}
A_B &= \amp \left( \frac{\etastep}{\rm 1 Gpc} \right) = \frac{6c}{\epsilonstep+ 3c} \left( \frac{\etastep}{\rm 1 Gpc} \right) \nonumber\\
&= A_C\left( \frac{\etastep}{\rm 1 Gpc} \right)^{3/2},
\end{align} 
which takes into account the dependence of the amplitude on the step scale, $\etastep$. 
Compared with $A_C$ the analogous amplitude for the angular power spectrum, the bispectrum amplitude increases as $\etastep^{3/2}$.   Recalling that the observational errors  on $A_C$ are only weakly dependent on $\etastep$, the SNR is maximized by placing the feature at the largest observable scales, 
near the current horizon.
 
With a finite $d$ there is an effective maximum multipole beyond which the SNR
saturates.  This scale depends on $\etastep$ as
\begin{align}
\ell_{d} =\frac{2 D}{  \etastep}x_d  = \frac{2D}{\etastep} \frac{\sqrt{2\epsilonstep+6c}}{\pi d} .
\label{eqn:ld}
\end{align}
In what follows we simply scale out the multipole space bispectrum amplitude $A_B$, and present the square of the SNR per unit $A_B$ as a function of the multipole space damping scale, $\ell_d$ or $\ell_{\rm max}$. These redefinitions render our results independent of the choice of the model parameters $\{ \phi_f,c,d \}$. Of course to make connection with a particular model, one must rescale by the appropriate factors of $A_{B}$ and choose the correct damping scale.

\begin{figure}
\centerline{\psfig{file=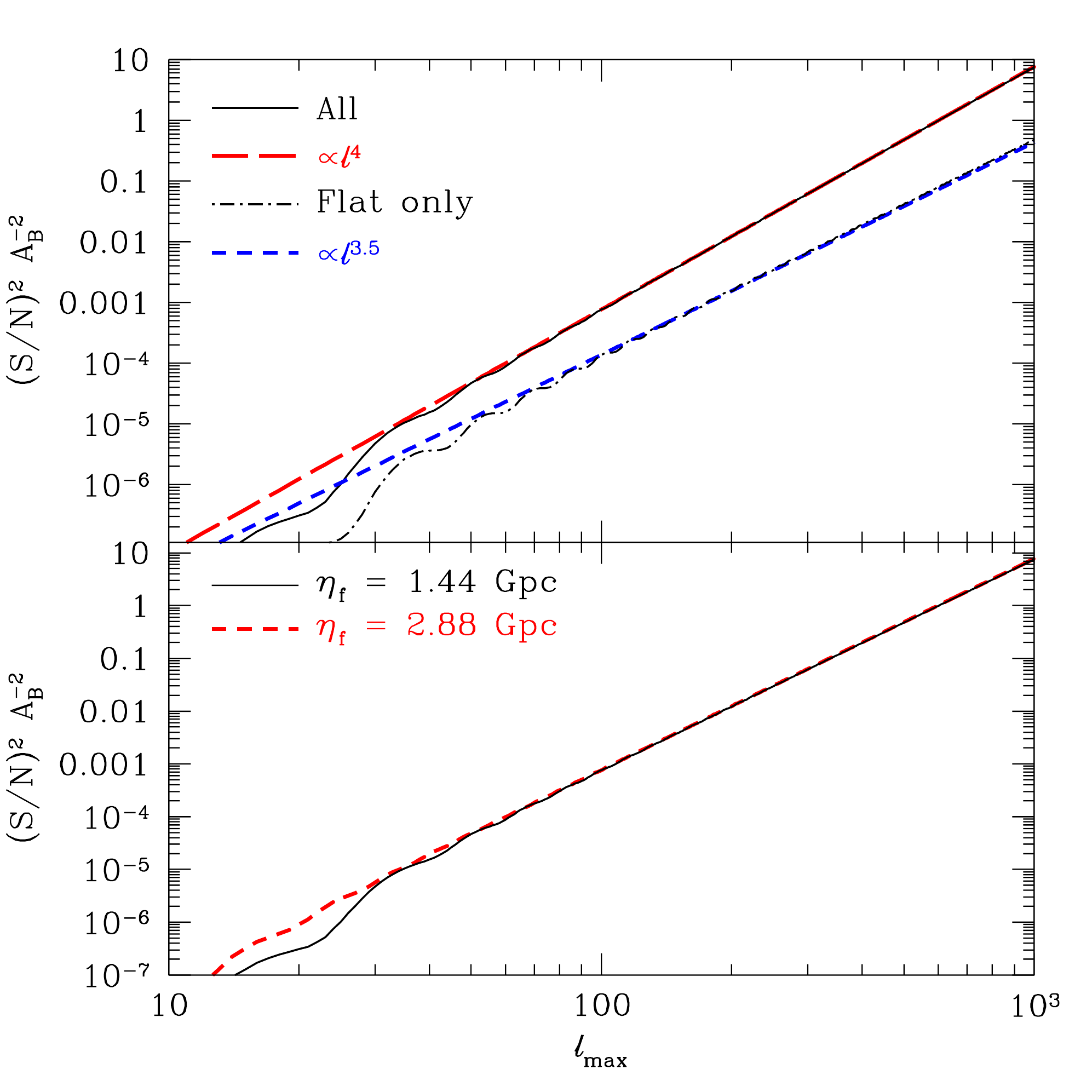, width = 3.45in}}
\caption{The square of the SNR per unit bispectrum amplitude ($A_{B}$) for $\ell_d = 8514$ for all triangles, and for flat configurations only. We also plot the scaling relations from \S \ref{sec:scaling}. In the lower panel, we demonstrate that changing $\eta_f$ while fixing  $A_{B}$ and $\ell_d$ leaves the SNR fixed at high $\ell$. \label{fig:SNRscal}}
\end{figure}

We verify the SNR scaling behaviors and obtain the proportionality coefficient in the scaling relations by numerically evaluating 
Eq.~(\ref{eqn:sn}). In the upper panel of Fig.~\ref{fig:SNRscal} we show the contribution to the SNR of all triangles and the contribution from triangles that are exactly flat. 
The close agreement of  $\ell^{4}$ and $\ell^{3.5}$   curves with the numerical results verifies our scaling relations, confirming that the SNR is dominated by triangles whose sides are all of comparable length. For this test,  the feature to occurs at  a scale $\etastep=1.44$ Gpc.

The lower panel in Fig.~\ref{fig:SNRscal} demonstrates that our results are independent 
of the specific choice of $\etastep$; plotted is the square of the SNR per unit bispectrum amplitude under a shift of the feature from 1.44 Gpc (black, solid curve)
to $2.88$ Gpc (red, dashed curve), while simultaneously reducing the width of the step as to hold the damping scale, $\ell_d$, constant.


\begin{figure}
\centerline{\psfig{file=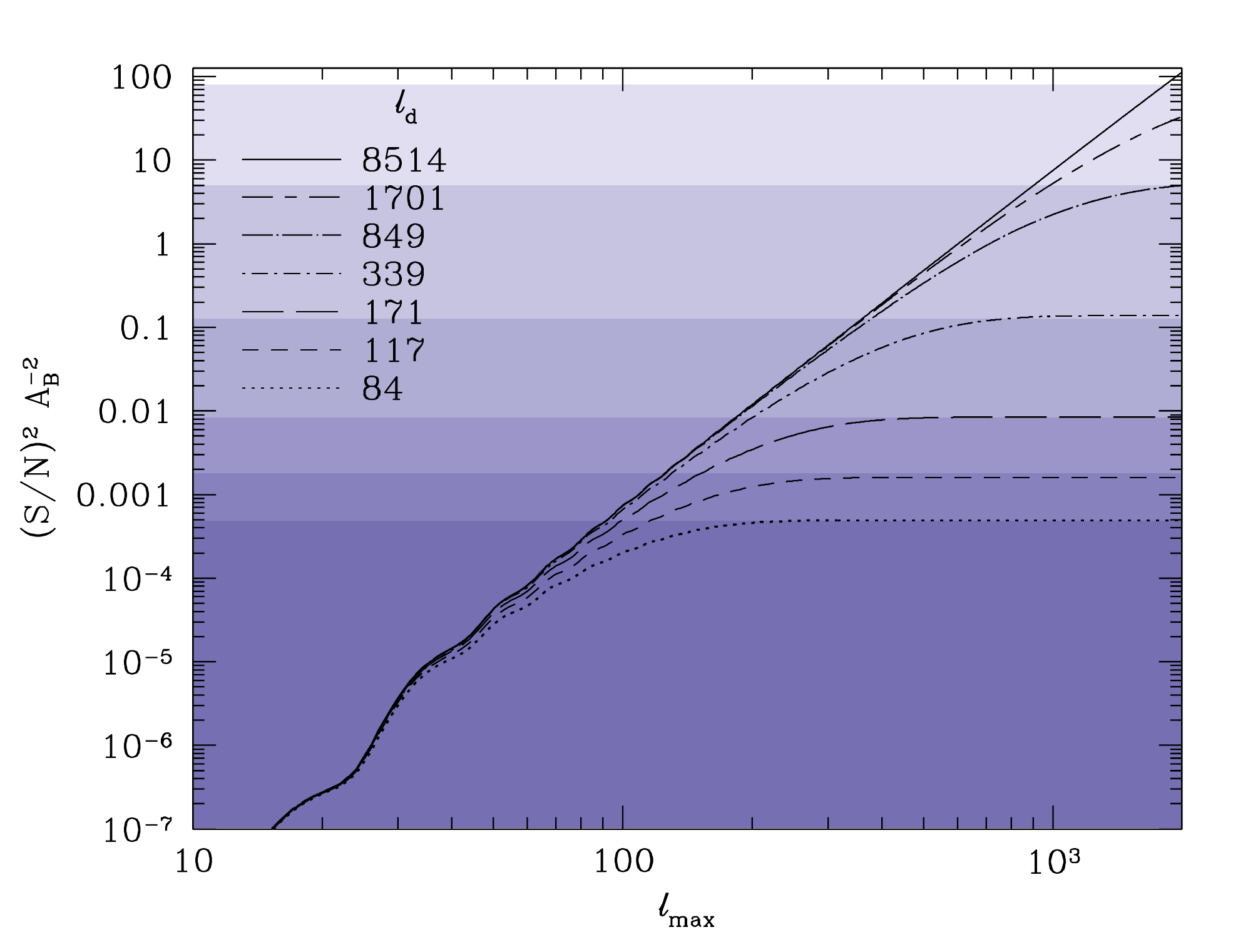, width = 3.45in}}
\caption{The square of the SNR per unit bispectrum amplitude ($A_{B}$) for various damping scales $\ell_d$. The shaded regions indicate the prediction of our scaling formula in Eq.~(\ref{eqn:SNRScaling}) and match the asymptotic high $\ell \gg \ell_d$ saturation point for each case. \label{fig:SNRoverc}}
\end{figure}

In Fig.~\ref{fig:SNRoverc}, we show the dependence of the SNR on the
angular damping scale $\ell_d$.  Notice the SNR saturates once the maximum multipole $\ell_{\rm max}$ exceeds the damping scale. 
In fact, to excellent approximation, the value at saturation can be approximated from the
$d=0$ results by setting $\ell_{\rm max}=1.06 \ell_d$.    However, for values of $d$ where   
$\ell_{d}>2000$ secondary anisotropy contributions such as gravitational lensing and the Sunyaev-Zeldovich effect become important.   Furthermore the Planck satellite will be limited by its instrumental beam to lower multipoles.

Thus our rough approximation of the maximal signal to noise accessible to the CMB temperature bispectrum can be expressed in terms of our scaling parameters and the 
amplitude of the features in the temperature power spectrum as
\begin{equation}\label{eqn:SNRScaling}
\left( \frac{S}{N} \right)^2 \approx 1.2 \left( \frac{A_C}{0.1} \right)^2\left( \frac{\etastep}{\rm 1 Gpc} \right)^{3}
\left(\frac{\ell_{\rm max}}{2000}\right)^4,
\end{equation}
where $\ell_{\rm max} = {\rm min}(1.06 \ell_d,2000)$.
Combined with the upper limit of $A_C \lesssim 0.05$ for a typical value of $\etastep$,
(see Fig.~\ref{plot:posterior}), one infers that sharp steps with $\etastep \gtrsim 2$ Gpc have potentially observable features in the bispectrum while still being consistent with current power spectrum results.  For the maximum likelihood model, where $A_C=0.11$ and 
$\etastep=8.163$ Gpc,   the maximal $(S/N)^2 \approx 790$.
In fact even taking $\ell_{\rm max}=800$, the $(S/N)^2 \approx 20$ and so there may even be  information about this model in the WMAP data if an optimal analysis can be performed.

Our scaling relation makes it seem that 
we can  increase the SNR without bound 
by moving the feature to larger and larger scales.
However, just as was the case in the power spectrum, our flat-sky approximation overestimates the bispectrum as $\etastep$ approaches the current horizon due to projection effects on the curved sky.   Adopting the same
maximal scale of 
$\etastep < 10$Gpc as was found for the power spectrum, the SNR can be at most
comparable to that of the maximum likelihood model.

 Should a horizon scale feature be detected in the data then one should
also consider the sample (co)variance of the finite number of such features that can fit in 
our horizon volume when measuring the parameters associated with the feature.  
 For upper limits, the Gaussian noise variance suffices.

\section{Discussion}
\label{sec:discussion}

We have investigated the power spectrum and bispectrum of curvature fluctuations generated by a sharp-step feature in the inflationary potential. In the limit where the change in potential energy due to the step is small compared to the kinetic energy of the inflaton, we have demonstrated that the background system can be solved perturbatively in this small
parameter.

Using this solution for the background evolution of the inflaton,  we obtained closed form, approximate analytic solutions for both the power spectrum and bispectrum of curvature fluctuations. The analytic solutions generically take the form of an oscillating window function times a damping function. The form of the window function is set by the behavior of the step in the infinitely sharp limit with its characteristic frequency determined by the step position in conformal time, $\etastep$. The  damping function on the other hand determines the maximal wavenumber out to which the oscillations persist, which scales with the inverse of the
finite width $d$, with a form factor that depends on the shape of the step.  In this work we have explicitly calculated the damping function for a step with a hyperbolic tangent shape. 

Our analytic power spectrum and bispectrum solutions pass several important consistency checks. 
They vanish for modes that are superhorizon size when the field crosses the step, and thus do not introduce spurious superhorizon effects.  They satisfy the consistency relation which relates the squeezed limit of the bispectrum to the tilt of the power spectrum, or scalar spectral index. Additionally, leading order corrections to this relationship scale only as the square of the squeezed side. Numerically, for very small perturbations to the kinetic energy, our solutions are accurate to approximately 10\%. This difference is largely due to slow roll corrections to the mode functions, which introduce additional phase shifts and thus alter the form of the non-Gaussianity. Somewhat surprisingly, rather than one percent level corrections as one might naively guess, these type of corrections generically lead to 10\% corrections to the bispectrum.

As the step height increases, and the perturbation to the kinetic energy becomes large, the analytic solutions begin to break down due to two effects. First, the perturbation to the kinetic energy becomes large and inflation ends at an appreciably different time for a potential with
and without the step.    This difference corresponds to a change in the matching of
inflationary field scales to physical scales and hence a 
phase shift between the analytic solution and its numerical counterpart. We  make use of the approximate time translation invariance of the inflationary background to absorb this phase shift into a redefinition of the conformal time of the unperturbed background. The second effect is that as the kinetic energy perturbation grows to order unity, our analytic solutions attain a larger amplitude while damping away at a faster rate compared with the numerical results. In our analytic solution these properties are controlled by a single parameter, $\epsilon_H$, which effectively measures the ratio of kinetic to potential energy of the inflaton. At leading order, our analytic solution takes into account only the kinetic energy of the inflaton on the background without the step. However, we have shown that the discrepancy between the numerical and analytic results can be substantially improved by including the first order kinetic energy perturbation in $\epsilon_{H}$.

As the step width becomes infinitely sharp, the high wavenumber limit of the spectra exhibits some interesting behavior. We have found that the perturbation to the curvature power spectrum approaches a constant amplitude oscillation, while the reduced bispectrum, or effective non-linearity, diverges quadratically in the perimeter of the momentum space triangle.
The 
reduced bispectrum or non-linearity  parameter $f_{\rm NL}$ can thus reach $10^4$ or more.
While one might naively infer  a strongly divergent non-Gaussianity on small spatial or angular scales,
the real space analogues of the power spectrum and bispectrum, the two- and three-point correlation functions remain tiny for most configurations, in particular all those involving only small separations. Oscillations largely prevent the various triangle configurations adding up to anything significant:\ For the most part they simply average to zero. However, for certain configurations, where the separation between points is comparable to the physical scale of the feature, resonance in the integral leads to logarithmically large values of the correlation.    High wavenumber
 oscillations in the spectra therefore are the result of Fourier transforming a sharp 
feature in the correlation functions on large scales.

The real space picture of the bispectrum suggests that the $k$-space triangles
are not all independent as they must sum to yield a sharp feature at a fixed, large scale.   This covariance of triangles must be accounted for if such a signal in the bispectrum is 
detected in the future.
In principle, one could use our methods to evaluate higher $N$-point functions which quantify this covariance and that of the power spectrum.   Furthermore, they would provide corroborating evidence for a step and are expected to grow even more strongly with the momentum perimeter than the bispectrum considered here.

On the other hand, to assess the detectability of the bispectrum, 
it is sufficient to compute the SNR
where the noise is attributed to Gaussian random fluctuations only.
We utilize our analytic bispectrum solutions to  assess  detectability as a function of the step location, height and width.  For simplicity, we neglect acoustic transfer and the sky curvature, working in the Sachs-Wolfe and flat-sky limits.  This over-simplification suffices for our order of magnitude estimate. Under this approximation, the signal-to-noise ratio is dominated by equilateral triangles and scales with the fourth power of the maximum multipole due to resolution or finite width $d$.   These scaling relations allow
us to characterize observability as a function of the step parameters. 

The SNR is maximized by placing a sharp feature  close to the horizon scale. Since the effective non-linearity grows quadratically with the ratio of the feature scale to the observed
wavelength, placing the feature at larger scales means it can grow to a larger amplitude within the window probed by the CMB. The maximum scale at which one can place the feature before it becomes unobservable is the horizon scale for  both the power spectrum and bispectrum. In real space, the high-$k$ behavior of either translates to sharp correlation features at separations comparable to the physical scale of the step. Because of 
causality, an observer confined to a single position cannot measure a correlation across a distance larger than the horizon.   Since  in the $d\rightarrow0$ limit oscillations in the power spectrum persist even for superhorizon scale features,
the momentum space results naively seem problematic.   However, as the horizon scale is approached, the frequency of the oscillations in the spectra approach the fundamental spacing of the spherical harmonics, $\Delta \ell = 1$. Thus, as this scale is exceeded, CMB spectra are rendered insensitive to the presence of the feature, except for an unobservable shift in the monopole. 

While one might guess that the angular power spectrum of the CMB would place severe restrictions on large, sharp steps due to their oscillations persisting into the strongly constrained $\ell \sim 200 -400$ region, in fact the constraints are considerably weaker. There is an additional source of damping when the oscillating features in the momentum space power spectrum are projected onto the sky. Additionally, effects such as gravitational lensing tend to hide the oscillations on very small scales. Curiously, the WMAP power spectrum data are in fact
better fit by such a feature than a pure power law spectrum with $2\Delta \ln L \approx 12$ for the 2 extra parameters.  While this improvement may be due to fitting excess noise in WMAP, it is formally more significant than a similar fit of the well-known glitch at $\ell =20-40$ with 3 extra parameters \cite{Peiris:2003ff,Covi:2006ci}.

If this improvement is due to a slow roll violating feature such as a step, there are testable
consequences in both the polarization power spectrum and the bispectrum.   The polarization power spectrum should carry matching oscillations whose amplitude is 
less affected by projection.  The bispectrum would be detectable as long as the
oscillations continue undamped to at least $\ell \sim 500$. In this case, the bispectrum may be able to confirm the primordial origin of such a signal.

\acknowledgements

We thank Andy Albrecht, Xingang Chen, Richard Easther, Raphael Flauger, and Mark Wyman for useful conversations.
This work was supported in part by the Kavli Institute for Cosmological Physics at the University of Chicago through grants NSF PHY-0114422 and NSF PHY-0551142 and an endowment from the Kavli Foundation and its founder Fred Kavli. CD was additionally supported by the Institute for Advanced Study through the NSF grant AST-0807444 and the Raymond and Beverly Sackler Funds. WH was additionally supported by U.S.~Dept.\ of Energy contract DE-FG02-90ER-40560 and the David and Lucile Packard Foundation. 
\appendix

\section{Real Space Correlation}
\label{sec:realspace}

Given that the step feature is localized in time, it is interesting to consider the correlation functions, the real space
analogs of the power spectrum and bispectrum.   While somewhat orthogonal to the main thrust of the paper, we will see that the real space correlation functions give insight into causality, sample vs.\ noise variance, 
consistency relation and the perturbative validity of our expansion of the action. In particular, we shall see that the $k \rightarrow \infty$ behavior of the power spectrum and bispectrum correspond to sharp features in the real space correlation functions. 

This Appendix is organized as follows. In \S \ref{sec:2ptreal}, we demonstrate that, in the limit in which $d= 0$, the perturbation to the curvature power spectrum leads to a sharp feature in real space, the modes as $k\to\infty$ summing to give a logarithmically divergent derivative at a single point associated with the physical scale at which the feature occurs. In \S \ref{sec:3ptreal}, we show that, despite attaining a large non-linearity, $f_{\rm NL}(k)$, the real space analogue of the bispectrum remains small everywhere, except in the vicinity special points, associated with the physical scale of the feature, where it becomes large, diverging as $\sim \ln(d)$. This behavior is shown to be  due to the fact that the momentum space triangles only coherently add for very special configurations of the spatial points which lead to resonances in the Fourier integrals.  In \S \ref{App:squeezed}, we show that at least some of this behavior can be anticipated from the consistency relation, which relates the slope of the two-point correlation function to a certain limit of the three-point correlation function.

\subsection{Two-point correlation}\label{sec:2ptreal}

The two-point correlation function is the Fourier transform of the power spectrum
\begin{eqnarray}
\xi_{2\curv}({\bf r}_1,{\bf r}_2) &\equiv& 
\langle \curv({\bf r}_1)\curv({\bf r}_2)\rangle \nonumber\\ 
&=&
 \left[ \prod_{i = 1}^{2}\int\frac{d^{3}k_{i}}{(2\pi)^3}e^{i{\bf k}_{i}\cdot{\bf r}_i}\right] P_{\curv}(k_1)\nonumber\\
&&\times (2\pi)^3\delta^3({\bf k}_1+{\bf k}_2)\nonumber\\
&=& \int {d k \over k} {\sin(k r) \over kr} \Delta_\curv^2(k),
\end{eqnarray}
where $r = | {\bf r}_1 -{\bf r}_2|$.   

With our analytic calculation for the power spectrum of the step feature in \S \ref{sec:powerspec}
we showed that 
in the limit in which the step width $d\rightarrow 0$, the power spectrum correction attains a constant amplitude oscillation which persists to $k \rightarrow \infty$. 
Corresponding to this high frequency behavior, there must be sharp features
in the correlation function.

The contribution of the step feature to $\xi(r)$ can be calculated from Eq.~(\ref{eqn:poweranalytic}) as
\begin{align}\label{eqn:realspacecorrelation}
\Delta\xi_{2\curv}(r)  = \frac{\amp}{3}\Delta_{{\cal R},0}^2\int \frac{d k}{k} \frac{\sin(kr)}{kr}
\damp\left( \frac{\pi d}{\sqrt{2\epsilonstep}} k\etastep \right) 
W'( k\etastep).
\end{align}
When $d = 0$, Eq.~(\ref{eqn:realspacecorrelation}) can be evaluated analytically, and one obtains,
\begin{equation}
\Delta\xi_{2\curv}(r)  =  - \amp \Delta_{{\cal R},0}^2 J(r/2\etastep),
\end{equation}
with
\begin{align}
J(x) = \frac{1}{3} -\frac{x^2}{2}   +\frac{x}{2}  ( x^2- 1  )
 \left\{  
\begin{array}{cc}
 \coth
   ^{-1} x ,  & x>1, \\
 \tanh
   ^{-1} x ,  & x<1.
\end{array}
\right.
\end{align}

\begin{figure}
\centerline{\psfig{file=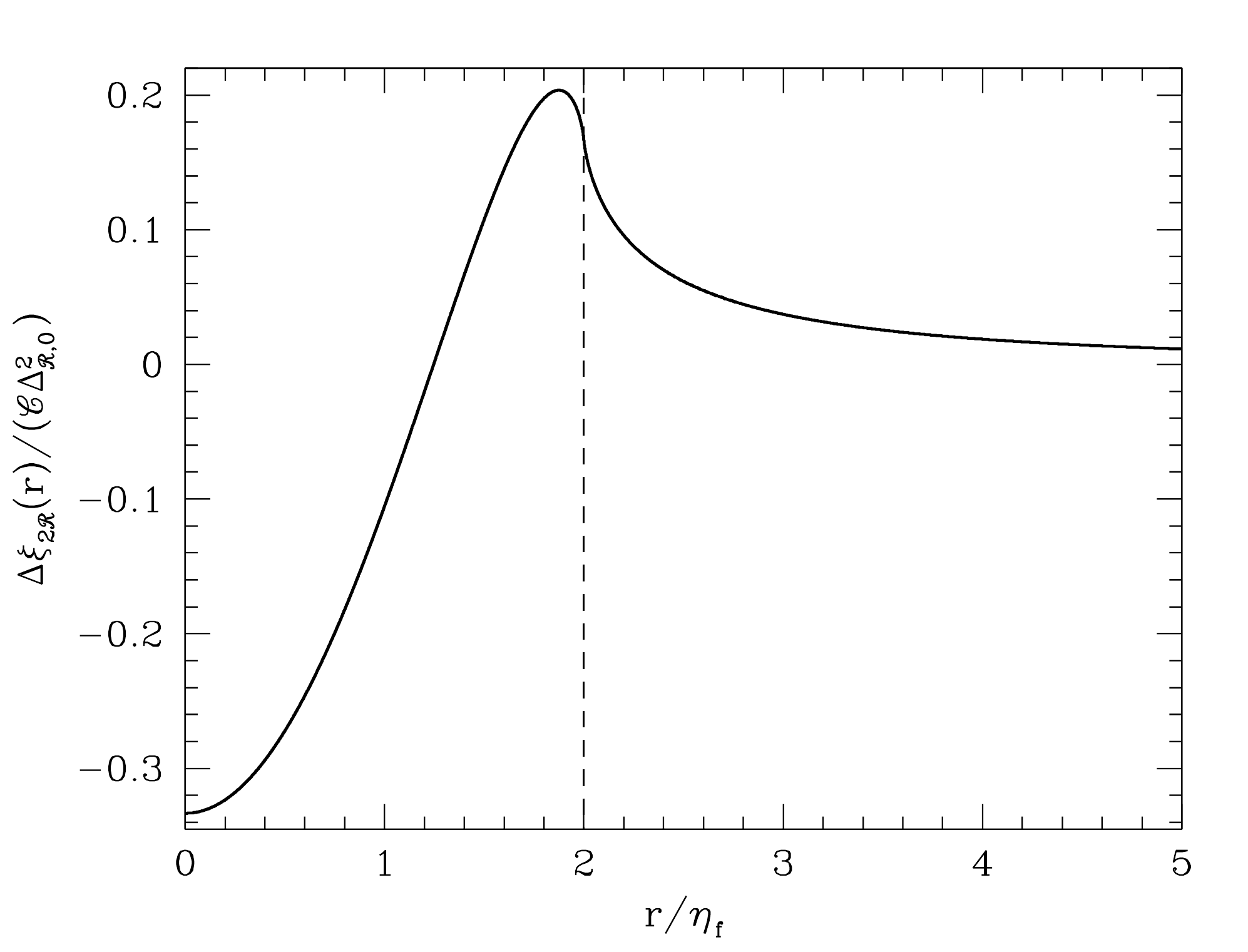, width=3.45in}}
\caption{Real space two-point correlation function changes due to the addition of
 a step feature ($d \rightarrow 0$).   At $r/\etastep=2$, the slope of the correlation function diverges as the dashed line indicates.}
\label{fig:2pt}
\end{figure}

We plot this function in Fig. \ref{fig:2pt}. In this case,  the sum over all of the modes out to infinity results in a sharp feature in the slope at $r = 2\etastep$ while the function remains finite, and order $\mathcal{O}(\amp)$ everywhere.
The fact that $r=2\etastep$ is a special point can in fact be read directly off of the integrand.   
Since $W' \propto \cos(2 k \etastep)$ at high $k$ and the exponentials in the transform give
$\sin(k r)$, contributions from different $k$-modes oscillate away except for the stationary
phase point $r = 2\etastep$.   We will use this type of reasoning to deduce the 
behavior of the three-point function in \S \ref{sec:resonances}.

The slope of the correlation function $\xi_{2\curv}(r)$ formally diverges for $d=0$  at the point $ r = 2\etastep$.   More generally 
\begin{align}\label{eqn:realspacecorrelationderiv}
\frac{ d\Delta\xi_{2\curv}}{d\ln r} = & \frac{\amp}{3}\Delta_{{\cal R},0}^2\int \frac{d k}{k}  
\left( \cos(kr) - \frac{\sin(kr)}{kr} \right)
\\ & \times
\damp\left( \frac{\pi d}{\sqrt{2\epsilonstep}} k\etastep \right) 
W'( k\etastep). \nonumber
\end{align}
Given that the damping envelope supplies a $k_{\rm max} \propto d^{-1}$, 
for $k_{\rm min} \etastep \gg 1$
\begin{align}
\frac{ d\Delta\xi_{2\curv}}{d\ln r} \propto \int_{k_{\rm min}}^{\rm k_{\rm max}} \frac{dk }{k} \cos(k r) \cos(2k\etastep) ,
\end{align}
and this integral
is logarithmically divergent  with $d$ at $r=2\etastep$.
 We will relate this divergence to the bispectrum behavior in \S \ref{App:squeezed}.

Examination of Figs.\ \ref{plot:DeltaPofk_test_pertcond1a} and  \ref{fig:2pt} also raises an interesting question about causality. As we have demonstrated, a sharp feature in the inflationary potential gives rise to oscillating features in the power spectrum of curvature fluctuations and as the step becomes sharper and sharper, the oscillating features ring to larger and larger wavenumbers, $k$, before decaying. 
One might then be tempted to conclude that an observer who only had access to a region of space $\eta  < \eta_f$, could measure superhorizon perturbations, by observing oscillations in his power spectra on small scales.

We have, however, also shown that, in real space, these modes simply sum up to a sharp feature at the physical scale corresponding to the size of the physical horizon when the inflaton crosses the feature. One therefore comes to the conclusion that an observer who has access to only a region of space $\eta \ll \etastep$ cannot measure the feature.
The resolution to this apparent discrepancy is that the $k$-space features are spaced
by less than $1/\eta$ and so cannot be sampled in the finite volume.  
In the CMB this is imposed by the finite volume within the distance to recombination. In large scale structure, this is similarly imposed by the finite
survey volume.

More specifically,  in the case of the CMB the projection of these oscillations onto the angular sky  imposes a very physical cutoff on the frequency of the oscillations, corresponding to the fundamental spacing of the multipoles, $\Delta \ell = 1$. Oscillations in the power spectrum with a frequency
spacing smaller than this fundamental are rendered unobservable when projected onto the spherical sky. In the Sachs-Wolfe approximation,
\begin{equation}
\frac{ \ell (\ell+1) C_\ell}{2\pi} = \frac{1}{5^2} \int \frac{dk}{k} j_{\ell}^2(kD) \Delta_\curv^2(k).
\end{equation}
In the example at hand, at $k\etastep \gg 1$
\begin{equation}
\Delta_\curv^2(k) \approx \Delta_{\curv,0}^2(k) 
\left[ 1 - \amp \cos(2 k \etastep) \right] \,.
\end{equation}
Since the Bessel function oscillates only at a frequency $kD$, if $2\etastep \gg D$ then
the oscillatory piece cancels out of the integral.
This cutoff prevents an observer from obtaining information about sharp features on scales larger than their horizon. However, we note that this cutoff is only exactly enforced by the full projection onto the spherical sky.
 In particular, in the periodic flat-sky approximation despite enforcing a fundamental spacing of
$\Delta \ell =1$, features with oscillations at $\Delta \ell < 1$ 
can imprint beat frequencies on the remaining modes making $\delta C_\ell/C_\ell 
\propto (\ell \etastep/D)^{-1/2}$ as in Eq.~(\ref{eqn:projection}). Fig.~\ref{plot:DeltaClTT_c1dm5_dm5_kfseries_nolensing} verifies that this scaling breaks in the
full sky calculation for $\etastep \gtrsim D$ leaving a suppressed contribution to the
anisotropy.

\subsection{Three-point correlation}\label{sec:3ptreal}

Similarly, the three-point correlation function is related to the bispectrum by a Fourier transform of
its arguments
\begin{eqnarray}
\xi_{3\curv}({\bf r}_1,{\bf r}_2,{\bf r}_3) &\equiv& 
\langle \curv({\bf r}_1)\curv({\bf r}_2)\curv({\bf r}_3)\rangle \nonumber\\
&=& \left[ \prod_{i = 1}^{3}\int\frac{d^{3}k_{i}}{(2\pi)^3}e^{i{\bf k}_{i}\cdot{\bf r}_i}\right] B_{\curv}(k_1,k_2,k_3)\nonumber\\
&&\times (2\pi)^3\delta^3({\bf k}_1+{\bf k}_2 + {\bf k}_3 ) \,.
\label{eqn:threepoint}
\end{eqnarray}
For a general configuration and the step potential bispectrum, this integral is difficult to compute.    As was the case with two-point function, we can gain insight on its behavior by
first studying resonances in the integrand where the phase is stationary.  We can then
confirm this behavior by explicitly evaluating simple configurations.

\begin{figure}
\centerline{\psfig{file=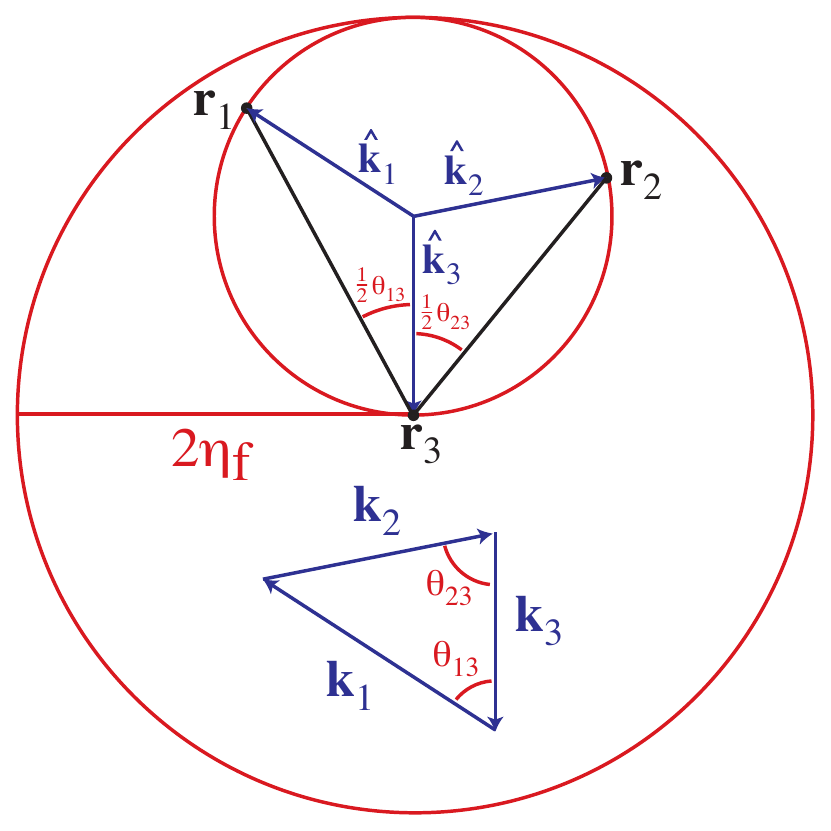, width=3.25in}}
\caption{Relationship between bispectrum triangle shapes and resonances in the three-point real space correlation function.  As $d \rightarrow 0$ each set of self-similar bispectrum triangles determined by $({\bf k}_1,{\bf k}_2,{\bf k}_3)s$, where $s$ is some scale factor,  causes a resonance at $({\bf r}_1,{\bf r}_2,{\bf r}_3)$ and its mirror image.  As the overall size of the triangle $s \rightarrow \infty$ this resonance
becomes sharper and sharper in real space but remain associated with 
separations  ${\cal O}(\etastep)$.  Flat triangles are responsible for the
largest scale resonance for ${\bf r}_1-{\bf r}_3={\bf r}_2-{\bf r}_3 = 2\etastep$.} \label{fig:resonance}
\end{figure}

\subsubsection{Resonances}
\label{sec:resonances}

The bispectrum of the step feature as its width $d \rightarrow 0$ and the wavenumbers
$k_i \rightarrow \infty$ have oscillatory behavior given by Eq.~(\ref{eqn:I0limit})
\begin{equation}
B_{\cal R}(k_1,k_2,k_3) \propto \cos[ (k_1+k_2+k_3)\eta_f] \,.
\end{equation}
This implies that in the real space correlation function most of the contributions
will integrate away except for special resonant points where the frequencies in the transform match
those of the bispectrum.  

Without loss of generality, we can set ${\bf r_3}=0$ and integrate out ${\bf k}_3$ in
Eq.~(\ref{eqn:threepoint})
\begin{equation}\label{eqn:fourierxfm}
\xi_{3\curv}({\bf r}_1, {\bf r}_2, 0)  =  \prod_{i = 1}^{2}\int\frac{d^{3}k_{i}}{(2\pi)^3}e^{i{\bf k}_{i}\cdot{\bf r}_i}B_{\cal R}(k_1,k_2,k_3),
\end{equation}
where $k_3$ is now defined by the cosine law.  The integrals are thus sums of
bispectrum triangles defined by the 6-dimensional vector ${\bf K} = ({\bf k}_1,{\bf k}_2)$.
In this 6D vector notation, the phase function
\begin{align}
\phi &\equiv 
{\bf k}_1 \cdot {\bf r}_1 +
{\bf k}_2 \cdot {\bf r}_2 \pm
(k_1+k_2+k_3)\etastep\\
&=
{\bf K}^{\rm T}{\bf R} \pm \left( \sqrt{{\bf K}^{\rm T}\mathbb{1}_{1}{\bf K} } +  \sqrt{{\bf K}^{\rm T}\mathbb{1}_{2}{\bf K} }+ \sqrt{{\bf K}^{\rm T}\mathbb{U}{\bf K} }\right) \eta_{f} ,
\nonumber
\end{align}
where ${\bf R}=({\bf r}_1,{\bf r}_2)$ and we have defined
\begin{align}
\mathbb{1}_{1} &= \left[\begin{array}{cc} \mathbb{1}_{3\times3} & 0 \\ 0 & 0 \end{array}\right], \quad \mathbb{1}_{2} = \left[\begin{array}{cc}0 & 0 \\ 0 &  \mathbb{1}_{3\times3} \end{array}\right], \nonumber \\
 \mathbb{U} &= \left[\begin{array}{cc}\mathbb{1}_{3\times3}  & \mathbb{1}_{3\times3}  \\ \mathbb{1}_{3\times3}  &  \mathbb{1}_{3\times3} \end{array}\right] .
\end{align}
Now, we can look for points where this is stationary with respect to triangle configuration.  Setting $d\phi/d{\bf K} =0$, we obtain
\begin{align}
{\bf R} \pm \left(\frac{\mathbb{1}_{1}{\bf K} }{ \sqrt{{\bf K}^{\rm T}\mathbb{1}_{1}{\bf K} }} + \frac{\mathbb{1}_{2}{\bf K} }{\sqrt{{\bf K}^{\rm T}\mathbb{1}_{2}{\bf K} }}+ \frac{\mathbb{U}{\bf K} }{\sqrt{{\bf K}^{\rm T}\mathbb{U}{\bf K} }}\right) \eta_{f} = 0 ,
\end{align}
which can be written,
\begin{equation}
{\bf r}_i = \mp \etastep( \hat{\bf k}_i - \hat{\bf k}_3 ) ,
\end{equation}
where hats denote unit vectors.    Note that the real space resonances occur for triangles
that are coplanar with the $k$-space triangles.  At these solutions, the phase
function reaches
\begin{align}
\frac{\phi}{\etastep} &= \mp (k_1+k_2)  \pm ({\bf k}_1 +{\bf k}_2) \cdot \hat{\bf k}_3 
\pm (k_1+k_2) \pm k_3 \nonumber\\ &= 0,
\label{eqn:phasesolution}
\end{align}
and so satisfies the constancy condition.

For each $k$-space triangle geometry there is
a resonance in real space triangles that
is fixed by the $k$-space geometry (see Fig.~\ref{fig:resonance}).  
The orientations of ${\bf r}_i$ relative to ${\bf k}_3$ are determined by 
the half angles of the respective $k$-space triangle angles.
 The physical length $r_i$ varies
from a maximum of $2\etastep$, if ${\bf k}_i \parallel -{\bf k_3}$; through 
$\sqrt{2}\etastep$, if they are orthogonal; to $0$, if they are parallel ${\bf k}_i \parallel {\bf k_3}$.  Note, however,
that if $r_1 \rightarrow 0$ then $r_2 \rightarrow 2\etastep$ so that resonant
triangles always have characteristic size $\etastep$.  
The maximum size of the correlation region of $2\etastep$ is achieved for
flat triangles.
Finally, there  is no resonance at ${\bf r}_1={\bf r}_2={\bf r}_3=0$ since 3 coparallel 
$k$-vectors cannot close.

The allowed resonances all correspond to points inscribed on
a sphere of radius $\etastep$ where   rotation in or around the ${\bf k}_3$ direction 
corresponds to a rotation of the real space resonance.

We can also show that these points are extrema rather than unstable saddle points
by considering the Hessian
\begin{align}
H_{ij} = \frac{\partial^{2}\phi}{\partial{K_i}\partial K_j} \,.
\end{align}
A similar but more tedious exercise shows that this matrix is positive semi-definite.
Namely for any vector ${\bf Q}=({{\bf q}_1,{\bf q}_2})$ defining a direction in the 6-dimensional space of bispectrum triangles with a closure relation ${\bf q}_3 = -({\bf q}_1+{\bf q}_2$)
\begin{align}
{\bf Q}^{\rm T} {\bf H} {\bf Q}
= & \sum_{i=1}^3 \frac{\eta_{f}}{k_{i}}\left[ q_{i}^2 - ({\bf q}_{i}\cdot\hat{\bf k}_{i})^{2}\right]
 \ge 0 .
\end{align}
The eigenvalues vanish along the special directions of ${\bf q}_i\parallel {\bf k}_{i}$.  There are two cases where this happens.  The first corresponds to fixing the triangle geometry while changing its overall size.   The second is for the family of flat triangles where
the sides are all coparallel but the ratio of lengths $k_1/k_2$ can vary.  For these triangles
there then two flat directions whereas non-degenerate triangles have only one.
Equivalently, Eq.~(\ref{eqn:phasesolution}) shows for both of these cases,
the phase function is strictly fixed.

Given the behavior of trajectories in the space of triangles, we can estimate the volume of phase space that satisfies the resonance condition, and thus contributes to the integrals.  Except for these special flat directions in the 6-dimensional space, the typical width of the resonances is $\Delta q \sim \sqrt{k/\etastep}$. Putting this together, we can estimate the degree of convergence or divergence of the three-point correlation function  in Eq.~(\ref{eqn:fourierxfm}). 
For a typical real space configuration where ${\bf r}_1 \ne {\bf r}_2$, there is only one flat direction $k_{s}\sim  k_i$ where the triangle size is
 multiplied by some scale factor $s$
\begin{align}\nonumber
\xi_{3\curv}({\bf r}_1, {\bf r}_2, 0)  & \propto \etastep^2 \int dk_s \left(\frac{k_s}{\etastep}\right)^{\frac{5}{2}} k_s^{-4}   \damp\\ &\propto  \frac{1}{\sqrt{k_{\rm max}\etastep}},
\end{align}
where $k_{\rm max} \propto d^{-1}$ from the damping function $\damp$.

On the other hand, for the special configuration where ${\bf r}_1 = {\bf r}_2 = 2\etastep$, there are two flat directions, $k_s$ and $R = k_2/k_1$ which implies the three-point correlation function will scale as
\begin{align}\nonumber\label{eqn:reallnUV}
\xi_{3\curv}({\bf r}_1, {\bf r}_2, 0)  & \propto {\etastep^2}\int k_s dk_s\,  \left(\frac{k_s}{\etastep}\right)^{2}
k_s^{-4} \damp \\ &\propto \ln(
k_{\rm max}\etastep),
\end{align}
where we assume that $k_i > \etastep^{-1}$.

The Hessian also identifies another special case where there is an approximately flat
direction if $k_i \ll \etastep^{-1}$.  This is the case of squeezed triangles where
one of the sides does not contribute significantly to the phase function.  
In this case $k_3 \ll k_2 \approx k_1 \equiv k$
\begin{equation}
\phi =  {\bf k}\cdot ({\bf r}_1-{\bf r}_2) \pm 2 k \etastep,
\end{equation}
for which the stationary solutions are
\begin{equation}
{\bf r}_1-{\bf r}_2 = \pm 2\etastep \hat{\bf k},
\end{equation}
such that there is a resonance whenever the separation between any two points
is $2\etastep$, regardless of the position of the third. 
In this case the dominant term in the bispectrum goes as $\etastep k^{-2} k_3^{-3}$
[see Eq.~(\ref{eqn:squeezedscaling})]  and  
the resonance contains one flat direction in ${\bf k}$ parallel to ${\bf r}_1-{\bf r}_2$  and two orthogonal to 
 of width $\sqrt{k/\etastep}$ each.
Therefore the three-point correlation scales as
\begin{align}\nonumber
\xi_{3\curv}({\bf r}_1, {\bf r}_1+2\etastep \hat {\bf e}, 0)  
&\propto   \etastep \int dk\,  \left(\frac{k}{\etastep}\right) k^{-2} \damp  \int d^3 k_3  k_3^{-3}  \\ 
&\propto \ln(
k_{\rm max}\etastep)\ln (k_{\rm min} \etastep),
\label{eqn:realsqueezed}
\end{align}
where $\hat {\bf e}$ is any unit vector.
Here the correlation depends on a cutoff in the infrared $k_{\rm min}$ which 
in practice comes from the survey volume of the data, i.e.~the current horizon for
a cosmic variance limited statistic.

These results imply that, despite the naive $K^2$ divergence of the reduced bispectrum, 
the three-point correlation at most positions actually converges except if points are separated
by $2\etastep$ where flat and squeezed triangles cause a logarithmic divergence.

The real space picture of the non-Gaussianity of a sharp step raises an interesting
question as to the meaning of the noise term in our signal-to-noise calculation.
In real space, the increasing $(S/N)^2$ at high $k$ is associated with modes that
superimpose to give a sharp correlation feature on large scales.  Thus each triangle cannot be considered
an independent probe of the non-Gaussianity.   By including only the Gaussian contribution 
to the noise, we neglect this sample covariance of triangles.   
Our SNR therefore only quantifies
 detectability of the non-Gaussianity rather than the measurability of
parameters associated with the step potential.   If in the future, non-Gaussianity of this
type is detected, then the covariance must be considered by computing the 6pt functions.

\subsubsection{Configurations}

We can test these properties of the three-point function by explicitly evaluating it for a few 
simple configurations. The expression in Eq.~(\ref{eqn:threepoint}) involve six integrals over a highly oscillatory function, making it difficult to evaluate numerically. However, in the limit in which two of the points are coincident, the expression can be simplified to 
\begin{align}\nonumber
\xi_{3\curv}({ r} ,0,0)  
=& (2\pi)^2\int\frac{k_1 dk_{1}}{(2\pi)^3}\int\frac{k_2 dk_{2}}{(2\pi)^3}\int_{|k_{1}-k_{2}|}^{k_{1}+k_{2}}k_3 dk_{3} \\ &  \times B_{\curv}(k_1,k_2,k_3)\left(\frac{\sin(k_{1}r)}{k_{1}r}+\frac{\sin(k_{2}r)}{k_{2}r}\right).
\label{eqn:threepointzero}
\end{align}
In the limit $d = 0$, using the result  for the bispectrum in \S \ref{sec:GSRzero}, the integral over $k_{3}$ in Eq.~(\ref{eqn:threepointzero}) can be performed analytically (the result is messy and not particularly enlightening, and we omit it here) leaving a more tractable problem of evaluating only two integrals. In order to numerically evaluate the resulting expression, and render the result at $r/\etastep = 2$ finite, we need to impose a cutoff in the integral at high wavenumber. However, the integral needs to be cut off in a way that reflects the fact that triangles of different shapes experience different damping behavior. For non-zero values of $d$, we note that as long as $d$ is small, this result can be utilized to integrate the bispectrum at finite $d$, by parts. We then need only keep the boundary term, since the remaining term will be suppressed by a factor of $d$, and will thus be negligible in the limit $d \rightarrow 0$. Moreover, since this explicitly retains the form of the damping function, we automatically apply the correct damping to each shape.

As well as divergences at the point $r/\etastep = 2$ arising from the ultra-violet parts of the phase space, there are also the usual divergences associated with the infrared parts of the scale invariant power spectra. In reality, these divergences are regulated by the finite size of the longest observable wavelength. To perform the numerical evaluation, we simply apply a hard cutoff on the minimum momenta, $k_{\rm min}$.

\begin{figure}
\centerline{\psfig{file=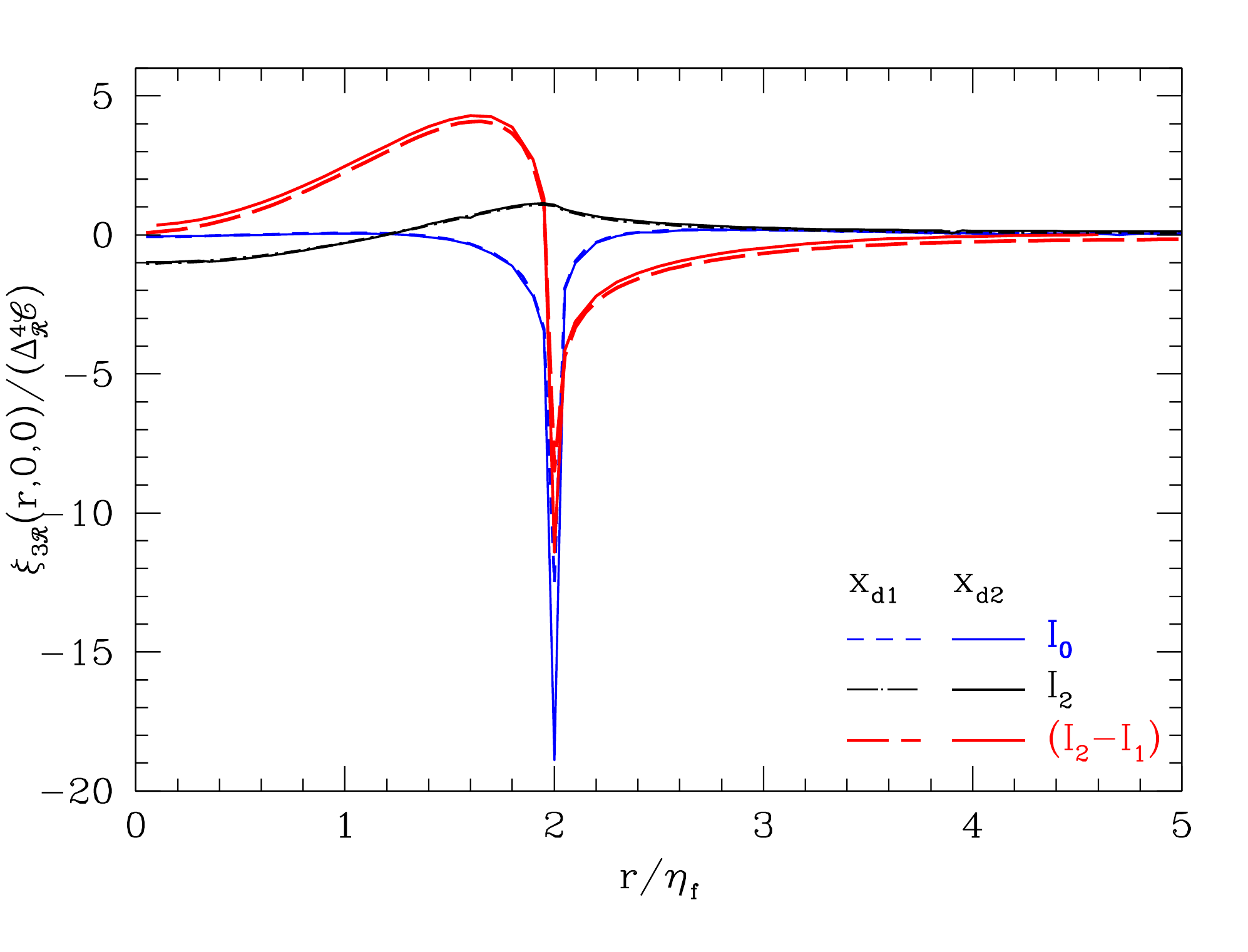, width=3.45in}}
\caption{ Contribution to the real space three-point correlation function from the three terms in Eq.~(\ref{eqn:biapp}). The thick dashed curves were computed for damping scale
  $x_d = x_{d1}  = 200$ with an infrared cutoff at $k_{\rm min} = 0.1/\eta_{f}$, while the corresponding thin solid curves were calculated with $x_{d2}=  1000$. \label{fig:3pt}}
\end{figure}

\begin{figure}
\centerline{\psfig{file=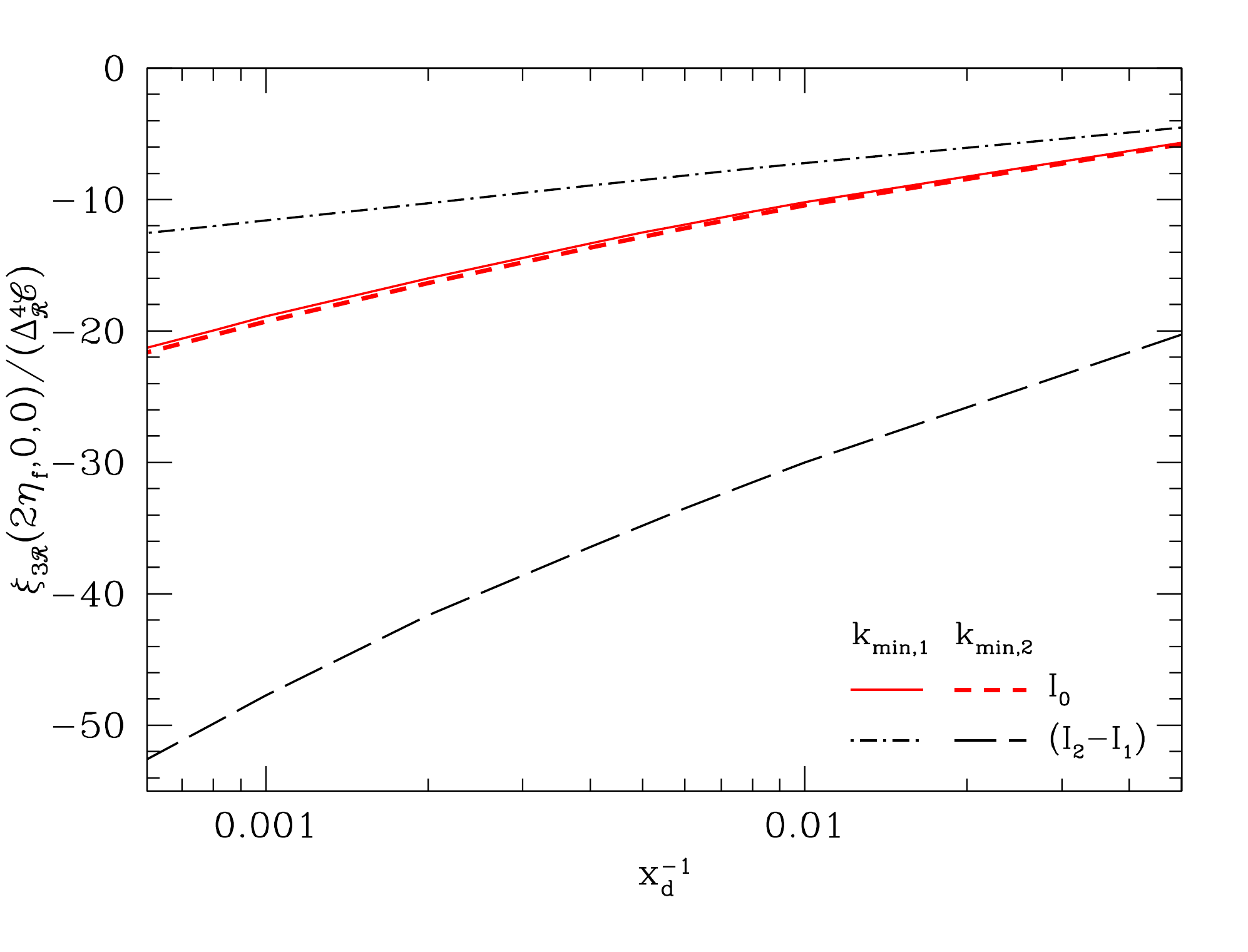, width=3.45in}}
\caption{ Behavior of the various contributions to the real space three-point correlation function $\xi_{3\curv}(r, 0, 0)$ in Eq.~(\ref{eqn:biapp}) at the point $r = 2\etastep$ as a function ultraviolet cutoff $x_{d}$. The dependence on the infrared cutoff is also demonstrated. Here  $k_{\rm min, 1} = 0.1/\eta_{f}$, while $k_{\rm min, 2} = 0.001/\eta_{f}$.
\label{fig:3ptr2behavior}}
\end{figure}

To perform the integration,  we write the bispectrum as
\begin{align}\label{eqn:biapp}
B_{\curv}(k_1,k_2,k_3) &\approx   {(2 \pi)^4  \over k_1^3 k_2^3 k_3^3} {\Delta_{\cal R}(k_1) 
\Delta_{\cal R}(k_2)
\Delta_{\cal R}(k_3)\over 4} \nonumber\\&\;\;
\times\!\!\Big[ -I_0(K) k_1 k_2 k_3 + I_2(K) (k_1^3+k_2^3 + k_3^3) 
\nonumber\\&\quad+(I_2(K)- I_1(K) )\sum_{i \ne j} k_i^2 k_j  \Big].
\end{align}
In Fig.\ \ref{fig:3pt} we show the contribution to the real space three-point correlation function due to each of the terms in Eq.~(\ref{eqn:biapp}) at two different values of the step width, or damping scale $x_{d}$. In particular, note that the skewness of the field  is finite, and small while $\xi_{3\curv}(r, 0, 0)$ vanishes quickly for $r/\etastep > 2$. The three-point function is also largely insensitive to the value of the damping scale, $x_{d}$, which regulates the UV, except at the special point, $r = 2\eta_{f}$, where it diverges. In Fig.\ \ref{fig:3ptr2behavior} we demonstrate that this divergence is no more pathological than $\ln(d)$. Fig. \ref{fig:3ptr2behavior} also demonstrates the scaling behavior of Eqs.\ (\ref{eqn:reallnUV}) and (\ref{eqn:realsqueezed}), note that the contribution to the three-point function due to the most dominant UV term, $I_{0}$, is insensitive to the infrared cutoff, $k_{\rm min}$, while the term which dominates the squeezed contribution, $I_{2} - I_{1}$, is logarithmically sensitive to the infrared cutoff. The remaining term, in Eq.~(\ref{eqn:biapp})  is not UV divergent and we omit it from the plot.

All of these properties and scalings are 
consistent with our resonance estimates from \S \ref{sec:resonances}.

\subsection{Real space consistency} \label{App:squeezed}

To shed light on why the two-point correlation reaches infinite slope at $r=2\etastep$ while
the three-point correlation logarithmically diverges there, we can extend the
familiar $k$-space consistency relation between the power spectrum
and squeezed bispectrum to real space correlation functions.

In fact the $k$-space consistency argument itself comes from a real space derivation. We follow the analysis of \cite{Creminelli:2004yq} here but make more explicit what is meant by the background. To this end, we split the field up into  low pass and high pass filtered pieces
\begin{align}
\curv({\bf x}) = & \bar\curv + \tilde\curv\\
= & \int_{k < k_{*}} \frac{d^{3}k}{(2\pi)^3} \curv_{\bf k} e^{i{\bf k}\cdot {\bf x}}+ \int_{k > k_{*}} \frac{d^{3}k}{(2\pi)^3} \curv_{\bf k}  e^{i{\bf k}\cdot {\bf x}}. \nonumber
\end{align}
That is, we split the contributions of the field at each point into contributions from modes shorter than some reference mode $k_{*}$, and long wavelength modes assumed to be much longer than this scale.  Next we consider the two-point function of the high pass field
in the fixed background. 
Since the background fluctuation is expected to be small, we can functionally expand
\begin{align}
\label{eqn:twoptresponse}
\langle\tilde\curv({\bf x}_{1})\tilde\curv({\bf x}_{2})\rangle  =&  \langle\tilde\curv({\bf x}_{1})\tilde\curv({\bf x}_{2})\rangle  \Big|_{\bar \curv = 0} \\ &  
+\bar \curv ({\bf x}_+) \frac{\delta}{\delta\bar\curv}\langle\tilde\curv({\bf x}_{1})\tilde\curv({\bf x}_{2})\rangle \Big|_{\bar \curv = 0} +\ldots,\nonumber
\end{align}
where the background field is slowly varying and
 evaluated in the vicinity of the points, which for definiteness we take 
 to be ${\bf x}_{+}=({\bf x}_1+{\bf x}_2)/2$.
  In the presence of the background mode, the metric is
\begin{align}
ds^2  = -dt^2 + a(t)^2e^{2\bar\curv({\bf x})}dx^2.
\end{align}
 Then, absorbing the effect of the background mode into a change in variable ${\bf x} \rightarrow {\bf x}' = e^{\bar\curv({\bf x})} {\bf x}$, we induce a change in the separation 
 between points
\begin{align}
\Delta x \rightarrow e^{\bar\curv({\bf x})}\Delta x .
\end{align}
Defining the two-point correlation function of the high and low pass fields
\begin{align}
\tilde \xi_{2\curv}(\Delta x_{12}) & = \langle\tilde\curv({\bf x}_{1})\tilde\curv({\bf x}_{2})\rangle \Big|_{\bar \curv = 0}  \nonumber\\
&= \int_{k>k_*} \frac{dk}{k}\Delta_{\curv}^2(k) \frac{ \sin{k \Delta x_{12}}}{k \Delta x_{12}},\\
 \bar\xi_{2\curv }(\Delta x_{+3})&= \langle \bar\curv({\bf x}_{+})\bar \curv({\bf x}_{3})\rangle \nonumber\\
&= \int_{k<k_*} \frac{dk}{k}\Delta_{\curv}^2(k) \frac{ \sin{k \Delta x_{+3}}}{k\Delta x_{+3}},
\end{align}
where $\Delta x_{ij} = |{\bf x}_i - {\bf x}_j|$,
we obtain
\begin{equation}
 \langle\tilde\curv({\bf x}_{1})\tilde\curv({\bf x}_{2})\rangle
 = \tilde \xi_{2\curv}(\Delta x_{12}) + \bar\curv({\bf x}_{+})  \frac{d\tilde \xi_{2\curv} }{d\ln \Delta x_{12}},
 \end{equation}
 which then correlates with a third point as
\begin{align}\nonumber
\big\langle \tilde\curv({\bf x}_{1})\tilde\curv({\bf x}_{2})\bar \curv({\bf x}_{3})\big\rangle =
\bar \xi_{2\curv }(\Delta x_{+3})
  \frac{d\tilde \xi_{2\curv} }{d\ln \Delta x_{12}} \,.
\end{align}
This defines the squeezed contribution to the analogous three-point function.  

Note that we can Fourier transform  this relation to obtain the bispectrum.  Given that
through integration by parts
\begin{equation}
 \frac{d\tilde \xi_{2\curv} }{d\ln r} = -\int_{k>k_*} {d^3 k \over (2\pi)^3 } e^{i{\bf k}\cdot {\bf r}}
 \left[  \frac{1}{k^3} {d( k^3 P_\curv) \over d\ln k} \right],
 \end{equation}
we obtain immediately,
\begin{equation} 
B_{\curv}(k_1,k_2,k_3) =  - P_\curv(k_{3}) \left[  \frac{1}{k^3} {d( k^3 P_\curv) \over d\ln k} \right]_{k \approx k_1\approx k_2},
\end{equation}
for 
$k_3 < k_*, k_1\approx k_2> k_*$ and zero otherwise. This is the well-known Fourier space consistency relation.

\begin{figure}
\centerline{\psfig{file=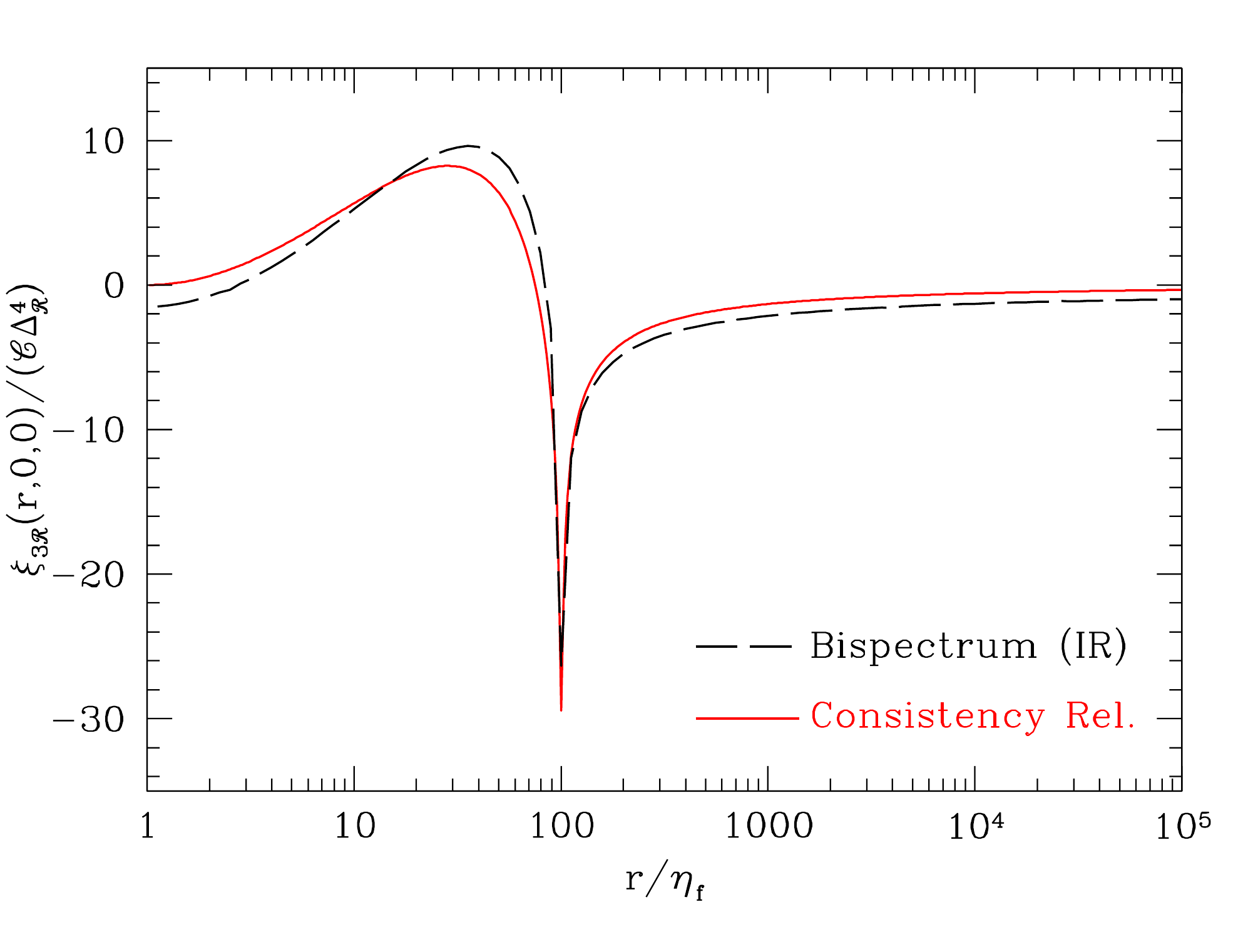, width=3.45in}}
\caption{ The contribution to the real space three-point correlation function from the infra-red terms (the second two terms) in Eq.~(\ref{eqn:biapp}) (dashed black curve) and the consistency relation of Eq.~(\ref{eqn:powersqueezed}) (red, solid). We take $x_d=100$ with an infrared cutoff at $k_{\rm min} = 0.001/\eta_{f}$. \label{fig:3ptsqueezed}}
\end{figure}

Thus we expect the three-point correlation in real space to have a contribution proportional to the slope of the two-point correlation function.
For the step potential, the slope of the two-point correlation function  at $r=2\etastep$ diverges
as  $\ln k_{\rm max}$ whereas
$\bar\xi_{2\curv}(\Delta x_{+3})$ diverges as $\ln k_{\rm min}$.
Thus, with a sufficiently small $k_{\rm min}$, we expect that the three full three-point
function will be dominated by these squeezed contributions.  Specifically, 
including the cyclic permutations which account for the pairings of high and low pass
filtered fields we expect
\begin{equation}\label{eqn:powersqueezed}
\xi_{3\curv}(r,0,0) \sim  -2 \ln(k_{\rm min}\etastep) \Delta_{\curv,0}^2 \frac{d \xi_{2\curv}}{d\ln r} ,
\end{equation}
near $r=2\etastep$ for nearly scale invariant spectra taking $k_*$ comparable to but smaller than $\etastep^{-1}$.   This explains the scaling of Eq.~(\ref{eqn:realsqueezed}). In Fig.\ \ref{fig:3ptsqueezed}  we demonstrate this relation numerically. Plotted is the contribution to the real space three-point correlation function due to the $I_{1}$ and $I_{2}$ terms of Eq.~(\ref{eqn:gsrlbi}) and the relation in Eq.~(\ref{eqn:powersqueezed}). We take $k_{\rm min}$ to be a factor of $10^{3}$ smaller than the feature scale.  The agreement is only approximate, however, this is possibly due to the fact that we have not omitted  the contribution of non-squeezed triangles to $\xi_{3\curv}$.

The $\ln k_{\rm min}$ divergence is the usual IR divergence of a scale invariant spectrum 
and is cured in any physical observable by the finite size of Hubble radius today.   
The UV divergence from the slope of the two-point function at $r=2\etastep$ however weak, signals a breakdown of perturbation theory.   
In the consistency relation context,  the assumed
change in the two-point correlation in response to the background mode in Eq.~(\ref{eqn:twoptresponse}) exceeds the total
change in the two-point correlation as a function of $r$.
This is consistent with the findings of \S \ref{sec:GSRzero}, where it was shown that perturbative validity
places a weak constraint of $c/d < 10^{4}$ such that a finite step cannot have infinitesimal
width.   

\section{Slow Roll Corrections to the Bispectrum}\label{app:slowrollcorrections}

In this Appendix we show that one of the largest and simplest corrections to the bispectrum can be associated with the difference in phase between the slow roll spacetime, and de Sitter space. This phase difference leads to a real component of the growing mode on super horizon scales. We show that these slow roll corrections are largely responsible for the 10\% error in the approximations in \S \ref{sec:bispectrumformal}.

In deriving the expression for the bispectrum, Eq.~(\ref{eqn:gsrlbi}), we used only the de Sitter form for the mode function $y(x)$, which is related to the curvature perturbation by
\begin{align}\label{eqn:curvpert}
\curv_k = \sqrt{\frac{2\pi^2}{k^3}}\frac{x y(x)}{f}.
\end{align}
Here $x = k\eta$ and $y(x)$ is the solution of the Mukhanov-Sasaki equation,
\begin{align} \label{eqn:MSeqn}
\frac{d^2 y}{dx^2}+\left(1-\frac{2+g(\ln x)}{x^2}\right)y = 0,
\end{align}
where
\begin{align}
g(\ln x) = \frac{f'' - 3f'}{f},
\end{align}
and primes denote derivatives with respect to $\ln x$. In the de Sitter limit, $g = 0$, and
\begin{align}
y(x) = \left(1+\frac{i}{x}\right)e^{ix}.
\end{align} 
In this approximation, for modes that are outside the horizon, $x \ll 1$, the growing mode of $y$ which yields the constant part of the curvature $\curv$ is purely imaginary, while the real component decays as $x^2$ in this limit. Writing the bispectrum in Eq.~(\ref{eqn:bispectrum1st}) as
\begin{align}\nonumber\label{eqn:bispectrumDom}
& B_{\curv}(k_1,k_2,k_3) \approx  -4 \Im\Bigg[ \mathcal{R}_{k_{1}}(\eta_{\sh})\mathcal{R}_{k_{2}}(\eta_{\sh})\mathcal{R}_{k_{3}}(\eta_{\sh})\Bigg]\\\nonumber
&\times \Re\Bigg[ \int_{\eta_{\sh}}^{\infty} {d\eta \over \eta^2}\,  {a^{2} \epsilon_{H}}(\epsilon_{H} - \eta_{H})' (\mathcal{R}^{*}_{k_{1}}\mathcal{R}^{*}_{ k_{2}}\mathcal{R}^{*}_{ k_{3}})'
 \Bigg] \\\nonumber  
 &  -4 \Re\Bigg[ \mathcal{R}_{k_{1}}(\eta_{\sh})\mathcal{R}_{k_{2}}(\eta_{\sh})\mathcal{R}_{k_{3}}(\eta_{\sh})\Bigg]\\
&\times \Im\Bigg[ \int_{\eta_{\sh}}^{\infty} {d\eta \over \eta^2}\,  {a^{2} \epsilon_{H}}(\epsilon_{H} - \eta_{H})' (\mathcal{R}^{*}_{k_{1}}\mathcal{R}^{*}_{ k_{2}}\mathcal{R}^{*}_{ k_{3}})'
 \Bigg], 
\end{align}
we expect that the bispectrum is dominated by the first term, whereas the second term should be of order $\mathcal{O}(\epsilon_H)$. Since $\epsilon_H \sim 0.01$, one may expect that neglecting the slow roll corrections amounts to a small percent level correction to the leading order bispectrum. However, as we will see the coefficient in front of these terms is such that the correction is on the order of 10\% (see also \cite{Burrage:2011hd}).

In the limit where the inflaton is slowly rolling down its potential, the function $g(\ln x)$ is smooth, slowly varying and  $\mathcal{O}(n_{s}-1)$. The scalar fluctuations, governed by Eq.~(\ref{eqn:MSeqn}), begin as plane waves $y\propto e^{i x}$ in Minkowski space at $x \gg \infty$ before they begin to feel the expansion when their wavelength becomes comparable to the horizon, $x\sim 1$, where they begin to grow as $ y \propto 1/x$ for $x \ll 1$. As long as $g$ is slowly varying, the only time it is important is when the mode crosses the horizon. Thus for a given wavenumber $k$, to a good approximation, we can Taylor expand about the time of horizon crossing, and to a first approximation, write
\begin{align}\label{eqn:MSconst}
\frac{d^2 y}{dx^2}+\left(1-\frac{2+g_0}{x^2}\right)y = 0,
\end{align}
where $g_{0} = g(x = 1)$, is a function of $k$, but is time independent. In this limit, Eq.\ \ref{eqn:MSconst} can be solved exactly. Demanding that $y$ is asymptotically a positive frequency plane wave as $x\to\infty$ implies that
\begin{align}
y(x) = -e^{-i\frac{\pi \nu}{2}+i\frac{3\pi}{2}} \sqrt{\frac{\pi x}{2}}H^{(1)}_{\nu}(x),
\end{align}
where $H^{(1)}_{\nu}(x)$ is the Hankel function of the first kind and 
\begin{align}
\nu = \frac{1}{2}\sqrt{9+4g_{0}} \approx \frac{3}{2}+ \frac{g_{0}}{3}.
\end{align}
Now, as described above in \S \ref{sec:bispectrumformal}, the bispectrum due to a step feature in the potential is dominated by modes that are deep inside the horizon when the inflaton crosses the feature. At early times, the modes are identical to the de Sitter modes, at leading order in $x$, while at late times, expanding the Hankel function one finds
\begin{align}\nonumber\label{eqn:hankelcorrection}
x\, y(x)& \to  i e^{-i\frac{\pi \nu}{2}+i\frac{3\pi}{2}}\Gamma(\nu)\sqrt{\frac{2}{\pi }} \left(\frac{2}{x}\right)^{\nu - \frac{3}{2}} \\ \nonumber
&\approx  i + i\frac{[2-\gamma-\ln(2)-\ln(x)]}{3}g_{0} +\frac{\pi g_{0}}{6} \\ &\quad +\mathcal{O}(x^{2}, g_{0}^2).
\end{align}
The growing mode on superhorizon scales is no longer purely imaginary. There is also a correction to the amplitude of the imaginary part of the growing mode, however, {note that we have already taken this correction into account by using the square root of Eq.~(\ref{eqn:powerintegral}) in Eq.~(\ref{eqn:gsrlbi}) which contains  slow roll corrections in $\Delta_{\curv, 0}^2$}.
 In this approximation, in the superhorizon limit, $xy(x)$ is not only not constant but
diverges as $\ln(x)$. However, it is easy to see that the curvature perturbation, defined in Eq.~(\ref{eqn:curvpert}), is actually constant in this limit. The assumption that $g = {\rm const.}$ implies that the function $f$ has time dependence 
\begin{align}
f  \approx f_{0} \left[ 1-{\frac{g_{0}}{3}}\ln\left(x\right)\right] +\mathcal{O}(g_{0}^2),
\end{align}
to leading order in $g_{0}$, which is precisely the right behavior to cancel out the time dependence of Eq.~(\ref{eqn:hankelcorrection}).

In the limit of large momenta, corresponding to modes well inside the horizon, the leading order (in $k$ and slow roll) correction to the bispectrum is given by
\begin{align}\label{eqn:1stOcorrection}\nonumber
B_{\curv}^{(1)}(k_{1}, k_{2}, k_{3}) \approx& -\frac{(2\pi)^4}{k_{1}^2k_{2}^2k_{3}^2}\frac{\Delta_\curv(k_{1})\Delta_\curv(k_{2})\Delta_\curv(k_{3})}{4}\\ &\qquad\times\frac{\pi}{2}g_{0} I_{3}(K),
\end{align}
where
\begin{align}
I_{3}(K) = \int_{\eta_{*}}^{\infty} \frac{d\eta}{\eta}G'_{B}(\ln\eta)K\eta \cos(K\eta).
\end{align}
In the same approximation as \S\ref{sec:bispectrumformal}, we can evaluate this integral, to obtain
\begin{align}
I_{3}(K) = \frac{\mathcal{C}}{2f_{0}}\damp\left(\frac{\pi d}{2\sqrt{2\epsilonstep}}K\etastep\right)X_{3}(K\etastep),
\end{align}
where
\begin{align}\nonumber\label{eqn:x3}
X_{3}(K) = &-\frac{ \left(x^4-9 x^2+54\right) \sin (x)}{x^2}\\ &+\frac{\left(-2 x^4+27 x^2-54\right) \cos (x)}{x^3}+\frac{54}{x^3}.
\end{align}
Relative to the dominant zeroth order term (the $X_{0}$ term of Eq.~(\ref{eqn:biwindows})) the main contribution from this slow roll correction is $\pi/2$ out of phase, but has the same envelope. Notice also that, in the limit $x\to 0$, this window function vanishes only as $X_3(x) \sim \mathcal{O}(x)$, compared to $ \mathcal{O}(x^2)$ for the window functions of Eq.~(\ref{eqn:biwindows}). This implies that, at sufficiently small $x$, this correction will eventually dominate the leading order result. This is not particularly surprising given that we have only considered the dominant correction in the high momentum limit. Near the horizon scale at feature crossing, $k\etastep = 1$, many other terms become equally, if not more important.

\begin{figure}[t]
\centerline{\psfig{file=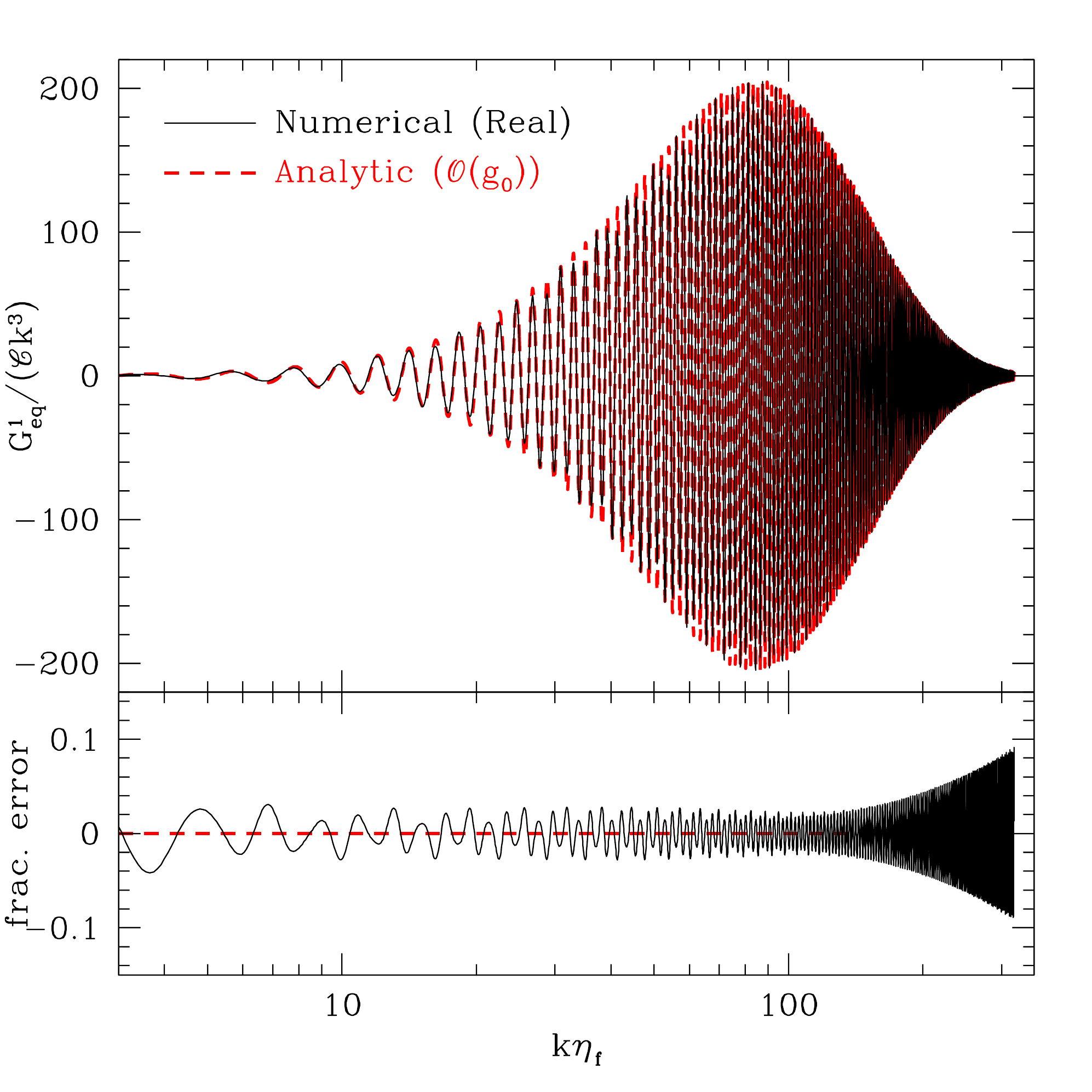, width=3.45in}}
\caption{\footnotesize
First order corrections to the equilateral bispectrum computed using the analytic approximation Eq.~(\ref{eqn:1stOcorrection}) and (\ref{eqn:x3}) compared with the numerical
evaluation of the second term in Eq.~(\ref{eqn:bispectrumDom}) for a step with height $c = 10^{-5}$ and width $d = 0.001$. The lower panel here shows the fractional error of the full bispectrum, once this first order correction is taken into account. We take $\pi g_{0}/2 = 0.1$ for this plot.}
\label{fig:cem5imaginary}
\end{figure}

The upper panel of Fig.\ \ref{fig:cem5imaginary} shows a comparison between the analytic approximation to this first order term, Eqs.\ (\ref{eqn:1stOcorrection}) and (\ref{eqn:x3}) and the second term in Eq.~(\ref{eqn:bispectrumDom}) for a step with height $c = 10^{-5}$ and width $d = 0.001$.  {The prefactor for this term varies slowly over the range of wavenumbers considered, and we take the value $\pi g_{0}/2 = 0.1$ which is near the middle of its range for this plot.} The lower panel in Fig.\ \ref{fig:cem5imaginary} shows the fractional error on the full bispectrum when this term is added to the analytic approximation of \S \ref{sec:bispectrumformal}. While the approximation is improved significantly at large $k$, the improvement gets steadily worse as the horizon scale is approached.   This is to be expected, given that we have not attempted to calculate all of the terms that are important for modes that are near the horizon as the feature is crossed. 

We also note that these corrections are important for the power spectrum of \S\ref{sec:powerspec} \cite{Dvorkin:2009ne}. In particular, note that the squeezed limit consistency relation (see \S \ref{sec:ksqueezedlimit}) implies that there should be a correction to the power spectrum of the same order of magnitude. We leave the calculation of further slow roll corrections to future work.

Before we end this section, we clarify one point that may confuse the reader. In deriving the correction Eq.~(\ref{eqn:hankelcorrection}), at first glance it appears that the prefactor is simply an irrelevant global phase that will simply cancel once all terms are taken into account. Indeed, the bispectrum is constructed out of the two-point functions which always involve the pair 
\begin{align}
\langle \hat\curv_{\bf k}(\eta)\hat\curv_{{\bf k}'}(\tilde\eta)\rangle = \curv_{k}(\eta)\curv^{*}_{k}(\tilde\eta)(2\pi)^3\delta^{3}({\bf k}+{\bf k}').
\end{align}
However, if one were to choose to cancel the phase in each propagator, one would of course obtain the same answer. While the growing mode would be purely imaginary outside the horizon, at very early times the mode would look as if it initially started with a phase shift relative to the de Sitter space modes, leading to the same correction we have calculated here.

\section{Separability}\label{app:separability}

Separability of the bispectrum into products of functions that depend only on the individual
$k_i$'s is desirable in that the angular bispectrum can then also be constructed from 
separate $\ell_i$ calculations using the full radiation transfer function instead of the
flat-sky Sachs-Wolfe expressions used in the main text.  

The formulation of the eigenmode decompositions of Fergusson, Liguori and Shellard \cite{Fergusson:2009nv, Fergusson:2010dm} which allow for the projection of nonseparable bispectra onto a (band-limited) complete basis of separable functions has largely ameliorated the need for the bispectra themselves to be separable. However, for a polynomial basis of degree $n$, one can only accurately fit a function with $n$ zero crossings, and thus, it seems that projection of the highly oscillatory functions considered here might prove to be very inefficient with this method. Polynomials, however, are not the only such basis that has been proposed, the oscillatory basis of Meerburg \cite{Meerburg:2010ks} would perhaps be more suited for this type of bispectra. Nonetheless, in this Appendix, we point out that the analytic form of the bispectrum derived above, is to a very good approximation, separable.

The bispectrum from Eqs.~(\ref{eqn:gsrlbi}) and (\ref{eqn:bispecintegrals})-(\ref{eqn:biwindows}) appears to be inseparable due to the fact that the damping factor appears to depend on the perimeter of the triangle in momentum space
\begin{equation}
\damp(x) = {x \over \sinh x}, \qquad x= \sum_{i=1}^3 x_i,
\end{equation}
where $x_i = k_i\etastep (\pi d /2\sqrt{2\epsilonstep})$.  This cannot be exactly written in
the form
\begin{equation}
\damp(x) = \prod_{i=1}^3 \damp_i(x_i).
\end{equation}
We can, however, approximate the damping factor in this form.   First note that at $x\gg 1$
\begin{equation}
\lim_{x \gg 1} \damp(x) =  2 x e^{-x}(1+ e^{-2x}),
\end{equation}
which is separable.
In the opposite limit, the damping factor vanishes as $x^2$ whereas the expansion goes as
$4 x$.   To fix this problem we take
\begin{equation}
\damp(x) \approx  2 x e^{-x}(1+ e^{-2x}) + e^{-4 x},
\end{equation}
which is of the form of a sum of separable components since $e^{-x} = \prod_i e^{-x_i}$.
The remaining inseparable  $1/K^n$ type factors that appear in the window functions can be written in separable form by introducing Schwinger parameters as described in \cite{Smith:2006ud}. However, we note that the terms that require this treatment are subdominant, and to a reasonable approximation away from $K = 0$, the bispectrum is dominated by terms that are separable.

\bibliography{smalldbispectrum}

\end{document}